\documentclass[aps,prb,reprint,superscriptaddress,citeautoscript
]{revtex4-2}
\setcitestyle{super}
\usepackage{graphicx}
\usepackage{amsmath,amsfonts,amssymb}
\usepackage{xcolor}
\usepackage{epstopdf}
\usepackage[retain-explicit-plus]{siunitx}
\usepackage[version=4]{mhchem}
\usepackage{graphicx}
\usepackage{subfigure}
\usepackage{braket}
\usepackage{booktabs}
\usepackage{hyperref}
\hypersetup{
    colorlinks=true,
    linkcolor=blue,
    filecolor=magenta,      
    urlcolor=blue,
    citecolor=blue
    }

\renewcommand{\epsilon}{\varepsilon}
\renewcommand{\emph}{\textit}
\definecolor{alizarin}{rgb}{0.82, 0.1, 0.26} 

\draft 

\begin{document}


\title
  {Is Cu$_{3-x}$P a semiconductor, a metal, or a semimetal?}
  
  
  



\author{Andrea Crovetto}
\email[]{Electronic mail: ancro@dtu.dk}
\affiliation{Materials Science Center, National Renewable Energy Laboratory, Golden, Colorado 80401, United States}
\affiliation{Department of Structure and Dynamics of Energy Materials, Helmholtz-Zentrum Berlin f\"ur Materialien und Energie GmbH, 14019 Berlin, Germany}
\affiliation{National Centre for Nano Fabrication and Characterization (DTU Nanolab), Technical University of Denmark, 2800 Kongens Lyngby, Denmark}

\author{Thomas Unold}
\affiliation{Department of Structure and Dynamics of Energy Materials, Helmholtz-Zentrum Berlin f\"ur Materialien und Energie GmbH, 14019 Berlin, Germany}

\author{Andriy Zakutayev}
\email[]{Electronic mail: andriy.zakutayev@nrel.gov}
\affiliation{Materials Science Center, National Renewable Energy Laboratory, Golden, Colorado 80401, United States}

\begin{abstract}
Despite the recent surge in interest in Cu$_{3-x}$P for catalysis, batteries, and plasmonics, the electronic nature of Cu$_{3-x}$P remains unclear. Some studies have shown evidence of semiconducting behavior, whereas others have argued that Cu$_{3-x}$P is a metallic compound.
Here, we attempt to resolve this dilemma on the basis of combinatorial thin-film experiments, electronic structure calculations, and semiclassical Boltzmann transport theory. We find strong evidence that stoichiometric, defect-free Cu$_3$P is an intrinsic semimetal, i.e., a material with a small overlap between the valence and the conduction band. On the other hand, experimentally realizable Cu$_{3-x}$P films are always p-type semimetals natively doped by copper vacancies regardless of $x$. It is not implausible that Cu$_{3-x}$P samples with very small characteristic sizes (such as small nanoparticles) are semiconductors due to quantum confinement effects that result in opening of a band gap.
We observe high hole mobilities (\SI{276}{cm^2/Vs}) in Cu$_{3-x}$P films at low temperatures, pointing to low ionized impurity scattering rates in spite of a high doping density.
We report an optical effect equivalent to the Burstein-Moss shift, and we assign an infrared absorption peak to bulk interband transitions rather than to a surface plasmon resonance. From a materials processing perspective, this study demonstrates the suitability of reactive sputter deposition for detailed high-throughput studies of emerging metal phosphides.


\end{abstract}


\maketitle 

\section{Introduction}

Most binary compounds consisting of Cu(I) and a moderately electronegative element are semiconductors of high interest for optoelectronics and/or thermoelectrics. Cu$_2$O-based solar cells have reached photovoltaic efficiencies close to 10\%.~\cite{Minami2016} Cu$_2$O itself is the host of the largest excitonic wavefunctions ever discovered.~\cite{Kazimierczuk2014}
Cu$_2$S was one of the prominent thin-film solar cell materials in the 1970s~\cite{Hall1979} and later regained prominence in nanoparticle form as a semiconductor system with localized surface plasmon resonances in the near infrared.~\cite{Luther2011} Cu$_2$Se and Cu$_2$Te are among the best p-type thermoelectric materials known today,~\cite{Liu2012a,Zhao2019} with record $zT$ values above 1 at high temperatures.
CuI is currently the p-type transparent conductor with the highest figure of merit~\cite{Yang2016a,Crovetto2020d} and one of the highest-performing transparent thermoelectric materials~\cite{Yang2017b}. Cu$_3$N is a defect-tolerant semiconductor~\cite{Zakutayev2014} with an ideal band gap for photovoltaics.~\cite{Caskey2014} CuP$_2$ is an unconventional semiconductor containing P-P bonds, with optimal band gap and doping density for photovoltaics.~\cite{Crovetto2022b}

Cu$_3$P is another copper(I) phosphide, which has been synthesized in the form of single crystals,~\cite{Olofsson1972,Robertson1980} powders,~\cite{Juza1956,AnnAitken2005,Wolff2016,Wolff2018} thin films,~\cite{Fritz1992,Read2016,Pawar2019,Kuwano2021,Lee2021} and especially nanoparticles and other nanostructured forms.~\cite{Manna2013,Tian2014,DeTrizio2015,Hou2016,Yue2016,Mu2020,Fu2021a} Cu$_3$P is usually found to be strongly p-type due to substantial Cu deficiency,~\cite{Olofsson1972,Wolff2016} and is therefore often referred to as Cu$_{3-x}$P. In recent years, Cu$_{3-x}$P has been the subject of intense research as an electro- and photocatalyst for various reactions,~\cite{Tian2014,Hou2016,Read2016,Yue2016,Pawar2019,Fu2021a} as well as a battery anode.~\cite{Stan2013,Liu2016d} However, the current understanding of the electrical and optical properties of Cu$_{3-x}$P is surprisingly poor. Various authors have identified Cu$_{3-x}$P as a metal,~\cite{Robertson1980,Wolff2018,Kuwano2021} but many others have classified it as a semiconductor instead~\cite{AnnAitken2005,Manna2013,DeTrizio2015,Yue2016,Mu2020,Fu2021a,Peng2021}
and one study labeled it a semimetal.~\cite{Hou2016}
Device-level applications of Cu$_{3-x}$P as a semiconducting photovoltaic absorber~\cite{Manna2013} and as a metallic contact~\cite{Kuwano2021} have both been claimed. Furthermore, it is not clear if the electrical properties of Cu$_{3-x}$P are modified by changing the composition, and in particular whether Cu$_{3-x}$P can be doped n-type by non-stoichiometry. The single-phase region of Cu$_{3-x}$P has only been determined for bulk samples,~\cite{Olofsson1972,Wolff2016,Wolff2018} with different results obtained in different studies. Finally, the effect of growth temperature and other parameters on the properties of Cu$_{3-x}$P has not been investigated.


Polycrystalline Cu$_{3-x}$P thin films are an ideal platform to answer these questions. Unlike the case of nanoparticles, the properties of thin films are not affected by quantum confinement, and electrical and optical properties can be measured more precisely. Compared to single crystals and bulk powders, an advantage of thin films is that their properties can rapidly be characterized as a function of elemental composition and process conditions using high-throughput methods.~\cite{Talley2019,Talley2021a,Zakutayev2018} Previous thin-film work on Cu$_{3-x}$P is, however, limited to non-reactive sputter deposition from a Cu$_{3-x}$P target,~\cite{Kuwano2021} chemical vapor deposition,~\cite{Read2016}
electrodeposition~\cite{Fritz1992,Pawar2019}, and phosphorization of metallic Cu~\cite{Lee2021} with only basic materials characterization. A deposition technique amenable to high-throughput experiments is reactive RF sputtering in a PH$_3$-containing atmosphere. Although this technique has rarely been applied to deposit metal phosphides, we have recently shown its feasibility for various phosphide compounds of current interest.~\cite{Schnepf2021,Crovetto2022,Willis2022,Crovetto2022b}


In this work, we combine high-throughput experiments on reactively sputtered Cu$_{3-x}$P, first-principles calculations, and semiclassical transport theory. The goal is to investigate the electronic nature of Cu$_{3-x}$P films as a function of composition and growth conditions. 
We present strong evidence that Cu$_{3-x}$P is a p-type semimetal natively doped by Cu vacancies regardless of overall composition and growth conditions.
Its single-phase region is likely limited to a narrow range of Cu-deficient compositions. We find that the density of Cu vacancies increases with deposition temperature.
We observe an anisotropic electrical conductivity in excellent agreement with the anisotropy of the hole effective masses.
Finally, we identify a near-infrared absorption feature somewhat similar to a localized surface plasmon resonance previously reported in Cu$_{3-x}$P nanoparticles. In thin-film samples, this peak is likely to be an intrinsic feature of bulk Cu$_{3-x}$P due to interband transitions.


\begin{figure}[t!]
\centering%
\includegraphics[width=\columnwidth]{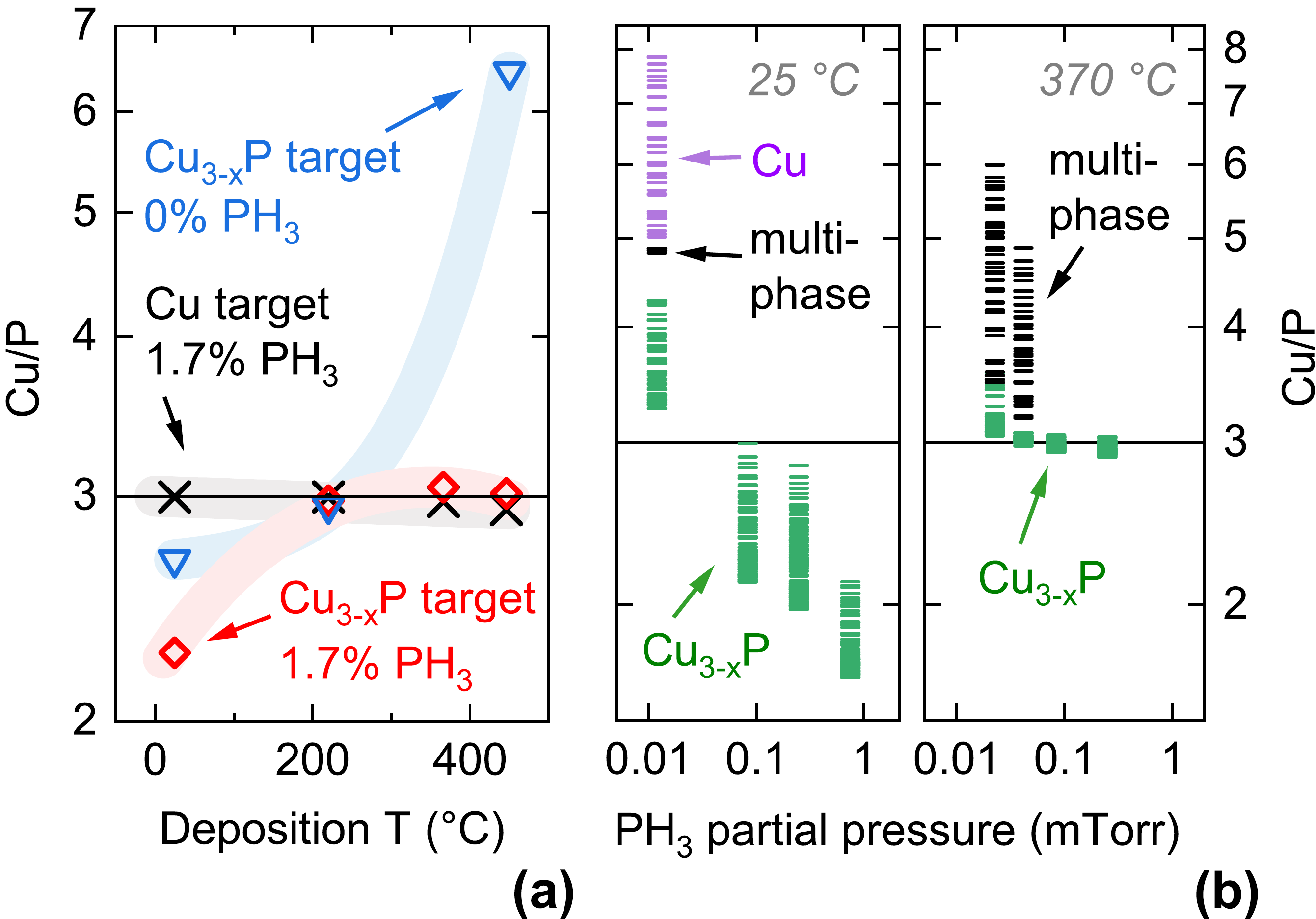}
\caption{Composition and phase analysis of Cu$_{3-x}$P films under different deposition conditions. \textbf{(a):} Film composition versus deposition temperature. Each data set is indicated by a guiding line. \textbf{(b):} Combinatorial film compositions obtained at different PH$_3$ partial pressures by co-sputtering a Cu and a Cu$_{3-x}$P target. Left panel: Room temperature deposition. Right panel: \SI{370}{\celsius} deposition. Data points are colored green, purple, and black when their corresponding XRD patterns contain Cu$_{3-x}$P peaks only, Cu peaks only, and both types of peaks, respectively.}
\label{fig:composition}
\end{figure}

\begin{figure}[h!]
\centering%
\includegraphics[width=\columnwidth]{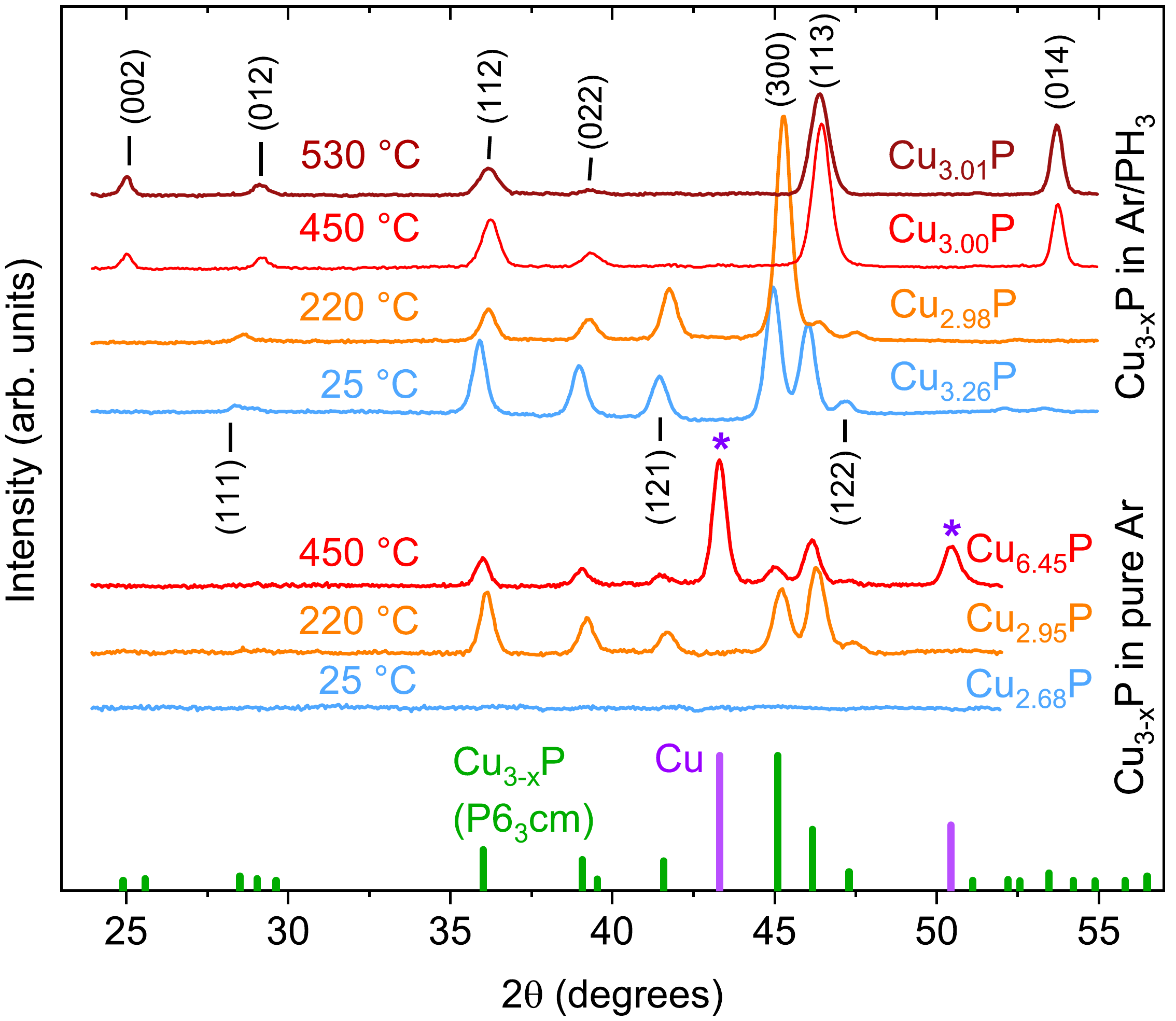}
\caption{XRD patterns of Cu$_{3-x}$P films deposited in pure Ar and in a PH$_3$-containing atmosphere. Film compositions, deposition temperatures, and reference peaks from powder samples~\cite{Olofsson1972} are indicated. At room temperature, the inclusion of PH$_3$ yields polycrystalline (rather than amorphous) films. Including PH$_3$ also prevents P losses at high temperatures.}
\label{fig:xrd}
\end{figure}


\section{Results}
\subsection{Growth, composition, and structure}
\subsubsection{Growth routes and composition} \label{sec:composition}
\begin{figure}[t!]
\centering%
\includegraphics[width=\columnwidth]{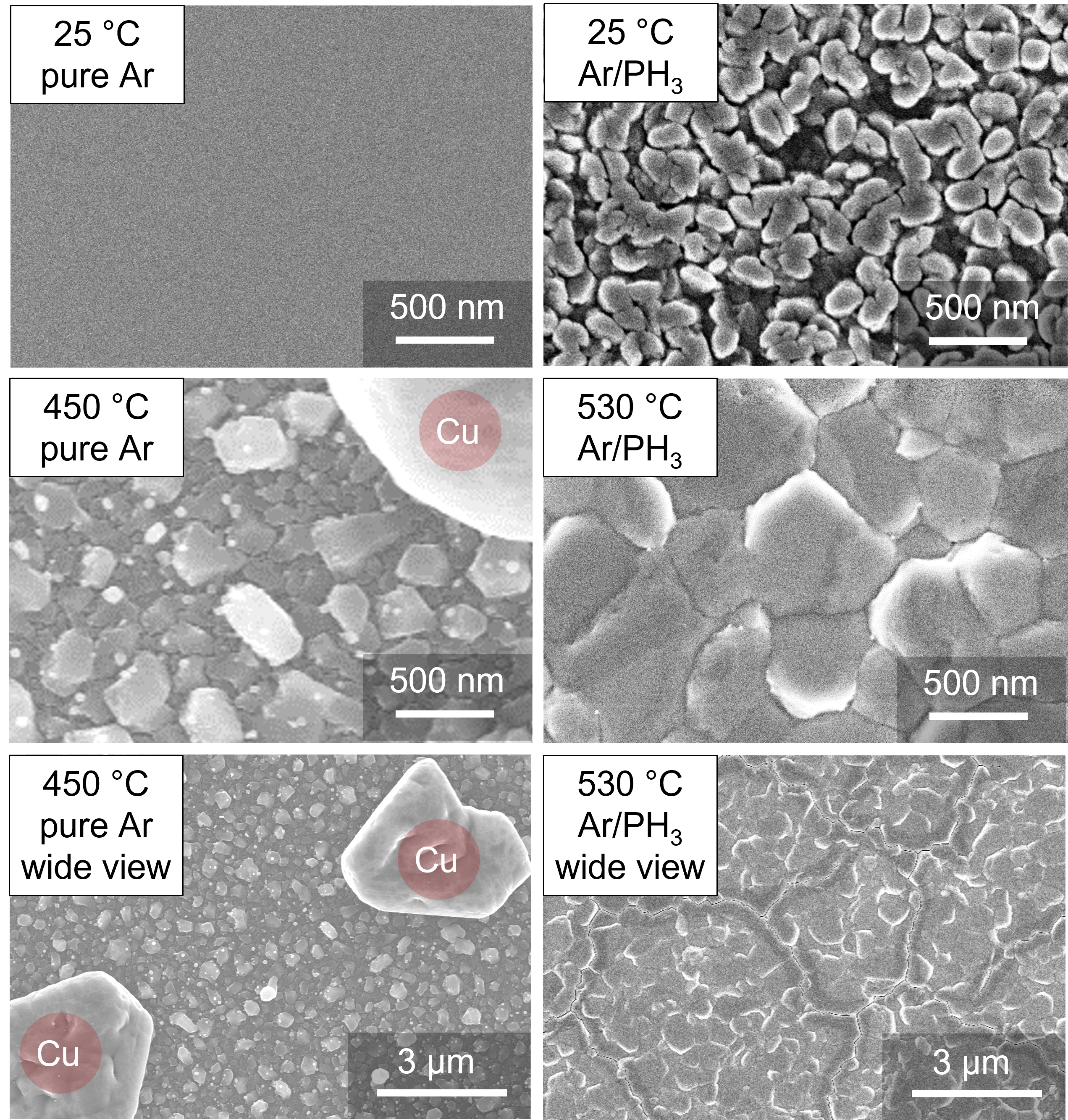}
\caption{SEM images of Cu$_{3-x}$P films deposited in pure Ar (left) and in a PH$_3$-containing atmosphere (right). A polycrystalline rather than amorphous morphology is found when PH$_3$ is included in the growth atmosphere at room temperature (top row). When depositing in pure Ar, most of the phosphorus is already lost at \SI{450}{\celsius}, resulting in formation of metallic Cu islands (middle- and bottom left). When depositing in Ar/PH$_3$, a homogeneous Cu$_{3-x}$P film with grains larger than \SI{500}{nm} can be grown at even higher temperatures (\SI{530}{\celsius}, middle- and bottom right).}
\label{fig:sem}
\end{figure}

We initially deposited Cu$_{3-x}$P films by RF sputtering of a Cu$_{3-x}$P target in pure Ar, but this process route does not allow the exploration of a broad parameter space. The films rapidly lose P with increasing deposition temperature (Fig.~\ref{fig:composition}(a)) and Cu/P ratios below 3 are only achievable below $\sim$\SI{250}{\celsius}. Without intentional heating, films are x-ray amorphous (Fig.~\ref{fig:xrd}). At \SI{220}{\celsius}, polycrystalline Cu$_{3-x}$P grows without detectable secondary phases (Fig.~\ref{fig:xrd}). At \SI{450}{\celsius}, polycrystalline Cu$_{3-x}$P is still present but the film has lost most of its P (Cu/P$\simeq 6.5$) and mainly consists of large Cu islands (Fig.~\ref{fig:sem}). The changes in film morphology versus temperature are shown in greater detail in Fig.~\ref{fig:sem_ar}.


Adding PH$_3$ to the growth atmosphere significantly expands the conditions under which polycrystalline Cu$_{3-x}$P can be grown. First, the deposition temperature can be increased up to at least \SI{530}{\celsius} (Figs.~\ref{fig:xrd},~\ref{fig:sem}). Second, deposition of Cu$_{3-x}$P by reactive sputtering of a metallic Cu target becomes possible in the same temperature range as for the Cu$_{3-x}$P target (Fig.~\ref{fig:composition}(a)). This is a significant process advantage due to the lower cost, higher purity, and higher sputter rate of a metal target. Finally, polycrystalline (rather than amorphous) Cu$_{3-x}$P is obtained when depositing at room temperature (Figs.~\ref{fig:xrd},~\ref{fig:sem}). This unexpected finding is an indication of a plasma-assisted crystallization process driven by the PH$_3$ dissociation products~\cite{Bruno1995} in an RF plasma. In the remainder of the article, we will discuss the properties of Cu$_{3-x}$P deposited in a PH$_3$-containing atmosphere, unless otherwise specified.

When depositing films at room temperature, a broad range of Cu/P ratios is accessible by tuning the PH$_3$ partial pressure (Fig.~\ref{fig:composition}(b)). By simultaneous sputtering of a Cu and a Cu$_{3-x}$P target facing two opposite sides of the substrate, combinatorial films with a gradient in the Cu/P ratio can be obtained at a single PH$_3$ partial pressure (Fig.~\ref{fig:composition}(b)).
When depositing at \SI{370}{\celsius}, P-rich compositions with Cu/P ratios significantly less than 3 are no longer accessible (Fig.~\ref{fig:composition}(b)). The lowest Cu/P ratios obtained at \SI{370}{\celsius}, \SI{450}{\celsius}, and \SI{530}{\celsius} are 2.89, 2.87, and 2.91, respectively. Compatibly with our previous study on CuP$_2$,~\cite{Crovetto2022b} we assume that all Cu-P phases with higher P content than the Cu$_{3-x}$P phase are unstable at these temperatures, due to the high driving force for P evaporation.
All Cu$_{3-x}$P films were found to be air-stable.


\subsubsection{Phase analysis}
Regardless of deposition temperature, the x-ray diffraction (XRD) patterns of all our Cu$_{3-x}$P films close to the stoichiometric point are consistent with the P6$_3$cm structure~\cite{Olofsson1972,Wolff2018} previously described for bulk samples (see Fig.~\ref{fig:xrd}).
We do not observe any peaks extraneous to the P6$_3$cm structure in any of the films with Cu/P $< 3$, not even in the extremely P-rich films (up to Cu$_{1.67}$P) deposited at room temperature.
Although polycrystalline secondary phases are absent, we note that CuP$_2$ and elemental phosphorus tend to form amorphous phases in bulk samples.~\cite{Wolff2016,Ruck2005} Even for the case of thin films reactively sputtered at room temperature, we confirmed elsewhere that CuP$_2$~\cite{Crovetto2022b} and elemental phosphorus~\cite{Crovetto2022} are in amorphous form. Thus, amorphous CuP$_2$ and/or P secondary phases are very likely to exist when Cu/P $< 2.9$, which is the most P-rich composition obtainable at temperatures where CuP$_2$ and P are no longer stable in the solid state.

In the Cu-rich region (Cu/P $> 3$), all films have a composition threshold beyond which XRD peaks from metallic Cu are observed.
The range of Cu-rich compositions free of metallic Cu peaks becomes narrower with increasing PH$_3$ partial pressure (Fig.~\ref{fig:composition}(b)). As argued in the Supporting Information, this is related to a shift in the prevalent formation mechanism of the Cu$_{3-x}$P film, from deposition of Cu$_{3-x}$P vapor to incomplete phosphorization of metallic Cu. In general, XRD results indicate that moderately Cu-rich Cu$_{3-x}$P films can be grown without precipitation of polycrystalline Cu. However, SEM characterization reveals that some secondary phases are already present at Cu$_{3.00}$P composition in films deposited above \SI{370}{\celsius} (Fig.~\ref{fig:sem_phases}). These phases are likely to be non-crystalline Cu. They are no longer visible by SEM at Cu$_{2.95}$P composition (Fig.~\ref{fig:sem_phases}).
Thus, the single-phase region of Cu$_{3-x}$P films deposited at \SI{370}{\celsius} and above is probably not wider than the Cu$_{2.9}$P--Cu$_{3.0}$P range, based on our SEM results. Films deposited at room temperature might have an extended single-phase region on the Cu-rich side, since we could not clearly distinguish secondary phases by SEM in these films. In general, formation of amorphous CuP$_2$ (at low temperatures) or P loss (at high temperatures) constrains the single-phase region on the P-rich side.
Formation of a Cu secondary phase constrains the single-phase region on the Cu-rich side.

In qualitative agreement with our results, previous studies of bulk Cu$_{3-x}$P found that single-phase samples always had Cu/P $< 3$,~\cite{Schlenger1971,Olofsson1972,Wolff2018,Wolff2016} although different single-phase stability ranges were found in each study. The stability of these P-rich compositions was generally attributed to a high concentration of Cu vacancies (V$_\mathrm{Cu}$), responsible for the high p-type conductivity in Cu$_{3-x}$P. Previous first-principles calculations confirmed that one V$_\mathrm{Cu}$ per 24-atom unit cell should be thermodynamically stable under a wide range of chemical potentials.~\cite{DeTrizio2015} In the following sections, we will assume that V$_\mathrm{Cu}$ defects generate most of the charge carriers in Cu$_{3-x}$P films. This assumption will, however, be justified throughout the article based on our own data.

\subsubsection{Structural properties}
\begin{figure}[t!]
\centering%
\includegraphics[width=\columnwidth]{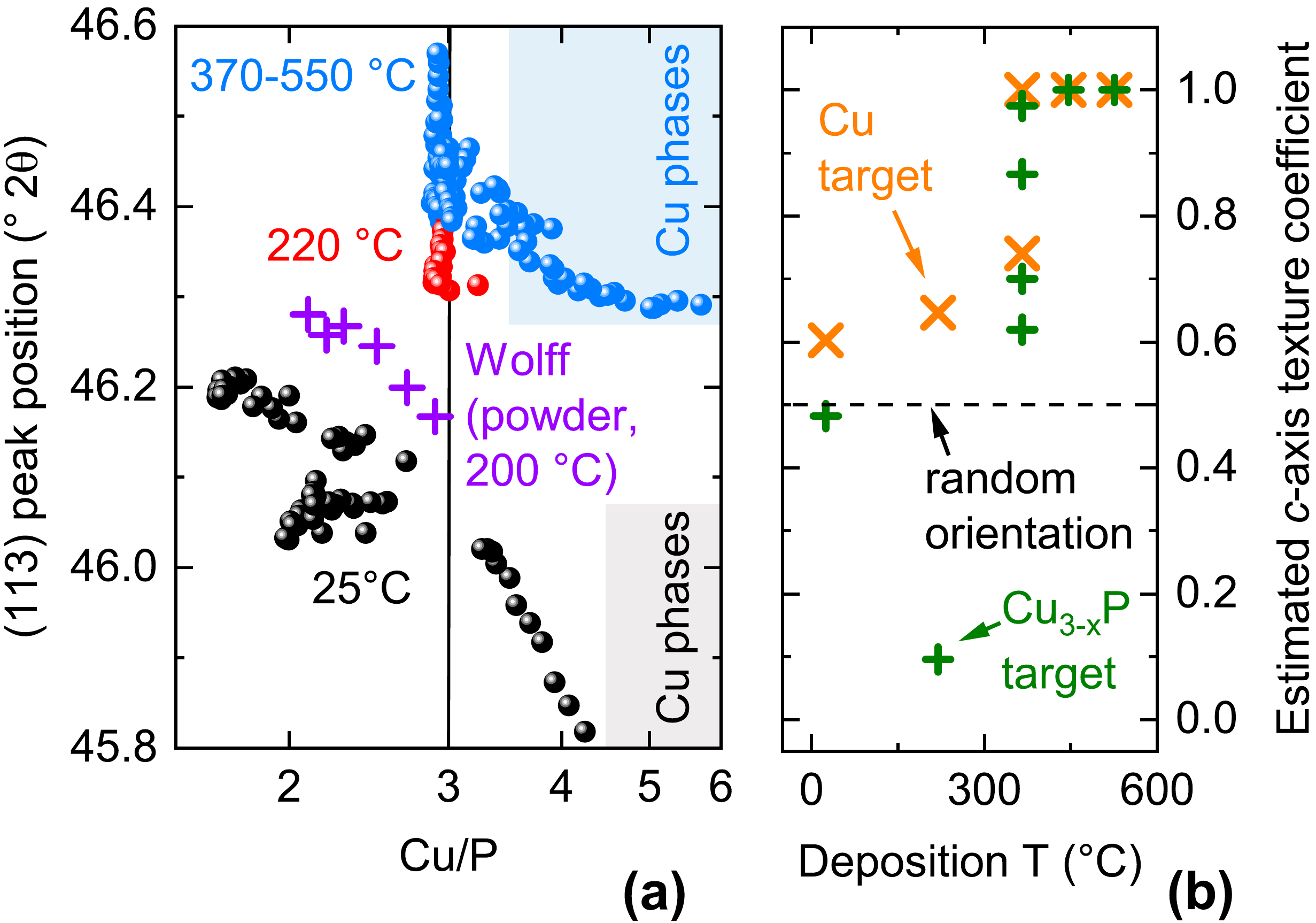}
\caption{Structural parameters of Cu$_{3-x}$P films. \textbf{(a):} Position of (113) XRD peak as a function of film composition, grouped by deposition temperature. Analogous data from Cu$_{3-x}$P powders grown at \SI{200}{\celsius} by Wolff et al.~\cite{Wolff2018} are shown. The composition regions where metallic Cu peaks are observed in XRD are shaded in grey (room-temperature deposition) and in blue (\SI{370}{\celsius} deposition), following the data in Fig.~\ref{fig:composition}(b). \textbf{(b):} Estimated $c$-axis texture coefficient versus deposition temperature, grouped by the target used for deposition. A coefficient of one indicates that the $c$-axis is perpendicular to the substrate plane. A coefficient of zero indicates that the $c$-axis lies in the substrate plane. The method used to estimate the texture coefficient is described in the Supporting Information.}
\label{fig:xrd_extra}
\end{figure}

To evaluate structural changes in Cu$_{3-x}$P as a function of composition and deposition temperature, we analyze the position of the (113) XRD peak (Fig.~\ref{fig:xrd_extra}(a)).
In general, the (113) peak shifts to higher diffraction angles (shorter interplanar distances) with increasing deposition temperature and with decreasing Cu/P ratio.
The other peaks in the XRD patterns tend to shift in the same direction (see Fig.~\ref{fig:xrd} for some examples). The relative magnitude of the shifts in each XRD pattern is generally compatible with a multiplication of the three lattice constants by a common factor. Hence, we conclude that the (113) peak shifts are indicative of lattice contraction or expansion in all three dimensions. Our data confirm and extend the results by Wolff et al.,~\cite{Wolff2018} who reported contraction of the unit cell in all directions in Cu$_{3-x}$P powders with decreasing Cu/P ratios (Fig.~\ref{fig:xrd_extra}(a)).

For films deposited at \SI{370}{\celsius}, the (113) peak stops shifting at high Cu/P ratios, roughly corresponding to the region where we observe precipitation of polycrystalline Cu (blue shaded area in Fig.~\ref{fig:xrd_extra}(a)). This suggests that all excess Cu precipitates as a secondary phase rather than being incorporated in the Cu$_{3-x}$P lattice. For films deposited at room temperature, the (113) peak shift is generally more pronounced up to the precipitation threshold for polycrystalline Cu. Hence, low-temperature-grown Cu$_{3-x}$P may be able to incorporate substantial Cu excess in the form of defects.
This would be a qualitative difference with Cu$_{3-x}$P grown at higher temperatures, where evidence for Cu precipitation already at Cu/P = 3 has been shown in bulk samples~\cite{Schlenger1971,Olofsson1972,Wolff2018,Wolff2016} and thin-film samples (this work).
As will be shown later, changes in lattice constants are generally correlated to changes in the concentration of Cu vacancies in the different films.

The texture coefficient of Cu$_{3-x}$P films is shown in Fig.~\ref{fig:xrd_extra}(b) as a function of deposition temperature. The $c$-axis of the Cu$_{3-x}$P crystallites has an increasing tendency to align perpendicular to the substrate plane with increasing temperature.
Similar $c$-axis texturing effects are often observed in other uniaxially anisotropic materials.~\cite{Tsai2016,Crovetto2020a,Crovetto2016a} We also observe that films grown from the Cu target are generally more c-axis textured than films grown from the Cu$_{3-x}$P target at the same temperature (Fig.~\ref{fig:xrd_extra}(b) and Fig.~\ref{fig:xrd_220c}).

Finally, Raman spectra of Cu$_{3-x}$P films do not exhibit any peaks (Fig.~\ref{fig:raman}). Raman peaks compatible with a few previous studies~\cite{Liu2016d,Peng2021} can only be observed at very high laser excitation intensities above the ablation threshold of Cu$_{3-x}$P, raising questions on the validity of these previously reported spectra.

\subsection{Electrical properties}

\subsubsection{Composition dependence} \label{sec:resistivity_composition}
\begin{figure}[t!]
\centering%
\includegraphics[width=\columnwidth]{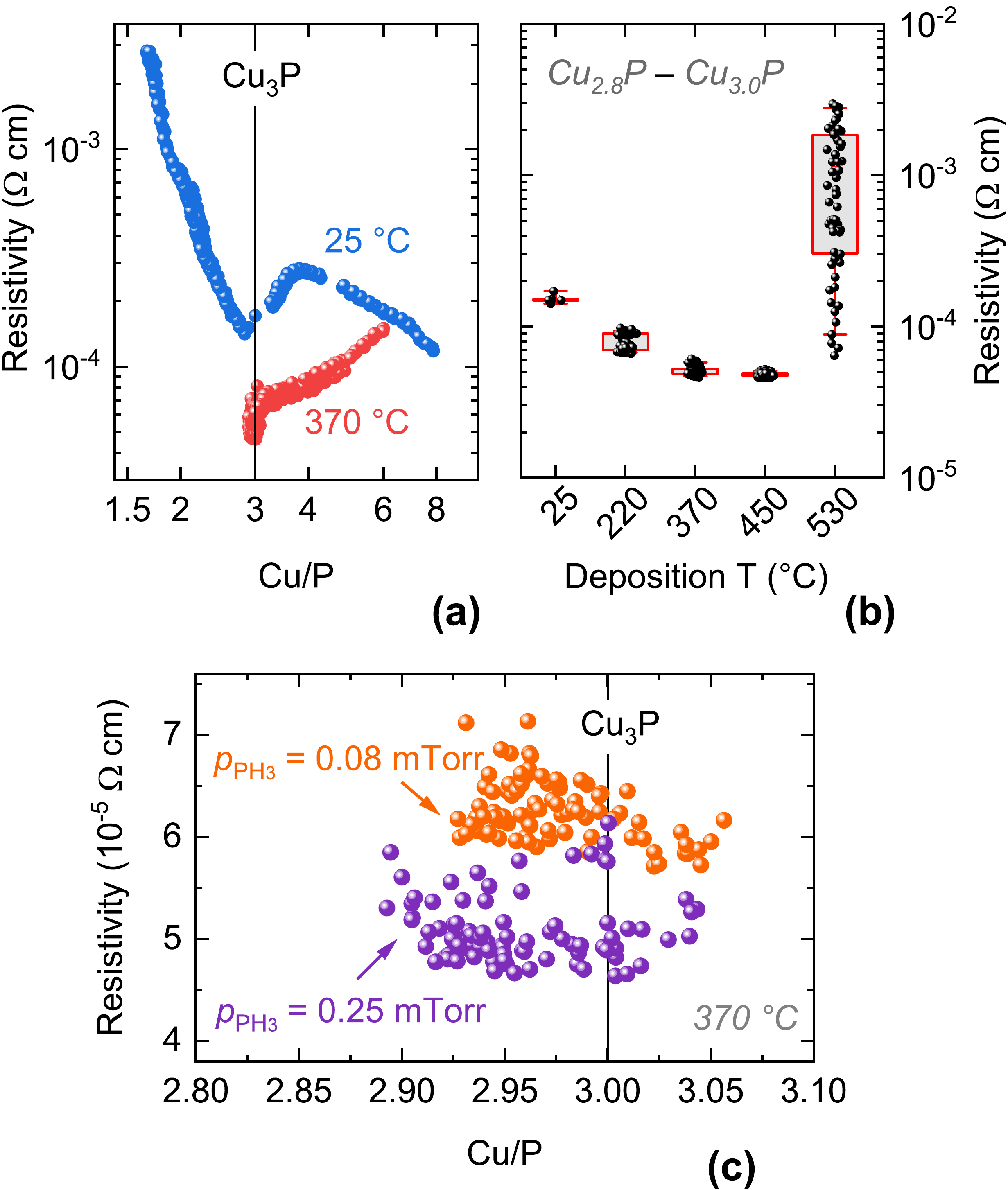}
\caption{Resistivity of Cu$_{3-x}$P films.\textbf{ (a)}: Resistivity versus composition for two different deposition temperatures. \textbf{(b):} Resistivity versus deposition temperature (with boxplots) for films in the $2.8 < \mathrm{Cu/P} < 3.0$ composition range. \textbf{(c):} Resistivity near the Cu$_3$P stoichiometric point as a function of PH$_3$ partial pressure at \SI{370}{\celsius} deposition temperature. The data comes from two combinatorial depositions with different PH$_3$ partial pressure. Other material properties (composition ranges, thicknesses, texture coefficients) are very similar in the two datasets.}
\label{fig:resistivity}
\end{figure}

The resistivity of the Cu-P system can be mapped over a broad region (from Cu$_{1.7}$P to Cu$_{8.0}$P) in films deposited at room temperature (Fig.~\ref{fig:resistivity}(a)). A local resistivity minimum (\SI{1.4e-4}{\ohm \cm}) exists at Cu/P $\simeq 2.83$. The position of the minimum is close to the most P-rich composition that could be obtained at temperatures above \SI{370}{\celsius} (Cu$_{2.87}$P). This result indicates that further electrical doping by nonstoichiometry is not possible for Cu/P ratios lower than this threshold. Thus, P excess is accommodated by secondary phases rather than defects, in line with the arguments in the previous section.
The steep increase in resistivity on the left side of the minimum is due to the much higher resistivity of CuP$_2$ (around \SI{1}{\ohm \cm}) with respect to Cu$_{3-x}$P, as shown elsewhere.~\cite{Crovetto2022b}

On the right side of the minimum, the resistivity increases until the Cu$_{3.8}$P composition is reached (Fig.~\ref{fig:resistivity}(a)). Since metallic Cu peaks begin to appear in XRD at a similar composition (Fig.~\ref{fig:composition}(b)), we assume that the resistivity increase is caused by changes in concentration of point defects up to the Cu precipitation threshold. These defects are also likely to be the cause of unit cell expansion up to a composition of roughly Cu$_{4.5}$P (Fig.~\ref{fig:xrd_extra}(a)). Beyond this threshold, a parallel transport channel along highly conductive Cu percolation paths causes the resistivity to decrease again.

The composition dependence of the resistivity in films deposited at \SI{370}{\celsius} (Fig.~\ref{fig:resistivity}(a)) has some qualitative differences. The minimum resistivity is still found in the vicinity of the stoichiometric point (\SI{4.6e-5}{\ohm \cm}). This value is very similar to the resistivity measured in Cu$_{3-x}$P single crystals (\SI{5e-5}{\ohm \cm})~\cite{Robertson1980} and a factor of two lower than in sintered powders (about \SI{1.0e-4}{\ohm \cm})~\cite{Wolff2018,Juza1956}.
Hall measurements on a Cu$_{2.95}$P film indicate p-type conductivity with hole concentration of \SI{3.81e21}{\cm^{-3}} and hole mobility of \SI{28.8}{cm^2/Vs} at room temperature (Fig.~\ref{fig:electrical}). P-type conductivity is confirmed by
measurement of a Seebeck coefficient of \SI{+11.2}{\micro V/K} in an analogously deposited film (Fig.~\ref{fig:seebeck}).

The Cu/P $< 2.9$ region is not experimentally available at \SI{370}{\celsius} deposition temperature, as explained in the previous section. When Cu/P $> 3.0$, the resistivity continues to increase up to the boundary of the investigated composition range (Cu/P $\simeq 6$), despite the fact that Cu secondary phase are already observed at much lower Cu/P ratios (Fig.~\ref{fig:composition}(b)). This indicates that transport along Cu$_3$P channels is a lower-resistance path in these films, even under a substantial presence of Cu secondary phases. 

Interestingly, the resistivity of films in the $3.2 < \mathrm{Cu/P} < 3.6$ range does not depend on whether polycrystalline Cu is detected by XRD (Fig.~\ref{fig:sem_phases}). This finding suggests that a certain amount of Cu secondary phases exist even when they are not detected by XRD, confirming the SEM observations in Fig.~\ref{fig:sem_phases}. Thus, formation of metallic Cu appears to be favored over formation of compensating donors in Cu-rich films. The moderate increase in (113) interplanar spacing for Cu/P $> 3$ (Fig.~\ref{fig:xrd_extra}(a)) may simply be due to a decrease in V$_\mathrm{Cu}$ concentration. As a consequence, the p-type doping effect of Cu vacancies is never fully compensated by native donors and Cu$_{3-x}$P films are always p-type, even when Cu/P $> 3.0$. Accordingly, we did not measure any negative Seebeck coefficients in any film in the $2.9 < \mathrm{Cu/P} < 3.6$ range.

The average resistivity of films in the vicinity of the stoichiometric point ($2.8 < \mathrm{Cu/P} < 3.0)$ decreases with increasing deposition temperature up to \SI{450}{\celsius} (Fig.~\ref{fig:resistivity}(b)). At a higher temperature (\SI{530}{\celsius}), a higher average resistivity and a much higher standard deviation are observed. This phenomenon is probably linked to an extrinsic effect, i.e., the microscopic cracks visible in SEM images (Fig.~\ref{fig:sem}), which are likely detrimental for electrical transport. Analysis of the optical properties (shown later) indicates that the general decrease in resistivity with increasing temperature is likely due to an increase in hole concentration rather than in hole mobility.

A detailed view of the composition-dependent resistivity data close to the stoichiometric point (Fig.~\ref{fig:resistivity}(c)) reveals that the resistivity depends on the PH$_3$ partial pressure during deposition, but not on film composition.
The overall decrease in resistivity with increasing PH$_3$ partial pressure (20\% lower resistivity with 3 times higher partial pressure) is expected, since a higher P chemical potential relative to the Cu chemical potential leads to formation energy lowering of acceptor defects such as Cu vacancies.~\cite{DeTrizio2015} The complete lack of correlation between resistivity and film composition is more difficult to rationalize, since a higher concentration of Cu vacancies (and thus a lower resistivity) should result in more P-rich compositions. Possible explanations are provided in the Discussion section.

\subsubsection{Temperature dependence}
\begin{figure}[t!]
\centering%
\includegraphics[width=\columnwidth]{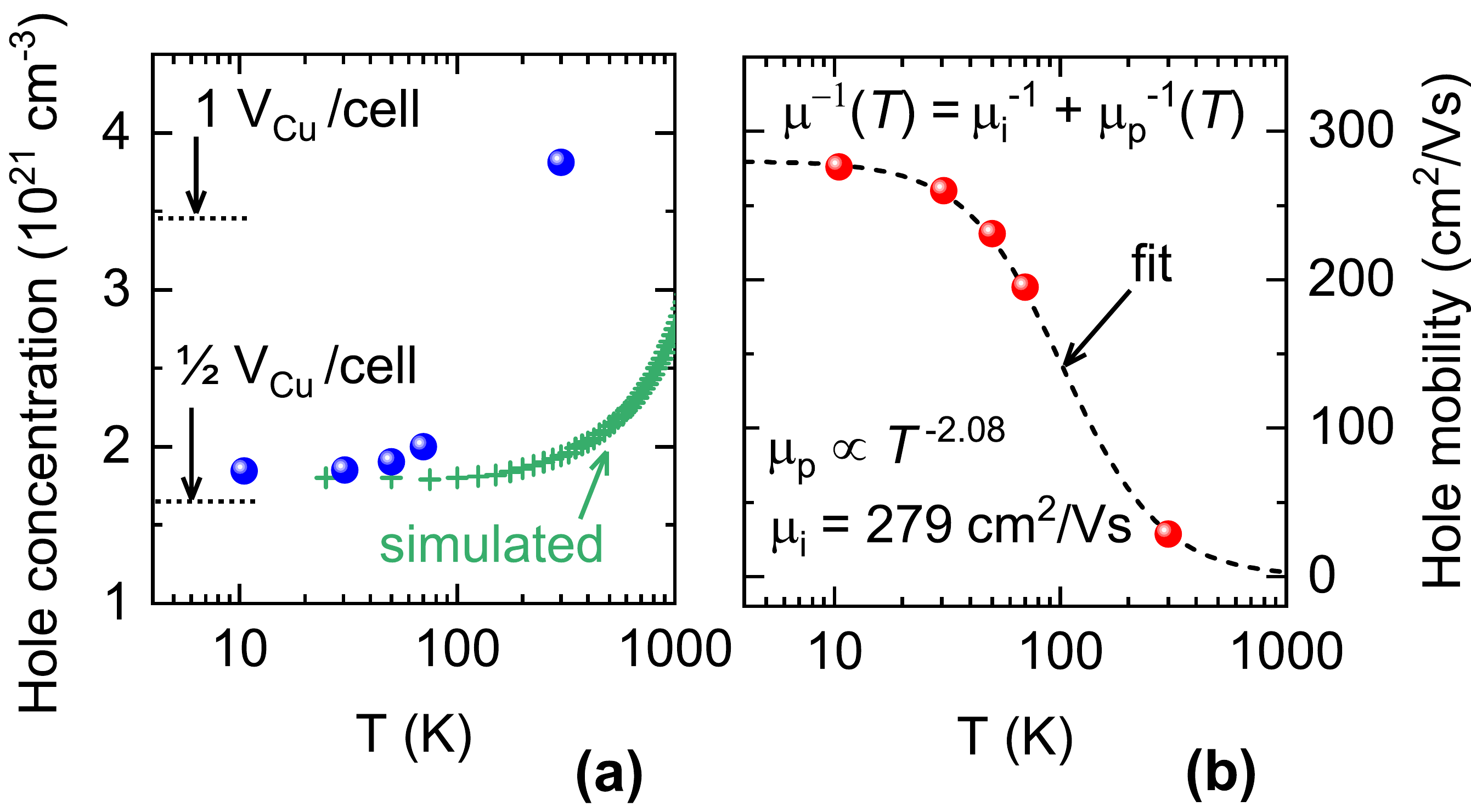}
\caption{Temperature-dependent Hall carrier concentration\textbf{ (a)} and Hall mobility\textbf{ (b)} of a Cu$_{2.95}$P film deposited at \SI{370}{\celsius}.
The hole concentration expected from half- and one uncompensated Cu vacancy per 24-atom unit cell (single acceptor) is indicated in (a). Also shown in (a) is the simulated temperature dependence of the Hall hole concentration, as calculated by Boltzmann transport theory on Cu$_{3}$P with a Fermi level \SI{0.3}{eV} below its intrinsic value. The temperature-dependent mobility in (b) is fitted considering two parallel scattering channels (ionized impurities and optical phonons) resulting in the fitting equation shown in the upper part of (b). The results of the fit are shown in the lower part of (b).
}
\label{fig:electrical}
\end{figure}

Hall effect measurements show that the hole concentration of a low-resistivity Cu$_{3-x}$P film decreases by a factor 2 from \SI{300}{\kelvin} to \SI{10}{\kelvin} (Fig.~\ref{fig:electrical}(a)), in quantitative agreement with previous measurements on powder samples.~\cite{Wolff2018}
On the other hand, the hole mobility increases with decreasing temperature up to \SI{276}{cm^2/Vs} at \SI{10}{\kelvin}, indicating that phonon scattering is the main mobility-limiting mechanism above a few tens of K. At lower temperatures, we expect ionized impurity scattering to be responsible for flattening of the mobility, due to the high density of acceptor defects in Cu$_{3-x}$P.

We fit the experimental hole mobility $\mu(T)$ with the expression $\mu^{-1}(T) = \mu_\mathrm{i}^{-1} + \mu_\mathrm{p}^{-1}(T)$. Here, $\mu_\mathrm{i}$ is the mobility resulting from the ionized impurity scattering channel and $\mu_\mathrm{p}(T)$ is the mobility resulting from the phonon scattering channel. We assume that ionized impurity scattering is roughly temperature-independent in a highly doped material like Cu$_{3-x}$P~\cite{Dingle1955,Willis2022} and that phonon-limited mobility can be described by a power law.~\cite{Wiley1970} The fit yields $\mu_\mathrm{i} = \SI{279}{cm^{2}\per Vs}$ and $\mu_\mathrm{p} = a T^{-2.08}$, where $a$ is a temperature-independent factor (Fig.~\ref{fig:electrical}(b)). The low-temperature mobility of this Cu$_{3-x}$P film is remarkably high for a non-epitaxial polycrystalline thin-film material with such a high carrier concentration. An explicit comparison with other materials is provided in the Supporting Information. The mobility of the present Cu$_{3-x}$P~ film is also much higher than the mobility of sintered Cu$_{3-x}$P powders presented in a previous study (\SI{7.4}{cm^2/Vs} at \SI{300}{\kelvin} and \SI{38}{cm^2/Vs} at \SI{2}{\kelvin}).~\cite{Wolff2018} Thus, we conclude that Cu$_{3-x}$P films have very low ionized impurity scattering rates, despite the high density of such impurities (V$_\mathrm{Cu}$ acceptors). The ionized impurity scattering rate in a highly doped material is expected to scale with $(m^*/\epsilon_s)^2$, where $m^*$ if the carrier effective mass and $\epsilon_s$ is the static dielectric constant.~\cite{ellmerTransparentConductiveZinc2011} Since the calculated $m^*$ for holes in Cu$_{3-x}$P at the observed carrier concentration is not unusually low (see Discussion section), it is likely that Cu$_3$P has a high static dielectric constant. In fact, high values of $\epsilon_s$ are often encountered in narrow band gap semiconductors and semimetals.~\cite{ravichScatteringCurrentCarriers1971a}
We emphasize that the electrical properties extracted from Hall measurements are derived under the assumption of a single carrier type, i.e., with holes being much more abundant than electrons. The next sections show that this assumption is probably justified for Cu$_{3-x}$P.

\subsubsection{Metal, semimetal, or degenerately doped semiconductor?}
The electrical properties of the present Cu$_{3-x}$P films could be compatible with a metal or a degenerately doped semiconductor. However, previous studies do not agree on which of the two classes Cu$_{3-x}$P belongs to. Some authors identified it as a metal,~\cite{Robertson1980,Wolff2018,Kuwano2021} others as a semiconductor.~\cite{AnnAitken2005,Manna2013,DeTrizio2015,Yue2016,Fu2021a,Mu2020,Peng2021}

Density functional theory (DFT) calculations with the PBE exchange correlation functional~\cite{Perdew1996} yield metallic band structures, both for stoichiometric Cu$_{3}$P~\cite{Jain2013} and for Cu$_{3-x}$P with 1 V$_\mathrm{Cu}$/unit cell.~\cite{DeTrizio2015} This is not conclusive evidence that Cu$_{3-x}$P is a metal, because semiconducting compounds are often incorrectly predicted to be metals at the PBE level due to the well-known "band gap problem" of semilocal functionals.~\cite{Sham1985} However, stoichiometric Cu$_{3}$P is also predicted to be a metal using the hybrid HSE06 functional,\cite{Kim2020b} with which band gaps are not systematically underestimated.~\cite{Heyd2005} This substantiates the hypothesis that Cu$_{3-x}$P is intrinsically a metal rather than a degenerately doped semiconductor.

\begin{figure}[t!]
\centering%
\includegraphics[width=\columnwidth]{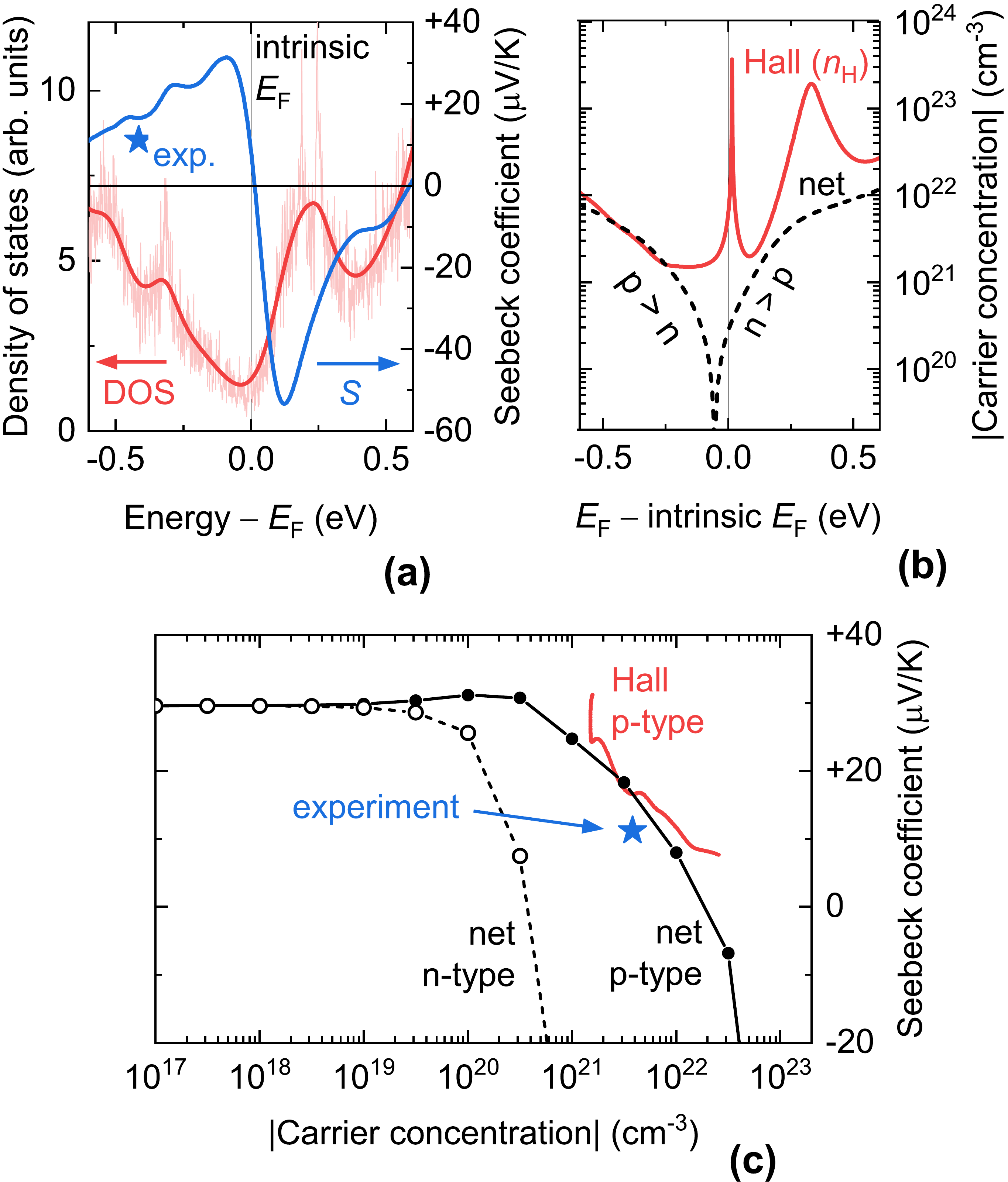}
\caption{Simulated electronic and transport properties of Cu$_3$P at various doping levels. \textbf{(a):} Total electronic DOS of pristine Cu$_3$P by DFT (PBE level), extracted from the Materials Project database.~\cite{Jain2013} The calculated DOS with and without thermal broadening is shown. The low -- but not zero -- DOS at the Fermi level indicates semimetallicity. Also shown is the simulated Seebeck coefficient as a function of Fermi level position with respect to the intrinsic Fermi level of pristine Cu$_3$P (zero of the energy scale). The experimental Seebeck coefficient of a Cu$_{3-x}$P film (star marker) is shown for comparison. \textbf{(b):} Simulated absolute value of the net carrier concentration $p-n$ and of the Hall carrier concentration $n_H$ as a function of Fermi level position with respect to the intrinsic Fermi level of pristine Cu$_3$P (zero of the energy scale). \textbf{(c):} Simulated Pisarenko plots based on the data in (a) and (b). The black data are plotted with the net carrier concentration as the abscissa (p-type for solid data points, n-type for hollow data points). The red data are plotted with the Hall carrier concentration as the abscissa. The star marker indicates the experimental Seebeck coefficient versus experimental Hall hole concentration.}
\label{fig:boltztrap}
\end{figure}

The total density of states (DOS) calculated for stoichiometric Cu$_{3}$P at the PBE level (Materials Project ID: mp-7463)~\cite{Jain2013} is shown in Fig.~\ref{fig:boltztrap}(a). The DOS calculated at the HSE level is similar.~\cite{Kim2020b} The DOS at the Fermi level is low, with a minimum in its immediate vicinity. Metallic materials with these features (such as graphite, As, and Bi) are often referred to as semimetals, to indicate that there is a small overlap between hole-like bands and electron-like bands at the Fermi level. The limited availability of states at the Fermi level results in smaller charge carrier concentrations than in conventional metals.~\cite{Ashcroft1976} In addition, semimetals may shift between n-type and p-type behavior with temperature or under relatively small changes in the Fermi level driven by perturbations such as doping, strain, or biasing.~\cite{Zhang2005}

\subsubsection{Semiclassical simulation of transport properties}
It is desirable to check if the experimentally-determined properties of Cu$_{3-x}$P films are compatible with a semimetal. For a direct comparison, we simulated transport properties of Cu$_{3}$P at \SI{300}{K} from the calculated band structure of Cu$_{3}$P at the PBE level (Materials Project ID: mp-7463).~\cite{Jain2013} Net carrier concentrations were derived from the DFT-computed density of states (Fig.~\ref{fig:boltztrap}(a)) and the Fermi distribution function. Other transport quantities (electrical conductivity, conductivity effective masses, Hall carrier concentration, Seebeck coefficient) were calculated by semiclassical Boltzmann transport theory based on an interpolation of the DFT-calculated band structure.~\cite{Madsen2018,Ricci2017} Two important approximations of the present implementation of Boltzmann transport theory are: (i) the carrier scattering time is constant across all bands and energies (we chose \SI{10}{fs} for reasons explained in the Supporting Information); (ii) doping affects the Fermi level position, but not the band structure itself. The net carrier concentration is defined as $|p-n|$, where $p$ is the free hole concentration and $n$ is the free electron concentration (both are positive quantities). These carriers can be either intrinsic to the material due to a non-zero DOS at the Fermi level, or can be generated by doping.

The Fermi level of intrinsic, defect-free Cu$_3$P is predicted to lie very close to the DOS minimum (Fig.~\ref{fig:boltztrap}(a)). This intrinsic Fermi level is indicated as the zero of the energy scale in Fig .~\ref{fig:boltztrap}(a,b). The net carrier concentration in this hypothetical Cu$_3$P without any dopants is \SI{3e20}{cm^{-3}} (about one net electron per 250 atoms, see dashed line in Fig.~\ref{fig:boltztrap}(b)). This intrinsic electron concentration arises from the non-zero DOS (and thus, non-zero free electron density) of Cu$_3$P at the Fermi level. As the Fermi level shifts down through p-type doping, the net carrier concentration vanishes around \SI{60}{meV} below the intrinsic Fermi level. At this energy, the intrinsic electrons are fully compensated by holes generated by doping. Upon further p-type doping, holes dominate the conductivity rather than electrons. This net carrier concentration is not directly measured by the Hall effect. Instead, the Hall carrier concentration (positive or negative) is defined as $n_H = 1/ e R_H$, where $R_H$ is the Hall coefficient and $e$ is the elementary charge. When both electrons and holes contribute to the conductivity

\begin{equation}
R_H = \frac{p\mu_h^2 - n\mu_e^2}{e(p\mu_h + n\mu_e)^2}
\label{eq:hall}
\end{equation}

where $\mu_h$ and $\mu_e$ are the (positive) mobilities of holes and electrons, respectively.~\cite{May2012} In the transport calculations, the mobilities are obtained from the conductivity effective masses $m_{e,h}^*$\cite{Madsen2018,Ricci2017} and the assumed scattering time $\tau$ as $\mu_{h,e} = e \tau / m_{e,h}^*$. From Eq.~\ref{eq:hall}, it is apparent that the Hall concentration deviates from the simple net carrier concentration $(p-n)$, unless electrons and holes have very different concentrations. This discrepancy is highlighted by calculating $(p-n)$ and $n_H$ in Cu$_3$P as a function of Fermi level (Fig.~\ref{fig:boltztrap}(b)). The discrepancy between $(p-n)$ and $n_H$ is particularly evident in materials predicted to be semimetallic, (like Cu$_3$P) because of the comparable concentrations of electrons and holes. Some counterintuitive features arise from this effect: 1) the Hall carrier concentration is predicted to always be above \SI{1.5e21}{cm^{-3}} even when the net carrier concentration is very low; 2) the Hall carrier concentration is expected to diverge about \SI{10}{meV} above the intrinsic Fermi level; 3) there is a small Fermi level range where the simulations predict that Cu$_3$P will appear as p-type from a Hall measurement, even though electrons are more abundant than holes. These points are further discussed in the Supporting Information.


The calculated Seebeck coefficient $S$ changes sign at roughly the same Fermi level as $R_H$ (Fig.~\ref{fig:boltztrap}(a)). When both electrons and holes contribute to the conductivity
\begin{equation}
S = \frac{p\mu_h S_h + n\mu_e S_e}{p\mu_h + n\mu_e}
\label{eq:seebeck}
\end{equation}
where $S_h$ (positive) and $S_e$ (negative) are the Seebeck coefficients for holes and electrons alone, respectively.~\cite{May2012} As expected, lowering and raising the Fermi level in the simulation leads to positive and negative Seebeck coefficients, respectively. In summary, a Hall effect measurement and a thermovoltage measurement on a Cu$_{3-x}$P sample are expected to yield the same conductivity type.

\subsubsection{Comparison with experiment}
The experimentally measured Hall carrier concentration at \SI{300}{K} (\SI{3.81e21}{\cm^{-3}}) corresponds to a Fermi level \SI{0.41}{eV} below the intrinsic Fermi level according to Fig.~\ref{fig:boltztrap}(b). The experimental and calculated Seebeck coefficients at this Fermi level (\SI{+11.2}{\micro V/K} and \SI{+16.6}{\micro V/K}, respectively) are in reasonable agreement. This is also visualized in the Pisarenko plot (Seebeck coefficient versus carrier concentration) in Fig.~\ref{fig:boltztrap}(c).

\begin{figure}[h!]
\centering%
\includegraphics[width=0.98\columnwidth]{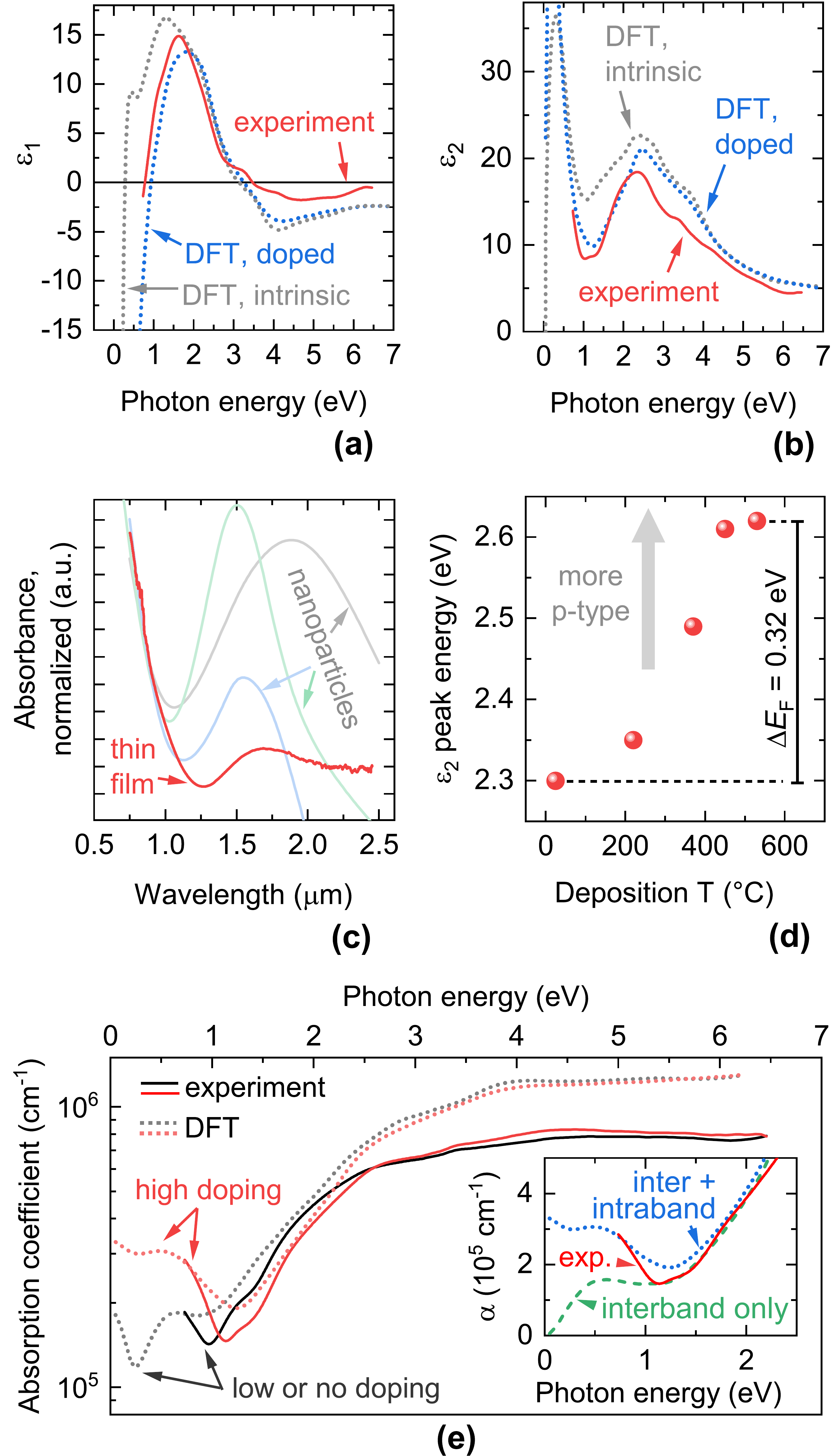}
\caption{Optical properties of Cu$_{3-x}$P films. \textbf{(a,b):} Real ($\epsilon_1$) and imaginary part ($\epsilon_2$) of the dielectric function. Experimental spectra were measured by ellipsometry. Simulated spectra were calculated by DFT on intrinsic Cu$_3$P and on doped Cu$_{2.83}$P with 1 V$_\mathrm{Cu}$/unit cell.
\textbf{(c):} NIR absorbance derived from transmission and reflection measurements.
Previously reported absorbances of Cu$_{3-x}$P nanoparticles~\cite{Manna2013,DeTrizio2015,Mu2020}
are also shown.
\textbf{(d):} Photon energy of the peak found in experimental $\epsilon_2$ spectra (see Fig.~\ref{fig:epsilon_2}) as a function of deposition temperature.
The shift in peak energy is attributed to a Burstein-Moss shift.
\textbf{(e):} Absorption coefficient $\alpha$.
The experimental $\alpha$ from a high-resistivity Cu$_{3-x}$P film is compared to the DFT-calculated $\alpha$ for intrinsic Cu$_3$P ("low or no doping" label). The experimental $\alpha$ from a low-resistivity Cu$_{3-x}$P film is compared to the DFT-calculated $\alpha$ for doped Cu$_{2.83}$P ("high doping" label). \textbf{(e, inset):} Comparison of experimental and calculated $\alpha$ spectra in the IR. One calculated spectrum (dashed line) only includes interband transitions.
The other calculated spectrum (dotted line) includes both intraband and interband transitions.
}
\label{fig:optical}
\end{figure}

Our experimental results are consistent with the presence of approximately one Cu vacancy per 24-atom unit cell, for two reasons. First, we expect the Fermi level of Cu$_{3-x}$P with 1~V$_\mathrm{Cu}$/unit cell to lie about \SI{0.3}{eV} lower than the Fermi level of pristine Cu$_3$P. This Fermi level position is estimated by aligning the localized Cu 3d states of the respective band structures,~\cite{DeTrizio2015} following a standard methodology used in band alignment measurements by photoemission spectroscopy.~\cite{Crovetto2017b} The magnitude of the Fermi level shift is confirmed by quantifying the blue shift of the absorption coefficient spectrum of doped Cu$_{3-x}$P versus pristine Cu$_3$P (Fig.~\ref{fig:optical}(e), discussed later).

Second, the hole concentration corresponding to one ionized Cu vacancy per unit cell (\SI{3.3e21}{\cm^{-3}}) minus the intrinsic free electrons concentration (\SI{3e20}{\cm^{-3}}) is \SI{3.0e21}{\cm^{-3}}. This value is close to the experimentally determined Hall carrier concentration of \SI{3.8e21}{\cm^{-3}} at \SI{300}{K} (Fig.~\ref{fig:electrical}(a)). At these p-type doping levels, $ p \gg n$ and the Hall carrier concentration closely follows the net carrier concentration (Fig.~\ref{fig:boltztrap}(b)). Hence, interpreting Hall effect data assuming a single carrier type is justified and the Hall mobility (Fig.~\ref{fig:electrical}(b)) should reflect the mobility of holes. The experimentally-observed decrease in Hall hole concentration with decreasing temperature is confirmed by Boltzmann transport theory, although not quantitatively. A more detailed discussion is given in the Supporting Information.
To conclude, the electrical properties of Cu$_{3-x}$P films are generally compatible with a semimetallic band structure.

\subsection{Optical properties}
\subsubsection{Optical signatures of semimetallicity}
The optical properties of Cu$_{3-x}$P films provide additional evidence that Cu$_{3-x}$P is a semimetal rather than a semiconductor.
The real and imaginary part of the dielectric function ($\epsilon_1$ and $\epsilon_2$, respectively) were measured on a low-resistivity films by ellipsometry (Fig.~\ref{fig:optical}(a,b)). When the photon energy decreases from the visible to the infrared (IR), $\epsilon_1$ becomes negative and $\epsilon_2$ increases, as expected for a material with a high concentration of free carriers.

We also calculated the dielectric function by DFT using linear response theory and the random phase approximation~\cite{Mortensen2005} (details in the Supporting Information). The calculation on intrinsic Cu$_3$P is in reasonably good agreement with the experimental spectra (Fig.~\ref{fig:optical}(a,b)). However, the sign change of $\epsilon_1$ in the IR occurs at $\sim$\SI{0.5}{eV} lower photon energy than in the experimental spectrum. This may be due to the higher carrier concentration in the experimental sample ($\sim$\SI{4e21}{\cm^{-3}}) than in the intrinsic Cu$_3$P system considered in the calculation ($\sim$\SI{3e20}{\cm^{-3}}, see previous section). To test this hypothesis, we repeated the dielectric function calculation on structurally-relaxed Cu$_{2.83}$P with one Cu vacancy per unit cell. Indeed, the sign change of $\epsilon_1$ shifts to higher photon energies, and the agreement with experiment becomes very good over the whole \SIrange{0.7}{6.5}{eV} spectral range (Fig.~\ref{fig:optical}(a,b)). Again, we emphasize that the calculated dielectric functions were derived from semimetallic Cu$_{3-x}$P. Thus, the agreement between experiment and theory is a further indication that Cu$_{3-x}$P films are intrinsically semimetallic rather than semiconducting. In addition, the calculation on defective Cu$_{2.83}$P confirms that Cu$_{3-x}$P thin films are p-type doped by about one Cu vacancy per unit cell.

\subsubsection{Near-infrared response}
To obtain more experimental information on the optical properties of Cu$_{3-x}$P films in the near infrared (NIR), we determined their absorbance from transmission and reflection measurements with an extended IR range (Fig.~\ref{fig:optical}(c)). Interestingly, the films exhibit a NIR absorption peak rather than the continuously increasing absorbance with decreasing photon energy characteristic of free carrier absorption. The peak maximum is around \SI{1.7}{\um} (\SI{0.73}{eV}). As shown in Fig.~\ref{fig:optical}(c), the peak position and width are compatible with previous reports of a localized surface plasmon resonance (LSPR) peak in Cu$_{3-x}$P nanoparticles,~\cite{Manna2013,DeTrizio2015,Bertoni2019} although its intensity relative to the main absorption band in the visible is lower than the LSPR peak.

Since our sample is a continuous film rather than disconnected nanoparticles, the NIR peak might be due to a (non-localized) surface plasmon polariton (SPP) instead of a LSPR. Surface plasmons cannot be excited by light at perfectly planar surfaces due to the requirement of momentum conservation between the plasmon and the photon.~\cite{Jasperson1969} However, the presence of substantial surface roughness can relax this requirement, so SPPs have been observed, for example, in rough Ag films by simple reflection measurements.~\cite{Jasperson1969,Kaspar1977}

A problem with the interpretation of the NIR peak as a SPP is that this peak is observed in all our measured Cu$_{3-x}$P samples -- even in thinner films processed at room temperature, which have very low surface roughness. An alternative interpretation of the NIR peak is offered by plotting DFT-calculated absorption coefficients (Fig.~\ref{fig:optical}(e)), which do not include any plasmonic effects and are only representative of the bulk optical properties of Cu$_{3-x}$P. Intriguingly, the calculations for both intrinsic Cu$_3$P and p-type doped Cu$_{2.83}$P reveal a peak centered at around \SI{0.6}{eV} photon energy.
When only interband transitions are included in the calculation on doped Cu$_{2.83}$P (inset of Fig.~\ref{fig:optical}(e), green dashed line), the NIR peak is still present but the absorption coefficient approaches zero at zero photon energy. When both interband and intraband transitions are included in the calculation (inset of Fig.~\ref{fig:optical}(e), blue dotted line), the NIR peak exists on top of an additional background of increasing absorption coefficient with decreasing photon energy. This background is due to free carrier absorption. Note that this free carrier absorption background occurs at much lower photon energies in intrinsic Cu$_3$P (below \SI{0.3}{eV} in intrinsic Cu$_3$P; below \SI{1.2}{eV} in doped Cu$_{2.83}$P) as expected from Drude theory.


To sum up, the NIR peak in Cu$_{3-x}$P is likely due to bulk interband transitions, which are independent of both plasmonic effects and the free carrier density. A simultaneous SPP response in the same spectral range cannot be excluded but it is difficult to deconvolve, due to overlap with the bulk response.
The findings presented in this section can help rationalize why the NIR peak in thin films is much less intense than in nanoparticles, and why the absorbance sharply decreases again at longer wavelengths in nanoparticles but not in films (Fig.~\ref{fig:optical}(c)). The first effect occurs because the NIR peak has a different origin in the two cases (bulk interband transitions in the films, LSPR in the nanoparticles). The second effect occurs because intraband transitions in very small nanoparticles are obscured by their intense surface plasmonic response. This does not occur in a thicker film free of quantum confinement effects.

\subsubsection{Burstein-Moss shift}
%

Comparing again the measured absorption coefficients of the two Cu$_{3-x}$P films shown in Fig.~\ref{fig:optical}(e), we notice that the higher-doping (lower-resistivity) film exhibits an overall blue-shift of the absorption coefficient spectrum at all measured photon energies (Fig.~\ref{fig:optical}(e)).
This effect is not related to absorption by free carriers or plasmons. Instead, it has a similar origin as the Burstein-Moss effect observed in degenerately doped semiconductors.~\cite{Burstein1954,Sernelius1988} It is caused by the lowering of the Fermi level with increasing p-type doping. The lower Fermi level causes optical transitions to collectively shift to higher photon energies, because the initial states of the transitions are shifted to deeper energies in the Cu$_{3-x}$P band structure, so higher-energy transitions are required to reach the same final states. Similar to other optical features of Cu$_{3-x}$P, the Burstein-Moss shift is also predicted by DFT, as evident by comparing the measured and calculated absorption coefficients in Fig.~\ref{fig:optical}(e).

We quantified the Burstein-Moss shift in the lowest-resistivity films at each deposition temperature by finding the photon energy at which their $\epsilon_2$ has a maximum (Fig.~\ref{fig:epsilon_2}). The Burstein-Moss shift become progressively larger as the temperature increases (Fig.~\ref{fig:optical}(d)), indicating that films deposited at higher temperatures are more p-type. This result shows that the electrically-probed decrease in resistivity with deposition temperature (Fig.~\ref{fig:resistivity}(b)) is at least partially due to an increase in hole concentration. It also shows that the increase in resistivity at \SI{530}{\celsius} deposition temperature (Fig.~\ref{fig:resistivity}(b)) is due to a decrease in mobility, probably due to cracks in the films. If it were due to a decrease in hole concentration, the Burstein-Moss shift would not keep increasing at this temperature. The continuously increasing hole concentration with deposition temperature follows the previously predicted increase in Cu vacancy concentration with temperature beyond 1~V$_\mathrm{Cu}$/unit cell using Boltzmann statistics.~\cite{DeTrizio2015}

Finally, we note that the difference in Burstein-Moss shift between the films with the smallest and largest shifts (deposited at room temperature and \SI{530}{\celsius} respectively) is \SI{0.32}{eV}. This result is compatible with our calculations of carrier concentration versus Fermi level, which predicted a lowering of the Fermi level by \SI{0.415}{eV} between intrinsic Cu$_3$P and Cu$_{3-x}$P with $n_H = \SI{3.8e21}{cm^{-3}}$ (Fig.~\ref{fig:boltztrap}(b)). It is also compatible with the $\sim$\SI{0.3}{eV} Fermi level down-shift extracted from the previously calculated band structure of Cu$_{2.83}$P with 1~V$_\mathrm{Cu}$/unit cell.~\cite{DeTrizio2015}

\subsection{Discussion}

\subsubsection{Effect of PH$_3$ plasma on sputter deposition of Cu$_{3-x}$P}
Five important roles of PH$_3$ in the sputter deposition of Cu$_{3-x}$P can be identified. First, PH$_3$ is a source of P that enables growth of Cu$_{3-x}$P from a metallic Cu target. Second, the presence of PH$_3$ counteracts the tendency of Cu$_{3-x}$P to decompose at elevated temperatures, enabling relatively high-temperature deposition where large crystal grains, $c$-axis texturing, and a higher V$_\mathrm{Cu}$ concentration can be achieved. Third, a very broad range of Cu/P ratios can be obtained at room temperature (Fig.~\ref{fig:resistivity}(a)), where P losses are minor. This possibility has allowed us to grow polycrystalline CuP$_2$ films by a two-step process.~\cite{Crovetto2022b}
Fourth, PH$_3$ assists the crystallization of Cu$_{3-x}$P even at room temperature, where films sputtered in pure Ar are amorphous (Fig.~\ref{fig:xrd}). This effect is presumably caused by bombardment of the growing film by energetic species formed from the dissociation of PH$_3$ in an RF plasma.~\cite{Bruno1995}
Fifth, the net concentration of electrically-active defects can be tuned by adjusting the PH$_3$ partial pressure, independently on the film composition (Fig.~\ref{fig:resistivity}(c)). This effect is a reminder that equilibrium defect concentrations ultimately depend on chemical potentials and not directly on film composition. These beneficial features demonstrate that reactive RF sputtering is a highly versatile route for both fundamental studies of new phosphides and their technological development.

\subsubsection{Structural-electrical property relationships}
\begin{figure}[t!]
\centering%
\includegraphics[width=\columnwidth]{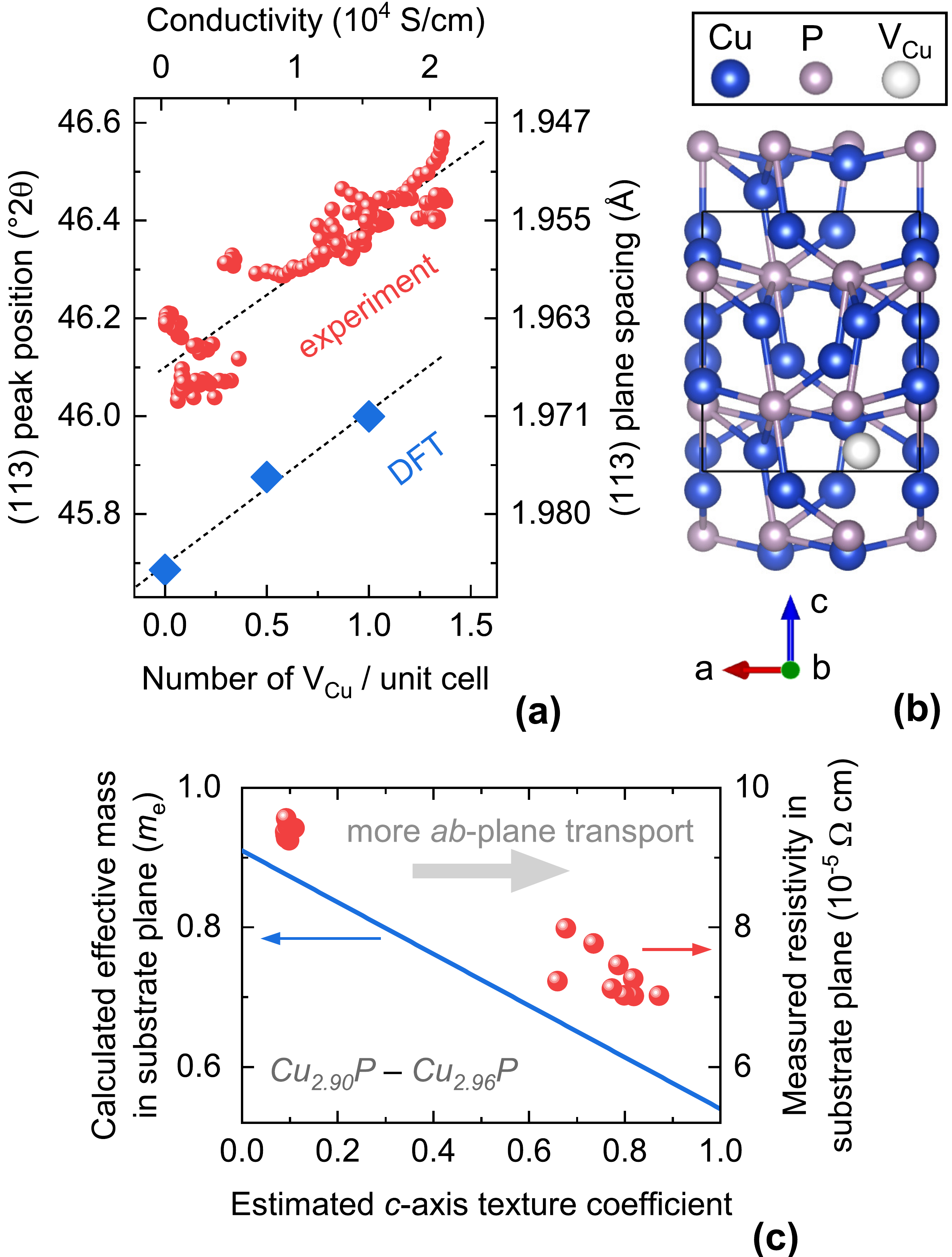}
\caption{Relationships between structural and electrical properties in Cu$_{3-x}$P films. \textbf{(a):} Spacing between (113) planes (determined from XRD peak position) as a function of measured conductivity. Also shown is the DFT-calculated (113) plane spacing for the case of zero, half, and one Cu vacancy per 24-atom unit cell. The conductivity axis and the V$_\mathrm{Cu}$ concentration axis are linked by assuming a constant mobility of \SI{29}{cm^2/Vs} (Fig.~\ref{fig:electrical}(b)). \textbf{(b):} The P6$_3$cm crystal structure of Cu$_3$P.~\cite{Momma2011} The unit cell and the energetically preferred site for V$_\mathrm{Cu}$ are shown. \textbf{(c):} Experimentally measured resistivity as a function of the estimated $c$-axis texture coefficient. A coefficient of one indicates that the resistivity in the $ab$-plane is probed. A coefficient of zero indicates that the resistivity in a $c$-axis-containing plane is probed. Only films deposited at \SI{220}{\celsius} with $2.90 < \mathrm{Cu/P} < 2.96$ are considered. The average conductivity effective mass of Cu$_{3-x}$P in the probed transport plane is also plotted versus the texture coefficient, showing that the changes in conductivity can be explained by changes in the effective mass with lattice direction.}
\label{fig:vacancies}
\end{figure}

In Fig.~\ref{fig:xrd_extra}, we identified large variations in lattice constants and preferential orientation in Cu$_{3-x}$P films processed under different conditions. Here, we investigate possible relationships between these structural trends and the electrical properties of the films. In general, the lattice constants of Cu$_{3-x}$P films tend to decrease with increasing electrical conductivity. This behavior is shown in Fig.~\ref{fig:vacancies}(a) by measuring the (113) plane spacing in combinatorial samples with different conductivities. As noted earlier in this article, changes in the interplanar spacing in other directions indicate an overall unit cell contraction or expansion in all directions, rather than strain in one particular direction.

A simple explanation for this phenomenon is that the unit cell contracts when Cu vacancies are formed. Unit cell contraction is expected since the atoms surrounding the Cu vacancy can be packed closer together due to the free space left by the missing Cu atom. Very limited charge transfer is expected between Cu and P, due to the semimetallic (rather than ionic) nature of Cu$_{3-x}$P. Hence, the counteracting effect of anion-anion repulsion, which may lead to unit cell expansion, is likely negligible.
To verify this behavior, we employed DFT to calculate the equilibrium lattice constants of Cu$_{3-x}$P in the presence of zero, half, and one Cu vacancy per unit cell. The Cu vacancy was placed in one of the symmetry-equivalent Cu(1) lattice sites following Olofsson's notation.~\cite{Olofsson1972} This site is indicated in Fig.~\ref{fig:vacancies}(b) and was previously found to be the most energetically favorable site for V$_\mathrm{Cu}$.~\cite{DeTrizio2015} As expected, the structure relaxes to lower lattice constants with increasing V$_\mathrm{Cu}$ concentration. For one vacancy per unit cell, the $a$, $b$, $c$ axes decrease by 0.56\%, 0.65\%, 0.63\% with respect to the intrinsic cell, respectively. The resulting (113) plane spacing for the calculated structures is shown in Fig.~\ref{fig:vacancies}(a) as a function of V$_\mathrm{Cu}$ concentration. The rate of change in the interplanar spacing is in quantitative agreement with the experimental behavior. The overall overestimation of lattice constants is a known problem of the PBE functional used for the relaxation.~\cite{Perdew2008} This result substantiates the hypotheses that (i) changes in V$_\mathrm{Cu}$ concentration are the main source of conductivity changes in Cu$_{3-x}$P films; and (ii) the estimate of 1--1.5 V$_\mathrm{Cu}$ per unit cell for the most electrically conductive films is reasonable.

Clearly, the decrease in lattice constants with increasing conductivity is only a rough experimental trend. Other factors such as defect compensation and variations in mobility are likely to affect the conductivity of a given film. For example, we may expect different hole mobilities $\mu$ in the $ab$-plane and along the $c$-axis, since the P6$_3$cm structure of Cu$_{3-x}$P is uniaxially anisotropic (Fig.~\ref{fig:vacancies}(b)). Under a p-type doping density of \SI{3e21}{cm^{-3}} at room temperature, we find a conductivity effective mass of \SI{0.54}{m_e} in the $ab$-plane and of \SI{1.28}{m_e} along the $c$-axis by applying Boltzmann transport theory. Assuming a direction-independent carrier scattering time $\tau$, one would expect the resistivity $\rho$ to be proportional to the effective mass $m^*$ in the transport direction, due to the $\rho \propto \mu^{-1} \propto m^*\tau^{-1}$ relationships. To test this hypothesis, we consider the Cu$_{3-x}$P samples deposited at \SI{220}{\celsius}. They were grown in the same deposition run, have similar composition (Cu$_{2.90}$P--Cu$_{2.96}$P), but have large differences in orientation with respect to the substrate depending on the prevalence of Cu$_{3-x}$P formation at the target or at the substrate (Fig.~\ref{fig:xrd_220c}). Indeed, we find that the resistivity measured in the plane of the substrate is proportional to the average effective mass in the plane of the substrate, derived from the estimated texture coefficient (Fig.~\ref{fig:vacancies}(c)). This correlation provides experimental evidence for enhanced hole mobility in the $ab$ plane, and is an indirect indication that the carrier scattering time at room temperature is not strongly direction-dependent.

\subsubsection{Consequences of nonstoichiometry}
We have seen that the net concentration of acceptor defects increases with deposition temperature (Figs.~\ref{fig:resistivity}(b),~\ref{fig:optical}(d)) and PH$_3$ partial pressure (Fig.~\ref{fig:resistivity}(c)). We have also gathered substantial evidence that these acceptors are Cu vacancies. However, an apparent discrepancy requires further discussion.
A large fraction of films are expected to have a V$_\mathrm{Cu}$ concentration between 1 and 1.5 per unit cell (Figs.~\ref{fig:electrical}(a),~\ref{fig:vacancies}(a)). These concentrations would correspond to Cu/P ratios between 2.75 and 2.83 in films without secondary phases or compensating defects. Yet, the most P-rich composition obtained in low-resistivity films is Cu$_{2.87}$P. This compositional discrepancy is only 4\% and may well be due to a small systematic error in the evaluation of elemental composition (see the Supporting Information).
However, we also find that the electrical resistivity and the overall composition in the $2.95 <$ Cu/P $ < 3.05$ range are completely uncorrelated. This finding cannot be explained by measurement errors, because we estimate the error of relative composition measurements to be lower than 1\% (more details in the Supporting Information).

A possible explanation is the existence of metallic Cu secondary phases for Cu/P $> 2.75$, so even in highly P-rich films. As discussed in the Supporting Information, this mechanism is incompatible with other experimental observations in this study.
The presence of other point defects beyond Cu vacancies could be an alternative explanation for the composition-independent resistivity around the stoichiometric point. As discussed in the Supporting Information, different types of defects could potentially explain the experimental results -- either extrinsic impurities or native defects, and either donor- or charge-neutral defects. Some defects that would be compatible with our results are Cu$_\mathrm{P}$ (either donors or neutrals), H$_\mathrm{i}$ donors, or H$_\mathrm{Cu}$ neutrals. Hydrogen-related defects may result from H incorporation in a reactive film deposition process involving PH$_3$ such as ours.~\cite{Weber1998} A more extensive qualitative discussion of possible defects is given in the Supporting Information.


Besides point defects, there are other mechanisms that may change the overall film composition without changing the net dopant concentration. They are intermediate between formation of isolated point defects and segregation of secondary phases. One mechanism is the formation of Cu-rich donor-acceptor defect clusters such as (V$_\mathrm{Cu}$ + Cu$_\mathrm{P}$). Another possible mechanism is extended clustering driven by entropy.~\cite{Zawadzki2015} A dedicated study of defect energetics in Cu$_{3-x}$P may help clarify which mechanism is dominant.

\subsubsection{On the previous identification of Cu$_{3-x}$P as a semiconductor}
As mentioned earlier in this article, many authors have identified Cu$_{3-x}$P as a degenerately doped semiconductor, rather than a (semi)metal.~\cite{AnnAitken2005,Manna2013,DeTrizio2015,Yue2016,Fu2021a,Mu2020,Peng2021} In many of these studies, the characterized samples were Cu$_{3-x}$P nanoparticles or nanoplatelets with characteristic size of less than \SI{10}{nm}. In these cases, it is possible that size effects (quantum confinement) result in the opening of a band gap in an otherwise (semi)metallic system. A similar effect is known for 1$T$-TiS$_2$.~\cite{Fang1997}
However, the results of our study demonstrate that many criteria previously used to assign semiconducting behavior to Cu$_{3-x}$P are not at all incompatible with a semimetal.

For example, several studies identified Cu$_{3-x}$P as a semiconductor with a band gap in the \SIrange{0.8}{1.7}{eV} range.~\cite{AnnAitken2005,Yue2016,Peng2021,Fu2021a} This assignment was justified by the observation of an absorption onset in this photon energy range. However, Fig.~\ref{fig:optical}(e) demonstrates that the photon energy of the absorption onset is not related to a band gap, but it depends instead on the energy at which interband absorption becomes dominant over intraband (free-carrier) absorption. The photon energy of this apparent "band gap" ultimately depends on the free carrier density in the semimetal. In a hypothetical Cu$_3$P sample without free carriers, the absorption coefficient only approaches zero at zero photon energy (inset of Fig.~\ref{fig:optical}(e), green dashed line).

In other cases, identification of Cu$_{3-x}$P as semiconductor was based on the detection of a positive Seebeck coefficient~\cite{DeTrizio2015} or of a photovoltaic effect.~\cite{Manna2013} 
However, Boltzmann transport theory predicts positive Seebeck coefficients in semimetallic Cu$_3$P, both under intrinsic- and p-type-doped conditions (Fig.~\ref{fig:boltztrap}(a)). Many elemental metals also have positive Seebeck coefficients~\cite{Ashcroft1976}. A photovoltaic effect was observed in a Cu$_{3-x}$P/CdS heterojunction.~\cite{Manna2013} However, CdS is itself a photovoltaic material and the reported photovoltaic parameters are compatible with a CdS-based solar cell in which Cu$_{3-x}$P acts as a contact. In particular, the reported short-current density of \SI{2.7}{mA/cm^2} is well within the Shockley-Queisser limit~\cite{Shockley1961} of \SI{7.5}{mA/cm^2} for a CdS cell assuming a \SI{2.4}{eV} band gap for CdS~\cite{Crovetto2018a}. In fact, Cu$_{3-x}$P has recently been incorporated into a ZnSnP$_2$ solar cell,~\cite{Kuwano2021} but rather as a hole contact than as an absorber.

Finally, one computational study found a small -- but non-zero -- band gap in pristine Cu$_3$P by first-principles calculations using hybrid exchange correlation functionals and a Gaussian basis set.~\cite{gaspariSemiconductingOpticalProperties2016} A band gap of \SI{0.68}{eV} was obtained with the PBE0 functional, and a \SI{0.18}{eV} gap was obtained with the HSE06 functional. However, unusually large values of the on-site Hubbard repulsion ($U$) were necessary to reproduce these band gaps by the DFT+U approach~\cite{Anisimov1991} using a plane wave basis set. It is also unusual that Hubbard repulsion had to be added on the P 3p orbitals rather than just on the Cu 3d orbtals. Most importantly, the complex dielectric function calculated from the DFT+U band structure in the same study~\cite{gaspariSemiconductingOpticalProperties2016} is incompatible with our present experimental results. As an example, the constant value of the calculated $\epsilon_1$ (around 4) up to \SI{4}{eV} photon energy~\cite{gaspariSemiconductingOpticalProperties2016} is in stark contrast with the high dispersion of the experimental $\epsilon_1$ (Fig.~\ref{fig:optical}(a)) in this spectral region (negative, up to 15, and negative again).

To summarize this section, there is no conclusive evidence in favor of identification of Cu$_{3-x}$P as a semiconductor without invoking quantum confinement effects.

\section{Conclusion}
We found strong evidence that reactively sputtered Cu$_{3-x}$P films are natively-doped p-type semimetals over a broad composition- and process parameter range. Unlike the case of non-reactive sputtering, Cu$_{3-x}$P films could be synthesized at temperatures significantly higher than \SI{200}{\celsius}, where a higher density of native acceptors can be stabilized and, therefore, the p-type conductivity can be higher. We achieved room-temperature electrical resistivities down to \SI{4.6e-5}{\ohm \cm}, the lowest reported for Cu$_{3-x}$P samples in any form.

In films deposited at temperatures above \SI{370}{\celsius}, the single-phase region of Cu$_{3-x}$P is likely not wider than the Cu$_{2.9}$P--Cu$_{3.0}$P composition range. However, we found some indications that polycrystalline Cu$_{3-x}$P films deposited at room-temperature may tolerate a higher deviation from stoichiometry on Cu-rich side ($x<0$).



The decreasing resistivity with increasing PH$_3$ partial pressure, the preference for Cu-deficient compositions, and the unit cell contraction with decreasing resistivity substantiated the hypothesis that Cu vacancies are responsible for p-type conductivity in Cu$_{3-x}$P. P-type conductivity was retained even when Cu/P $>3$, indicating that Cu vacancies are more stable than any compensating donor even under Cu-rich conditions.
Interestingly, stoichiometric defect-free Cu$_3$P is also predicted to be a semimetal by DFT, though with a much lower carrier concentration and a net excess of electrons rather than holes. This "intrinsic" Cu$_3$P may be very challenging to synthesize due to the high stability of Cu vacancies under all process conditions.

Despite the very high concentration of native dopants, the high mobility of Cu$_{3-x}$P films at low temperatures (\SI{276}{cm^2/Vs} at \SI{10}{K}) points to surprisingly low ionized impurity scattering rates. By studying transport properties as a function of lattice direction, we observed anisotropy in the electrical conductivity of Cu$_{3-x}$P, with higher conductivities achieved along the \textit{ab}-plane. This trend is in quantitative agreement with the expected inverse proportionality between hole effective masses (higher in the \textit{c}-axis direction) and hole mobility.

A NIR absorption feature somewhat similar to the previously identified LSPR peak of Cu$_{3-x}$P nanoparticles was found. Curiously, DFT calculations indicate that the interband transitions intrinsic to bulk Cu$_{3-x}$P also give rise to an absorption peak in that spectral region. In thin-film samples, interband transitions are a more likely explanation for the experimentally observed peak, rather than a plasmonic response. Finally, Fermi level lowering in increasingly doped Cu$_{3-x}$P films gave rise an optical effect equivalent to the Burstein-Moss shift. From this effect, the carrier concentration can in principle be estimated from a simple optical transmission measurement.

\section*{Acknowledgments}
This project has received funding from the European Union’s Horizon 2020 research and innovation programme under the Marie Sklodowska-Curie grant agreement No 840751 (synthesis, characterization, simulation, and data analysis work).
This work was authored in part at the National Renewable Energy Laboratory, operated by Alliance for Sustainable Energy, LLC, for the U.S. Department of Energy (DOE) under Contract No. DE-AC36-08GO28308. Funding supporting development and operation of synthesis and characterization equipment was provided by the Office of Science, Office of Basic Energy Sciences. We acknowledge Karen N. Heinselman for evaporation of metal contacts and RBS measurements.
A.C. is grateful to S\o ren Raza, Mark K. Svendsen, and Francesco Ricci for useful feedback on optical properties, DFT calculations, and transport calculations.

\bibliography{library}

\begin{thebibliography}{75}%
\makeatletter
\providecommand \@ifxundefined [1]{%
 \@ifx{#1\undefined}
}%
\providecommand \@ifnum [1]{%
 \ifnum #1\expandafter \@firstoftwo
 \else \expandafter \@secondoftwo
 \fi
}%
\providecommand \@ifx [1]{%
 \ifx #1\expandafter \@firstoftwo
 \else \expandafter \@secondoftwo
 \fi
}%
\providecommand \natexlab [1]{#1}%
\providecommand \enquote  [1]{``#1''}%
\providecommand \bibnamefont  [1]{#1}%
\providecommand \bibfnamefont [1]{#1}%
\providecommand \citenamefont [1]{#1}%
\providecommand \href@noop [0]{\@secondoftwo}%
\providecommand \href [0]{\begingroup \@sanitize@url \@href}%
\providecommand \@href[1]{\@@startlink{#1}\@@href}%
\providecommand \@@href[1]{\endgroup#1\@@endlink}%
\providecommand \@sanitize@url [0]{\catcode `\\12\catcode `\$12\catcode
  `\&12\catcode `\#12\catcode `\^12\catcode `\_12\catcode `\%12\relax}%
\providecommand \@@startlink[1]{}%
\providecommand \@@endlink[0]{}%
\providecommand \url  [0]{\begingroup\@sanitize@url \@url }%
\providecommand \@url [1]{\endgroup\@href {#1}{\urlprefix }}%
\providecommand \urlprefix  [0]{URL }%
\providecommand \Eprint [0]{\href }%
\providecommand \doibase [0]{https://doi.org/}%
\providecommand \selectlanguage [0]{\@gobble}%
\providecommand \bibinfo  [0]{\@secondoftwo}%
\providecommand \bibfield  [0]{\@secondoftwo}%
\providecommand \translation [1]{[#1]}%
\providecommand \BibitemOpen [0]{}%
\providecommand \bibitemStop [0]{}%
\providecommand \bibitemNoStop [0]{.\EOS\space}%
\providecommand \EOS [0]{\spacefactor3000\relax}%
\providecommand \BibitemShut  [1]{\csname bibitem#1\endcsname}%
\let\auto@bib@innerbib\@empty
\bibitem [{\citenamefont {Minami}\ \emph {et~al.}(2016)\citenamefont {Minami},
  \citenamefont {Nishi},\ and\ \citenamefont {Miyata}}]{Minami2016}%
  \BibitemOpen
  \bibfield  {author} {\bibinfo {author} {\bibfnamefont {T.}~\bibnamefont
  {Minami}}, \bibinfo {author} {\bibfnamefont {Y.}~\bibnamefont {Nishi}},\ and\
  \bibinfo {author} {\bibfnamefont {T.}~\bibnamefont {Miyata}},\ }\bibfield
  {title} {\bibinfo {title} {Efficiency enhancement using a {{Zn$_{1-x}$Ge$_x$-O}} thin
  film as an n-type window layer in {{Cu$_2$O-based}} heterojunction solar
  cells},\ }\href {https://doi.org/10.7567/APEX.9.052301/XML} {\bibfield
  {journal} {\bibinfo  {journal} {Applied Physics Express}\ }\textbf {\bibinfo
  {volume} {9}},\ \bibinfo {pages} {052301} (\bibinfo {year}
  {2016})}\BibitemShut {NoStop}%
\bibitem [{\citenamefont {Kazimierczuk}\ \emph {et~al.}(2014)\citenamefont
  {Kazimierczuk}, \citenamefont {Fr{\"o}hlich}, \citenamefont {Scheel},
  \citenamefont {Stolz},\ and\ \citenamefont {Bayer}}]{Kazimierczuk2014}%
  \BibitemOpen
  \bibfield  {author} {\bibinfo {author} {\bibfnamefont {T.}~\bibnamefont
  {Kazimierczuk}}, \bibinfo {author} {\bibfnamefont {D.}~\bibnamefont
  {Fr{\"o}hlich}}, \bibinfo {author} {\bibfnamefont {S.}~\bibnamefont
  {Scheel}}, \bibinfo {author} {\bibfnamefont {H.}~\bibnamefont {Stolz}},\ and\
  \bibinfo {author} {\bibfnamefont {M.}~\bibnamefont {Bayer}},\ }\bibfield
  {title} {\bibinfo {title} {Giant {{Rydberg}} excitons in the copper oxide
  {{Cu$_2$O}}},\ }\href {https://doi.org/10.1038/nature13832} {\bibfield
  {journal} {\bibinfo  {journal} {Nature}\ }\textbf {\bibinfo {volume} {514}},\
  \bibinfo {pages} {343} (\bibinfo {year} {2014})}\BibitemShut {NoStop}%
\bibitem [{\citenamefont {Hall}\ and\ \citenamefont {Meakin}(1979)}]{Hall1979}%
  \BibitemOpen
  \bibfield  {author} {\bibinfo {author} {\bibfnamefont {R.}~\bibnamefont
  {Hall}}\ and\ \bibinfo {author} {\bibfnamefont {J.}~\bibnamefont {Meakin}},\
  }\bibfield  {title} {\bibinfo {title} {The design and fabrication of high
  efficiency thin film {{CdS}}/{{Cu$_2$S}} solar cells},\ }\href
  {https://doi.org/10.1016/0040-6090(79)90127-5} {\bibfield  {journal}
  {\bibinfo  {journal} {Thin Solid Films}\ }\textbf {\bibinfo {volume} {63}},\
  \bibinfo {pages} {203} (\bibinfo {year} {1979})}\BibitemShut {NoStop}%
\bibitem [{\citenamefont {Luther}\ \emph {et~al.}(2011)\citenamefont {Luther},
  \citenamefont {Jain}, \citenamefont {Ewers},\ and\ \citenamefont
  {Alivisatos}}]{Luther2011}%
  \BibitemOpen
  \bibfield  {author} {\bibinfo {author} {\bibfnamefont {J.~M.}\ \bibnamefont
  {Luther}}, \bibinfo {author} {\bibfnamefont {P.~K.}\ \bibnamefont {Jain}},
  \bibinfo {author} {\bibfnamefont {T.}~\bibnamefont {Ewers}},\ and\ \bibinfo
  {author} {\bibfnamefont {A.~P.}\ \bibnamefont {Alivisatos}},\ }\bibfield
  {title} {\bibinfo {title} {Localized surface plasmon resonances arising from
  free carriers in doped quantum dots},\ }\href
  {https://doi.org/10.1038/nmat3004} {\bibfield  {journal} {\bibinfo  {journal}
  {Nature Materials}\ }\textbf {\bibinfo {volume} {10}},\ \bibinfo {pages}
  {361} (\bibinfo {year} {2011})}\BibitemShut {NoStop}%
\bibitem [{\citenamefont {Liu}\ \emph {et~al.}(2012)\citenamefont {Liu},
  \citenamefont {Shi}, \citenamefont {Xu}, \citenamefont {Zhang}, \citenamefont
  {Zhang}, \citenamefont {Chen}, \citenamefont {Li}, \citenamefont {Uher},
  \citenamefont {Day},\ and\ \citenamefont {Snyder}}]{Liu2012a}%
  \BibitemOpen
  \bibfield  {author} {\bibinfo {author} {\bibfnamefont {H.}~\bibnamefont
  {Liu}}, \bibinfo {author} {\bibfnamefont {X.}~\bibnamefont {Shi}}, \bibinfo
  {author} {\bibfnamefont {F.}~\bibnamefont {Xu}}, \bibinfo {author}
  {\bibfnamefont {L.}~\bibnamefont {Zhang}}, \bibinfo {author} {\bibfnamefont
  {W.}~\bibnamefont {Zhang}}, \bibinfo {author} {\bibfnamefont
  {L.}~\bibnamefont {Chen}}, \bibinfo {author} {\bibfnamefont {Q.}~\bibnamefont
  {Li}}, \bibinfo {author} {\bibfnamefont {C.}~\bibnamefont {Uher}}, \bibinfo
  {author} {\bibfnamefont {T.}~\bibnamefont {Day}},\ and\ \bibinfo {author}
  {\bibfnamefont {G.~J.}\ \bibnamefont {Snyder}},\ }\bibfield  {title}
  {\bibinfo {title} {Copper ion liquid-like thermoelectrics},\ }\href
  {https://doi.org/10.1038/nmat3273} {\bibfield  {journal} {\bibinfo  {journal}
  {Nature Materials}\ }\textbf {\bibinfo {volume} {11}},\ \bibinfo {pages}
  {422} (\bibinfo {year} {2012})}\BibitemShut {NoStop}%
\bibitem [{\citenamefont {Zhao}\ \emph {et~al.}(2019)\citenamefont {Zhao},
  \citenamefont {Liu}, \citenamefont {Yue}, \citenamefont {Wang}, \citenamefont
  {Song}, \citenamefont {Li}, \citenamefont {Guan}, \citenamefont {Xu},
  \citenamefont {Qiu}, \citenamefont {Zhu}, \citenamefont {Chen},\ and\
  \citenamefont {Shi}}]{Zhao2019}%
  \BibitemOpen
  \bibfield  {author} {\bibinfo {author} {\bibfnamefont {K.}~\bibnamefont
  {Zhao}}, \bibinfo {author} {\bibfnamefont {K.}~\bibnamefont {Liu}}, \bibinfo
  {author} {\bibfnamefont {Z.}~\bibnamefont {Yue}}, \bibinfo {author}
  {\bibfnamefont {Y.}~\bibnamefont {Wang}}, \bibinfo {author} {\bibfnamefont
  {Q.}~\bibnamefont {Song}}, \bibinfo {author} {\bibfnamefont {J.}~\bibnamefont
  {Li}}, \bibinfo {author} {\bibfnamefont {M.}~\bibnamefont {Guan}}, \bibinfo
  {author} {\bibfnamefont {Q.}~\bibnamefont {Xu}}, \bibinfo {author}
  {\bibfnamefont {P.}~\bibnamefont {Qiu}}, \bibinfo {author} {\bibfnamefont
  {H.}~\bibnamefont {Zhu}}, \bibinfo {author} {\bibfnamefont {L.}~\bibnamefont
  {Chen}},\ and\ \bibinfo {author} {\bibfnamefont {X.}~\bibnamefont {Shi}},\
  }\bibfield  {title} {\bibinfo {title} {Are {{Cu$_2$Te-Based Compounds
  Excellent Thermoelectric Materials}}?},\ }\href
  {https://doi.org/10.1002/adma.201903480} {\bibfield  {journal} {\bibinfo
  {journal} {Advanced Materials}\ }\textbf {\bibinfo {volume} {31}},\ \bibinfo
  {pages} {1903480} (\bibinfo {year} {2019})}\BibitemShut {NoStop}%
\bibitem [{\citenamefont {Yang}\ \emph {et~al.}(2016)\citenamefont {Yang},
  \citenamefont {Knei{$\beta$}}, \citenamefont {Lorenz},\ and\ \citenamefont
  {Grundmann}}]{Yang2016a}%
  \BibitemOpen
  \bibfield  {author} {\bibinfo {author} {\bibfnamefont {C.}~\bibnamefont
  {Yang}}, \bibinfo {author} {\bibfnamefont {M.}~\bibnamefont {Knei{$\beta$}}},
  \bibinfo {author} {\bibfnamefont {M.}~\bibnamefont {Lorenz}},\ and\ \bibinfo
  {author} {\bibfnamefont {M.}~\bibnamefont {Grundmann}},\ }\bibfield  {title}
  {\bibinfo {title} {Room-temperature synthesized copper iodide thin film as
  degenerate p-type transparent conductor with a boosted figure of merit},\
  }\href {https://doi.org/10.1073/pnas.1613643113} {\bibfield  {journal}
  {\bibinfo  {journal} {Proceedings of the National Academy of Sciences}\
  }\textbf {\bibinfo {volume} {113}},\ \bibinfo {pages} {12929} (\bibinfo
  {year} {2016})}\BibitemShut {NoStop}%
\bibitem [{\citenamefont {Crovetto}\ \emph
  {et~al.}(2020{\natexlab{a}})\citenamefont {Crovetto}, \citenamefont {Hempel},
  \citenamefont {Rusu}, \citenamefont {Choubrac}, \citenamefont {Kojda},
  \citenamefont {Habicht},\ and\ \citenamefont {Unold}}]{Crovetto2020d}%
  \BibitemOpen
  \bibfield  {author} {\bibinfo {author} {\bibfnamefont {A.}~\bibnamefont
  {Crovetto}}, \bibinfo {author} {\bibfnamefont {H.}~\bibnamefont {Hempel}},
  \bibinfo {author} {\bibfnamefont {M.}~\bibnamefont {Rusu}}, \bibinfo {author}
  {\bibfnamefont {L.}~\bibnamefont {Choubrac}}, \bibinfo {author}
  {\bibfnamefont {D.}~\bibnamefont {Kojda}}, \bibinfo {author} {\bibfnamefont
  {K.}~\bibnamefont {Habicht}},\ and\ \bibinfo {author} {\bibfnamefont
  {T.}~\bibnamefont {Unold}},\ }\bibfield  {title} {\bibinfo {title} {Water
  {{Adsorption Enhances Electrical Conductivity}} in {{Transparent P-Type
  CuI}}},\ }\href {https://doi.org/10.1021/acsami.0c11040} {\bibfield
  {journal} {\bibinfo  {journal} {ACS Applied Materials and Interfaces}\
  }\textbf {\bibinfo {volume} {12}},\ \bibinfo {pages} {48741} (\bibinfo {year}
  {2020}{\natexlab{a}})}\BibitemShut {NoStop}%
\bibitem [{\citenamefont {Yang}\ \emph {et~al.}(2017)\citenamefont {Yang},
  \citenamefont {Souchay}, \citenamefont {Knei{\ss}}, \citenamefont {Bogner},
  \citenamefont {Wei}, \citenamefont {Lorenz}, \citenamefont {Oeckler},
  \citenamefont {Benstetter}, \citenamefont {Fu},\ and\ \citenamefont
  {Grundmann}}]{Yang2017b}%
  \BibitemOpen
  \bibfield  {author} {\bibinfo {author} {\bibfnamefont {C.}~\bibnamefont
  {Yang}}, \bibinfo {author} {\bibfnamefont {D.}~\bibnamefont {Souchay}},
  \bibinfo {author} {\bibfnamefont {M.}~\bibnamefont {Knei{\ss}}}, \bibinfo
  {author} {\bibfnamefont {M.}~\bibnamefont {Bogner}}, \bibinfo {author}
  {\bibfnamefont {H.~M.}\ \bibnamefont {Wei}}, \bibinfo {author} {\bibfnamefont
  {M.}~\bibnamefont {Lorenz}}, \bibinfo {author} {\bibfnamefont
  {O.}~\bibnamefont {Oeckler}}, \bibinfo {author} {\bibfnamefont
  {G.}~\bibnamefont {Benstetter}}, \bibinfo {author} {\bibfnamefont {Y.~Q.}\
  \bibnamefont {Fu}},\ and\ \bibinfo {author} {\bibfnamefont {M.}~\bibnamefont
  {Grundmann}},\ }\bibfield  {title} {\bibinfo {title} {Transparent flexible
  thermoelectric material based on non-toxic earth-abundant p-type copper
  iodide thin film},\ }\href {https://doi.org/10.1038/ncomms16076} {\bibfield
  {journal} {\bibinfo  {journal} {Nature Communications}\ }\textbf {\bibinfo
  {volume} {8}},\ \bibinfo {pages} {16076} (\bibinfo {year}
  {2017})}\BibitemShut {NoStop}%
\bibitem [{\citenamefont {Zakutayev}\ \emph {et~al.}(2014)\citenamefont
  {Zakutayev}, \citenamefont {Caskey}, \citenamefont {Fioretti}, \citenamefont
  {Ginley}, \citenamefont {Vidal}, \citenamefont {Stevanovi{\'c}},
  \citenamefont {Tea},\ and\ \citenamefont {Lany}}]{Zakutayev2014}%
  \BibitemOpen
  \bibfield  {author} {\bibinfo {author} {\bibfnamefont {A.}~\bibnamefont
  {Zakutayev}}, \bibinfo {author} {\bibfnamefont {C.~M.}\ \bibnamefont
  {Caskey}}, \bibinfo {author} {\bibfnamefont {A.~N.}\ \bibnamefont
  {Fioretti}}, \bibinfo {author} {\bibfnamefont {D.~S.}\ \bibnamefont
  {Ginley}}, \bibinfo {author} {\bibfnamefont {J.}~\bibnamefont {Vidal}},
  \bibinfo {author} {\bibfnamefont {V.}~\bibnamefont {Stevanovi{\'c}}},
  \bibinfo {author} {\bibfnamefont {E.}~\bibnamefont {Tea}},\ and\ \bibinfo
  {author} {\bibfnamefont {S.}~\bibnamefont {Lany}},\ }\bibfield  {title}
  {\bibinfo {title} {Defect {{Tolerant Semiconductors}} for {{Solar Energy
  Conversion}}},\ }\href {https://doi.org/10.1021/jz5001787} {\bibfield
  {journal} {\bibinfo  {journal} {The Journal of Physical Chemistry Letters}\
  }\textbf {\bibinfo {volume} {5}},\ \bibinfo {pages} {1117} (\bibinfo {year}
  {2014})}\BibitemShut {NoStop}%
\bibitem [{\citenamefont {Caskey}\ \emph {et~al.}(2014)\citenamefont {Caskey},
  \citenamefont {Richards}, \citenamefont {Ginley},\ and\ \citenamefont
  {Zakutayev}}]{Caskey2014}%
  \BibitemOpen
  \bibfield  {author} {\bibinfo {author} {\bibfnamefont {C.~M.}\ \bibnamefont
  {Caskey}}, \bibinfo {author} {\bibfnamefont {R.~M.}\ \bibnamefont
  {Richards}}, \bibinfo {author} {\bibfnamefont {D.~S.}\ \bibnamefont
  {Ginley}},\ and\ \bibinfo {author} {\bibfnamefont {A.}~\bibnamefont
  {Zakutayev}},\ }\bibfield  {title} {\bibinfo {title} {Thin film synthesis and
  properties of copper nitride, a metastable semiconductor},\ }\href
  {https://doi.org/10.1039/C4MH00049H} {\bibfield  {journal} {\bibinfo
  {journal} {Mater. Horiz.}\ }\textbf {\bibinfo {volume} {1}},\ \bibinfo
  {pages} {424} (\bibinfo {year} {2014})}\BibitemShut {NoStop}%
\bibitem [{\citenamefont {Crovetto}\ \emph
  {et~al.}(2022{\natexlab{a}})\citenamefont {Crovetto}, \citenamefont {Kojda},
  \citenamefont {Yi}, \citenamefont {Heinselman}, \citenamefont {LaVan},
  \citenamefont {Habicht}, \citenamefont {Unold},\ and\ \citenamefont
  {Zakutayev}}]{Crovetto2022b}%
  \BibitemOpen
  \bibfield  {author} {\bibinfo {author} {\bibfnamefont {A.}~\bibnamefont
  {Crovetto}}, \bibinfo {author} {\bibfnamefont {D.}~\bibnamefont {Kojda}},
  \bibinfo {author} {\bibfnamefont {F.}~\bibnamefont {Yi}}, \bibinfo {author}
  {\bibfnamefont {K.~N.}\ \bibnamefont {Heinselman}}, \bibinfo {author}
  {\bibfnamefont {D.~A.}\ \bibnamefont {LaVan}}, \bibinfo {author}
  {\bibfnamefont {K.}~\bibnamefont {Habicht}}, \bibinfo {author} {\bibfnamefont
  {T.}~\bibnamefont {Unold}},\ and\ \bibinfo {author} {\bibfnamefont
  {A.}~\bibnamefont {Zakutayev}},\ }\bibfield  {title} {\bibinfo {title}
  {Crystallize {{It}} before {{It Diffuses}}: {{Kinetic Stabilization}} of
  {{Thin-Film Phosphorus-Rich Semiconductor CuP$_2$}}},\ }\href
  {https://doi.org/10.1021/jacs.2c04868} {\bibfield  {journal} {\bibinfo
  {journal} {Journal of the American Chemical Society}\ }\textbf {\bibinfo
  {volume} {144}},\ \bibinfo {pages} {13334} (\bibinfo {year}
  {2022}{\natexlab{a}})}\BibitemShut {NoStop}%
\bibitem [{\citenamefont {Olofsson}(1972)}]{Olofsson1972}%
  \BibitemOpen
  \bibfield  {author} {\bibinfo {author} {\bibfnamefont {O.}~\bibnamefont
  {Olofsson}},\ }\bibfield  {title} {\bibinfo {title} {The crystal structure of
  Cu$_3$P},\ }\href@noop {} {\bibfield  {journal} {\bibinfo  {journal}
  {Acta Chemica Scandinava}\ }\textbf {\bibinfo {volume} {26}},\ \bibinfo
  {pages} {2777} (\bibinfo {year} {1972})}\BibitemShut {NoStop}%
\bibitem [{\citenamefont {Robertson}\ \emph {et~al.}(1980)\citenamefont
  {Robertson}, \citenamefont {Snowball},\ and\ \citenamefont
  {Webber}}]{Robertson1980}%
  \BibitemOpen
  \bibfield  {author} {\bibinfo {author} {\bibfnamefont {D.~S.}\ \bibnamefont
  {Robertson}}, \bibinfo {author} {\bibfnamefont {G.}~\bibnamefont
  {Snowball}},\ and\ \bibinfo {author} {\bibfnamefont {H.}~\bibnamefont
  {Webber}},\ }\bibfield  {title} {\bibinfo {title} {The preparation and
  properties of single crystal copper phosphide},\ }\href
  {https://doi.org/10.1007/BF00552456} {\bibfield  {journal} {\bibinfo
  {journal} {Journal of Materials Science}\ }\textbf {\bibinfo {volume} {15}},\
  \bibinfo {pages} {256} (\bibinfo {year} {1980})}\BibitemShut {NoStop}%
\bibitem [{\citenamefont {Juza}\ and\ \citenamefont
  {B{\"a}r}(1956)}]{Juza1956}%
  \BibitemOpen
  \bibfield  {author} {\bibinfo {author} {\bibfnamefont {R.}~\bibnamefont
  {Juza}}\ and\ \bibinfo {author} {\bibfnamefont {K.}~\bibnamefont {B{\"a}r}},\
  }\bibfield  {title} {\bibinfo {title} {Leiter und {{Halbleiter}} unter den
  {{Phosphiden}} der ersten und zweiten {{Nebengruppe}}},\ }\href
  {https://doi.org/10.1002/zaac.19562830123} {\bibfield  {journal} {\bibinfo
  {journal} {Zeitschrift f\"ur anorganische und allgemeine Chemie}\ }\textbf
  {\bibinfo {volume} {283}},\ \bibinfo {pages} {230} (\bibinfo {year}
  {1956})}\BibitemShut {NoStop}%
\bibitem [{\citenamefont {Ann~Aitken}\ \emph {et~al.}(2005)\citenamefont
  {Ann~Aitken}, \citenamefont {{Ganzha-Hazen}},\ and\ \citenamefont
  {Brock}}]{AnnAitken2005}%
  \BibitemOpen
  \bibfield  {author} {\bibinfo {author} {\bibfnamefont {J.}~\bibnamefont
  {Ann~Aitken}}, \bibinfo {author} {\bibfnamefont {V.}~\bibnamefont
  {{Ganzha-Hazen}}},\ and\ \bibinfo {author} {\bibfnamefont {S.~L.}\
  \bibnamefont {Brock}},\ }\bibfield  {title} {\bibinfo {title} {Solvothermal
  syntheses of Cu$_3$P via reactions of amorphous red phosphorus with a
  variety of copper sources},\ }\href
  {https://doi.org/10.1016/j.jssc.2004.10.004} {\bibfield  {journal} {\bibinfo
  {journal} {Journal of Solid State Chemistry}\ }\textbf {\bibinfo {volume}
  {178}},\ \bibinfo {pages} {970} (\bibinfo {year} {2005})}\BibitemShut
  {NoStop}%
\bibitem [{\citenamefont {Wolff}\ \emph {et~al.}(2016)\citenamefont {Wolff},
  \citenamefont {Pallmann}, \citenamefont {Boucher}, \citenamefont {Weiz},
  \citenamefont {Brunner}, \citenamefont {Doert},\ and\ \citenamefont
  {Ruck}}]{Wolff2016}%
  \BibitemOpen
  \bibfield  {author} {\bibinfo {author} {\bibfnamefont {A.}~\bibnamefont
  {Wolff}}, \bibinfo {author} {\bibfnamefont {J.}~\bibnamefont {Pallmann}},
  \bibinfo {author} {\bibfnamefont {R.}~\bibnamefont {Boucher}}, \bibinfo
  {author} {\bibfnamefont {A.}~\bibnamefont {Weiz}}, \bibinfo {author}
  {\bibfnamefont {E.}~\bibnamefont {Brunner}}, \bibinfo {author} {\bibfnamefont
  {T.}~\bibnamefont {Doert}},\ and\ \bibinfo {author} {\bibfnamefont
  {M.}~\bibnamefont {Ruck}},\ }\bibfield  {title} {\bibinfo {title}
  {Resource-{{Efficient High-Yield Ionothermal Synthesis}} of
  Microcrystalline Cu$_{3-x}$P},\ }\href
  {https://doi.org/10.1021/acs.inorgchem.6b01397} {\bibfield  {journal}
  {\bibinfo  {journal} {Inorganic Chemistry}\ }\textbf {\bibinfo {volume}
  {55}},\ \bibinfo {pages} {8844} (\bibinfo {year} {2016})}\BibitemShut
  {NoStop}%
\bibitem [{\citenamefont {Wolff}\ \emph {et~al.}(2018)\citenamefont {Wolff},
  \citenamefont {Doert}, \citenamefont {Hunger}, \citenamefont {Kaiser},
  \citenamefont {Pallmann}, \citenamefont {Reinhold}, \citenamefont {Yogendra},
  \citenamefont {Giebeler}, \citenamefont {Sichelschmidt}, \citenamefont
  {Schnelle}, \citenamefont {Whiteside}, \citenamefont {Gunaratne},
  \citenamefont {Nockemann}, \citenamefont {Weigand}, \citenamefont {Brunner},\
  and\ \citenamefont {Ruck}}]{Wolff2018}%
  \BibitemOpen
  \bibfield  {author} {\bibinfo {author} {\bibfnamefont {A.}~\bibnamefont
  {Wolff}}, \bibinfo {author} {\bibfnamefont {T.}~\bibnamefont {Doert}},
  \bibinfo {author} {\bibfnamefont {J.}~\bibnamefont {Hunger}}, \bibinfo
  {author} {\bibfnamefont {M.}~\bibnamefont {Kaiser}}, \bibinfo {author}
  {\bibfnamefont {J.}~\bibnamefont {Pallmann}}, \bibinfo {author}
  {\bibfnamefont {R.}~\bibnamefont {Reinhold}}, \bibinfo {author}
  {\bibfnamefont {S.}~\bibnamefont {Yogendra}}, \bibinfo {author}
  {\bibfnamefont {L.}~\bibnamefont {Giebeler}}, \bibinfo {author}
  {\bibfnamefont {J.}~\bibnamefont {Sichelschmidt}}, \bibinfo {author}
  {\bibfnamefont {W.}~\bibnamefont {Schnelle}}, \bibinfo {author}
  {\bibfnamefont {R.}~\bibnamefont {Whiteside}}, \bibinfo {author}
  {\bibfnamefont {H.~Q.~N.}\ \bibnamefont {Gunaratne}}, \bibinfo {author}
  {\bibfnamefont {P.}~\bibnamefont {Nockemann}}, \bibinfo {author}
  {\bibfnamefont {J.~J.}\ \bibnamefont {Weigand}}, \bibinfo {author}
  {\bibfnamefont {E.}~\bibnamefont {Brunner}},\ and\ \bibinfo {author}
  {\bibfnamefont {M.}~\bibnamefont {Ruck}},\ }\bibfield  {title} {\bibinfo
  {title} {Low-{{Temperature Tailoring}} of Copper-Deficient Cu$_{3-x}$P\textemdash{{Electric Properties}}, {{Phase
  Transitions}}, and {{Performance}} in {{Lithium-Ion Batteries}}},\ }\href
  {https://doi.org/10.1021/acs.chemmater.8b02950} {\bibfield  {journal}
  {\bibinfo  {journal} {Chemistry of Materials}\ }\textbf {\bibinfo {volume}
  {30}},\ \bibinfo {pages} {7111} (\bibinfo {year} {2018})}\BibitemShut
  {NoStop}%
\bibitem [{\citenamefont {Fritz}\ and\ \citenamefont
  {R{\"u}ger}(1992)}]{Fritz1992}%
  \BibitemOpen
  \bibfield  {author} {\bibinfo {author} {\bibfnamefont {H.~P.}\ \bibnamefont
  {Fritz}}\ and\ \bibinfo {author} {\bibfnamefont {V.}~\bibnamefont
  {R{\"u}ger}},\ }\bibfield  {title} {\bibinfo {title} {Electrochemical
  deposition of copper phosphides},\ }\href
  {https://doi.org/10.1007/BF00817595} {\bibfield  {journal} {\bibinfo
  {journal} {Monatshefte f\"ur Chemie Chemical Monthly}\ }\textbf {\bibinfo
  {volume} {123}},\ \bibinfo {pages} {397} (\bibinfo {year}
  {1992})}\BibitemShut {NoStop}%
\bibitem [{\citenamefont {Read}\ \emph {et~al.}(2016)\citenamefont {Read},
  \citenamefont {Callejas}, \citenamefont {Holder},\ and\ \citenamefont
  {Schaak}}]{Read2016}%
  \BibitemOpen
  \bibfield  {author} {\bibinfo {author} {\bibfnamefont {C.~G.}\ \bibnamefont
  {Read}}, \bibinfo {author} {\bibfnamefont {J.~F.}\ \bibnamefont {Callejas}},
  \bibinfo {author} {\bibfnamefont {C.~F.}\ \bibnamefont {Holder}},\ and\
  \bibinfo {author} {\bibfnamefont {R.~E.}\ \bibnamefont {Schaak}},\ }\bibfield
   {title} {\bibinfo {title} {General {{Strategy}} for the {{Synthesis}} of
  {{Transition Metal Phosphide Films}} for {{Electrocatalytic Hydrogen}} and
  {{Oxygen Evolution}}},\ }\href
  {https://doi.org/10.1021/ACSAMI.6B02352/SUPPL_FILE/AM6B02352_SI_001.PDF}
  {\bibfield  {journal} {\bibinfo  {journal} {ACS Applied Materials and
  Interfaces}\ }\textbf {\bibinfo {volume} {8}},\ \bibinfo {pages} {12798}
  (\bibinfo {year} {2016})}\BibitemShut {NoStop}%
\bibitem [{\citenamefont {Pawar}\ \emph {et~al.}(2019)\citenamefont {Pawar},
  \citenamefont {Pawar}, \citenamefont {Babar}, \citenamefont {Aqueel~Ahmed},
  \citenamefont {Chavan}, \citenamefont {Jo}, \citenamefont {Cho},
  \citenamefont {Kim}, \citenamefont {Inamdar}, \citenamefont {Kim},
  \citenamefont {Kim},\ and\ \citenamefont {Im}}]{Pawar2019}%
  \BibitemOpen
  \bibfield  {author} {\bibinfo {author} {\bibfnamefont {S.~M.}\ \bibnamefont
  {Pawar}}, \bibinfo {author} {\bibfnamefont {B.~S.}\ \bibnamefont {Pawar}},
  \bibinfo {author} {\bibfnamefont {P.~T.}\ \bibnamefont {Babar}}, \bibinfo
  {author} {\bibfnamefont {A.~T.}\ \bibnamefont {Aqueel~Ahmed}}, \bibinfo
  {author} {\bibfnamefont {H.~S.}\ \bibnamefont {Chavan}}, \bibinfo {author}
  {\bibfnamefont {Y.}~\bibnamefont {Jo}}, \bibinfo {author} {\bibfnamefont
  {S.}~\bibnamefont {Cho}}, \bibinfo {author} {\bibfnamefont {J.~H.~J.}\
  \bibnamefont {Kim}}, \bibinfo {author} {\bibfnamefont {A.~I.}\ \bibnamefont
  {Inamdar}}, \bibinfo {author} {\bibfnamefont {J.~H.~J.}\ \bibnamefont {Kim}},
  \bibinfo {author} {\bibfnamefont {H.}~\bibnamefont {Kim}},\ and\ \bibinfo
  {author} {\bibfnamefont {H.}~\bibnamefont {Im}},\ }\bibfield  {title}
  {\bibinfo {title} {Electrosynthesis of copper phosphide thin films for
  efficient water oxidation},\ }\href
  {https://doi.org/10.1016/j.matlet.2019.01.118} {\bibfield  {journal}
  {\bibinfo  {journal} {Materials Letters}\ }\textbf {\bibinfo {volume}
  {241}},\ \bibinfo {pages} {243} (\bibinfo {year} {2019})}\BibitemShut
  {NoStop}%
\bibitem [{\citenamefont {Kuwano}\ \emph {et~al.}(2021)\citenamefont {Kuwano},
  \citenamefont {Katsube}, \citenamefont {Kazumi},\ and\ \citenamefont
  {Nose}}]{Kuwano2021}%
  \BibitemOpen
  \bibfield  {author} {\bibinfo {author} {\bibfnamefont {T.}~\bibnamefont
  {Kuwano}}, \bibinfo {author} {\bibfnamefont {R.}~\bibnamefont {Katsube}},
  \bibinfo {author} {\bibfnamefont {K.}~\bibnamefont {Kazumi}},\ and\ \bibinfo
  {author} {\bibfnamefont {Y.}~\bibnamefont {Nose}},\ }\bibfield  {title}
  {\bibinfo {title} {Performance enhancement of {{ZnSnP$_2$}} solar cells by a
  {{Cu$_3$P}} back buffer layer},\ }\href
  {https://doi.org/10.1016/j.solmat.2020.110891} {\bibfield  {journal}
  {\bibinfo  {journal} {Solar Energy Materials and Solar Cells}\ }\textbf
  {\bibinfo {volume} {221}},\ \bibinfo {pages} {110891} (\bibinfo {year}
  {2021})}\BibitemShut {NoStop}%
\bibitem [{\citenamefont {Lee}\ \emph {et~al.}(2021)\citenamefont {Lee},
  \citenamefont {Kim}, \citenamefont {Woo}, \citenamefont {Park}, \citenamefont
  {Yoon}, \citenamefont {Park}, \citenamefont {Kim}, \citenamefont {Shin},\
  and\ \citenamefont {Lee}}]{Lee2021}%
  \BibitemOpen
  \bibfield  {author} {\bibinfo {author} {\bibfnamefont {S.~W.}\ \bibnamefont
  {Lee}}, \bibinfo {author} {\bibfnamefont {J.~M.}\ \bibnamefont {Kim}},
  \bibinfo {author} {\bibfnamefont {S.-G.}\ \bibnamefont {Woo}}, \bibinfo
  {author} {\bibfnamefont {Y.}~\bibnamefont {Park}}, \bibinfo {author}
  {\bibfnamefont {J.~C.}\ \bibnamefont {Yoon}}, \bibinfo {author}
  {\bibfnamefont {H.~J.}\ \bibnamefont {Park}}, \bibinfo {author}
  {\bibfnamefont {N.~Y.}\ \bibnamefont {Kim}}, \bibinfo {author} {\bibfnamefont
  {H.~S.}\ \bibnamefont {Shin}},\ and\ \bibinfo {author} {\bibfnamefont
  {Z.}~\bibnamefont {Lee}},\ }\bibfield  {title} {\bibinfo {title} {Epitaxially
  grown copper phosphide ({{Cu$_3$P}}) nanosheets nanoarchitecture compared with
  film morphology for energy applications},\ }\href
  {https://doi.org/10.1016/j.surfin.2021.101369} {\bibfield  {journal}
  {\bibinfo  {journal} {Surfaces and Interfaces}\ ,\ \bibinfo {pages} {101369}}
  (\bibinfo {year} {2021})}\BibitemShut {NoStop}%
\bibitem [{\citenamefont {Manna}\ \emph {et~al.}(2013)\citenamefont {Manna},
  \citenamefont {Bose},\ and\ \citenamefont {Pradhan}}]{Manna2013}%
  \BibitemOpen
  \bibfield  {author} {\bibinfo {author} {\bibfnamefont {G.}~\bibnamefont
  {Manna}}, \bibinfo {author} {\bibfnamefont {R.}~\bibnamefont {Bose}},\ and\
  \bibinfo {author} {\bibfnamefont {N.}~\bibnamefont {Pradhan}},\ }\bibfield
  {title} {\bibinfo {title} {Semiconducting and {{Plasmonic Copper Phosphide
  Platelets}}},\ }\href {https://doi.org/10.1002/anie.201210277} {\bibfield
  {journal} {\bibinfo  {journal} {Angewandte Chemie International Edition}\
  }\textbf {\bibinfo {volume} {52}},\ \bibinfo {pages} {6762} (\bibinfo {year}
  {2013})}\BibitemShut {NoStop}%
\bibitem [{\citenamefont {Tian}\ \emph {et~al.}(2014)\citenamefont {Tian},
  \citenamefont {Liu}, \citenamefont {Cheng}, \citenamefont {Asiri},\ and\
  \citenamefont {Sun}}]{Tian2014}%
  \BibitemOpen
  \bibfield  {author} {\bibinfo {author} {\bibfnamefont {J.}~\bibnamefont
  {Tian}}, \bibinfo {author} {\bibfnamefont {Q.}~\bibnamefont {Liu}}, \bibinfo
  {author} {\bibfnamefont {N.}~\bibnamefont {Cheng}}, \bibinfo {author}
  {\bibfnamefont {A.~M.}\ \bibnamefont {Asiri}},\ and\ \bibinfo {author}
  {\bibfnamefont {X.}~\bibnamefont {Sun}},\ }\bibfield  {title} {\bibinfo
  {title} {Self-Supported Cu$_3$P Nanowire Arrays as an {{Integrated
  High-Performance Three-Dimensional Cathode}} for {{Generating Hydrogen}} from
  {{Water}}},\ }\href {https://doi.org/10.1002/anie.201403842} {\bibfield
  {journal} {\bibinfo  {journal} {Angewandte Chemie International Edition}\
  }\textbf {\bibinfo {volume} {53}},\ \bibinfo {pages} {9577} (\bibinfo {year}
  {2014})}\BibitemShut {NoStop}%
\bibitem [{\citenamefont {De~Trizio}\ \emph {et~al.}(2015)\citenamefont
  {De~Trizio}, \citenamefont {Gaspari}, \citenamefont {Bertoni}, \citenamefont
  {Kriegel}, \citenamefont {Moretti}, \citenamefont {Scotognella},
  \citenamefont {Maserati}, \citenamefont {Zhang}, \citenamefont {Messina},
  \citenamefont {Prato}, \citenamefont {Marras}, \citenamefont {Cavalli},\ and\
  \citenamefont {Manna}}]{DeTrizio2015}%
  \BibitemOpen
  \bibfield  {author} {\bibinfo {author} {\bibfnamefont {L.}~\bibnamefont
  {De~Trizio}}, \bibinfo {author} {\bibfnamefont {R.}~\bibnamefont {Gaspari}},
  \bibinfo {author} {\bibfnamefont {G.}~\bibnamefont {Bertoni}}, \bibinfo
  {author} {\bibfnamefont {I.}~\bibnamefont {Kriegel}}, \bibinfo {author}
  {\bibfnamefont {L.}~\bibnamefont {Moretti}}, \bibinfo {author} {\bibfnamefont
  {F.}~\bibnamefont {Scotognella}}, \bibinfo {author} {\bibfnamefont
  {L.}~\bibnamefont {Maserati}}, \bibinfo {author} {\bibfnamefont
  {Y.}~\bibnamefont {Zhang}}, \bibinfo {author} {\bibfnamefont {G.~C.}\
  \bibnamefont {Messina}}, \bibinfo {author} {\bibfnamefont {M.}~\bibnamefont
  {Prato}}, \bibinfo {author} {\bibfnamefont {S.}~\bibnamefont {Marras}},
  \bibinfo {author} {\bibfnamefont {A.}~\bibnamefont {Cavalli}},\ and\ \bibinfo
  {author} {\bibfnamefont {L.}~\bibnamefont {Manna}},\ }\bibfield  {title}
  {\bibinfo {title} {Cu$_{3-x}$P Nanocrystals as a {{Material Platform}} for
  {{Near-Infrared Plasmonics}} and {{Cation Exchange Reactions}}},\ }\href
  {https://doi.org/10.1021/cm5044792} {\bibfield  {journal} {\bibinfo
  {journal} {Chemistry of Materials}\ }\textbf {\bibinfo {volume} {27}},\
  \bibinfo {pages} {1120} (\bibinfo {year} {2015})}\BibitemShut {NoStop}%
\bibitem [{\citenamefont {Hou}\ \emph {et~al.}(2016)\citenamefont {Hou},
  \citenamefont {Chen}, \citenamefont {Wang}, \citenamefont {Liang},
  \citenamefont {Lin}, \citenamefont {Fu},\ and\ \citenamefont
  {Chen}}]{Hou2016}%
  \BibitemOpen
  \bibfield  {author} {\bibinfo {author} {\bibfnamefont {C.-C.}\ \bibnamefont
  {Hou}}, \bibinfo {author} {\bibfnamefont {Q.-Q.}\ \bibnamefont {Chen}},
  \bibinfo {author} {\bibfnamefont {C.-J.}\ \bibnamefont {Wang}}, \bibinfo
  {author} {\bibfnamefont {F.}~\bibnamefont {Liang}}, \bibinfo {author}
  {\bibfnamefont {Z.}~\bibnamefont {Lin}}, \bibinfo {author} {\bibfnamefont
  {W.-F.}\ \bibnamefont {Fu}},\ and\ \bibinfo {author} {\bibfnamefont
  {Y.}~\bibnamefont {Chen}},\ }\bibfield  {title} {\bibinfo {title}
  {Self-{{Supported Cedarlike Semimetallic Cu$_3$P Nanoarrays}} as a {{3D
  High-Performance Janus Electrode}} for {{Both Oxygen}} and {{Hydrogen
  Evolution}} under {{Basic Conditions}}},\ }\href
  {https://doi.org/10.1021/acsami.6b06251} {\bibfield  {journal} {\bibinfo
  {journal} {ACS Applied Materials and Interfaces}\ }\textbf {\bibinfo {volume}
  {8}},\ \bibinfo {pages} {23037} (\bibinfo {year} {2016})}\BibitemShut
  {NoStop}%
\bibitem [{\citenamefont {Yue}\ \emph {et~al.}(2016)\citenamefont {Yue},
  \citenamefont {Yi}, \citenamefont {Wang}, \citenamefont {Zhang},\ and\
  \citenamefont {Qiu}}]{Yue2016}%
  \BibitemOpen
  \bibfield  {author} {\bibinfo {author} {\bibfnamefont {X.}~\bibnamefont
  {Yue}}, \bibinfo {author} {\bibfnamefont {S.}~\bibnamefont {Yi}}, \bibinfo
  {author} {\bibfnamefont {R.}~\bibnamefont {Wang}}, \bibinfo {author}
  {\bibfnamefont {Z.}~\bibnamefont {Zhang}},\ and\ \bibinfo {author}
  {\bibfnamefont {S.}~\bibnamefont {Qiu}},\ }\bibfield  {title} {\bibinfo
  {title} {A novel and highly efficient earth-abundant Cu$_3$P with
  {{TiO}}$_2$ ``{{P}}\textendash{{N}}'' heterojunction nanophotocatalyst for
  hydrogen evolution from water},\ }\href {https://doi.org/10.1039/C6NR06620H}
  {\bibfield  {journal} {\bibinfo  {journal} {Nanoscale}\ }\textbf {\bibinfo
  {volume} {8}},\ \bibinfo {pages} {17516} (\bibinfo {year}
  {2016})}\BibitemShut {NoStop}%
\bibitem [{\citenamefont {Mu}\ \emph {et~al.}(2020)\citenamefont {Mu},
  \citenamefont {Liu}, \citenamefont {Bao}, \citenamefont {Wan}, \citenamefont
  {Liu}, \citenamefont {Li}, \citenamefont {Shao}, \citenamefont {Xing},
  \citenamefont {Shabbir}, \citenamefont {Li}, \citenamefont {Sun},
  \citenamefont {Li}, \citenamefont {Ma},\ and\ \citenamefont {Bao}}]{Mu2020}%
  \BibitemOpen
  \bibfield  {author} {\bibinfo {author} {\bibfnamefont {H.}~\bibnamefont
  {Mu}}, \bibinfo {author} {\bibfnamefont {Z.}~\bibnamefont {Liu}}, \bibinfo
  {author} {\bibfnamefont {X.}~\bibnamefont {Bao}}, \bibinfo {author}
  {\bibfnamefont {Z.}~\bibnamefont {Wan}}, \bibinfo {author} {\bibfnamefont
  {G.}~\bibnamefont {Liu}}, \bibinfo {author} {\bibfnamefont {X.}~\bibnamefont
  {Li}}, \bibinfo {author} {\bibfnamefont {H.}~\bibnamefont {Shao}}, \bibinfo
  {author} {\bibfnamefont {G.}~\bibnamefont {Xing}}, \bibinfo {author}
  {\bibfnamefont {B.}~\bibnamefont {Shabbir}}, \bibinfo {author} {\bibfnamefont
  {L.}~\bibnamefont {Li}}, \bibinfo {author} {\bibfnamefont {T.}~\bibnamefont
  {Sun}}, \bibinfo {author} {\bibfnamefont {S.}~\bibnamefont {Li}}, \bibinfo
  {author} {\bibfnamefont {W.}~\bibnamefont {Ma}},\ and\ \bibinfo {author}
  {\bibfnamefont {Q.}~\bibnamefont {Bao}},\ }\bibfield  {title} {\bibinfo
  {title} {Highly stable and repeatable femtosecond soliton pulse generation
  from saturable absorbers based on two-dimensional Cu$_{3-x}$P nanocrystals},\
  }\href {https://doi.org/10.1007/s12200-020-1018-y} {\bibfield  {journal}
  {\bibinfo  {journal} {Frontiers of Optoelectronics}\ }\textbf {\bibinfo
  {volume} {13}},\ \bibinfo {pages} {139} (\bibinfo {year} {2020})}\BibitemShut
  {NoStop}%
\bibitem [{\citenamefont {Fu}\ \emph {et~al.}(2021)\citenamefont {Fu},
  \citenamefont {Ma}, \citenamefont {Xia}, \citenamefont {Hu}, \citenamefont
  {Fan},\ and\ \citenamefont {Liu}}]{Fu2021a}%
  \BibitemOpen
  \bibfield  {author} {\bibinfo {author} {\bibfnamefont {Z.}~\bibnamefont
  {Fu}}, \bibinfo {author} {\bibfnamefont {X.}~\bibnamefont {Ma}}, \bibinfo
  {author} {\bibfnamefont {B.}~\bibnamefont {Xia}}, \bibinfo {author}
  {\bibfnamefont {X.}~\bibnamefont {Hu}}, \bibinfo {author} {\bibfnamefont
  {J.}~\bibnamefont {Fan}},\ and\ \bibinfo {author} {\bibfnamefont
  {E.}~\bibnamefont {Liu}},\ }\bibfield  {title} {\bibinfo {title} {Efficient
  photocatalytic {{H$_2$}} evolution over {{Cu}} and Cu$_3$P co-modified {{TiO$_2$}}
  nanosheet},\ }\href {https://doi.org/10.1016/j.ijhydene.2021.03.089}
  {\bibfield  {journal} {\bibinfo  {journal} {International Journal of Hydrogen
  Energy}\ }\textbf {\bibinfo {volume} {46}},\ \bibinfo {pages} {19373}
  (\bibinfo {year} {2021})}\BibitemShut {NoStop}%
\bibitem [{\citenamefont {Stan}\ \emph {et~al.}(2013)\citenamefont {Stan},
  \citenamefont {Kl{\"o}psch}, \citenamefont {Bhaskar}, \citenamefont {Li},
  \citenamefont {Passerini},\ and\ \citenamefont {Winter}}]{Stan2013}%
  \BibitemOpen
  \bibfield  {author} {\bibinfo {author} {\bibfnamefont {M.~C.}\ \bibnamefont
  {Stan}}, \bibinfo {author} {\bibfnamefont {R.}~\bibnamefont {Kl{\"o}psch}},
  \bibinfo {author} {\bibfnamefont {A.}~\bibnamefont {Bhaskar}}, \bibinfo
  {author} {\bibfnamefont {J.}~\bibnamefont {Li}}, \bibinfo {author}
  {\bibfnamefont {S.}~\bibnamefont {Passerini}},\ and\ \bibinfo {author}
  {\bibfnamefont {M.}~\bibnamefont {Winter}},\ }\bibfield  {title} {\bibinfo
  {title} {Cu$_3$P Binary Phosphide: {{Synthesis}} via a {{Wet
  Mechanochemical Method}} and {{Electrochemical Behavior}} as {{Negative
  Electrode Material}} for {{Lithium-Ion Batteries}}},\ }\href
  {https://doi.org/10.1002/aenm.201200655} {\bibfield  {journal} {\bibinfo
  {journal} {Advanced Energy Materials}\ }\textbf {\bibinfo {volume} {3}},\
  \bibinfo {pages} {231} (\bibinfo {year} {2013})}\BibitemShut {NoStop}%
\bibitem [{\citenamefont {Liu}\ \emph {et~al.}(2016)\citenamefont {Liu},
  \citenamefont {He}, \citenamefont {Zhu}, \citenamefont {Xu},\ and\
  \citenamefont {Tong}}]{Liu2016d}%
  \BibitemOpen
  \bibfield  {author} {\bibinfo {author} {\bibfnamefont {S.}~\bibnamefont
  {Liu}}, \bibinfo {author} {\bibfnamefont {X.}~\bibnamefont {He}}, \bibinfo
  {author} {\bibfnamefont {J.}~\bibnamefont {Zhu}}, \bibinfo {author}
  {\bibfnamefont {L.}~\bibnamefont {Xu}},\ and\ \bibinfo {author}
  {\bibfnamefont {J.}~\bibnamefont {Tong}},\ }\bibfield  {title} {\bibinfo
  {title} {Cu$_3$P/{{RGO}} nanocomposite as a new anode for lithium-ion
  batteries},\ }\href {https://doi.org/10.1038/srep35189} {\bibfield  {journal}
  {\bibinfo  {journal} {Scientific Reports}\ }\textbf {\bibinfo {volume} {6}},\
  \bibinfo {pages} {1} (\bibinfo {year} {2016})}\BibitemShut {NoStop}%
\bibitem [{\citenamefont {Peng}\ \emph {et~al.}(2021)\citenamefont {Peng},
  \citenamefont {Lv}, \citenamefont {Fu}, \citenamefont {Chen}, \citenamefont
  {Su}, \citenamefont {Li}, \citenamefont {Zhang},\ and\ \citenamefont
  {Zhao}}]{Peng2021}%
  \BibitemOpen
  \bibfield  {author} {\bibinfo {author} {\bibfnamefont {X.}~\bibnamefont
  {Peng}}, \bibinfo {author} {\bibfnamefont {Y.}~\bibnamefont {Lv}}, \bibinfo
  {author} {\bibfnamefont {L.}~\bibnamefont {Fu}}, \bibinfo {author}
  {\bibfnamefont {F.}~\bibnamefont {Chen}}, \bibinfo {author} {\bibfnamefont
  {W.}~\bibnamefont {Su}}, \bibinfo {author} {\bibfnamefont {J.}~\bibnamefont
  {Li}}, \bibinfo {author} {\bibfnamefont {Q.}~\bibnamefont {Zhang}},\ and\
  \bibinfo {author} {\bibfnamefont {S.}~\bibnamefont {Zhao}},\ }\bibfield
  {title} {\bibinfo {title} {Photoluminescence properties of cuprous phosphide
  prepared through phosphating copper with a native oxide layer},\ }\href
  {https://doi.org/10.1039/D1RA07112B} {\bibfield  {journal} {\bibinfo
  {journal} {RSC Advances}\ }\textbf {\bibinfo {volume} {11}},\ \bibinfo
  {pages} {34095} (\bibinfo {year} {2021})}\BibitemShut {NoStop}%
\bibitem [{\citenamefont {Talley}\ \emph {et~al.}(2019)\citenamefont {Talley},
  \citenamefont {Bauers}, \citenamefont {Melamed}, \citenamefont {Papac},
  \citenamefont {Heinselman}, \citenamefont {Khan}, \citenamefont {Roberts},
  \citenamefont {Jacobson}, \citenamefont {Mis}, \citenamefont {Brennecka},
  \citenamefont {Perkins},\ and\ \citenamefont {Zakutayev}}]{Talley2019}%
  \BibitemOpen
  \bibfield  {author} {\bibinfo {author} {\bibfnamefont {K.~R.}\ \bibnamefont
  {Talley}}, \bibinfo {author} {\bibfnamefont {S.~R.}\ \bibnamefont {Bauers}},
  \bibinfo {author} {\bibfnamefont {C.~L.}\ \bibnamefont {Melamed}}, \bibinfo
  {author} {\bibfnamefont {M.~C.}\ \bibnamefont {Papac}}, \bibinfo {author}
  {\bibfnamefont {K.~N.}\ \bibnamefont {Heinselman}}, \bibinfo {author}
  {\bibfnamefont {I.}~\bibnamefont {Khan}}, \bibinfo {author} {\bibfnamefont
  {D.~M.}\ \bibnamefont {Roberts}}, \bibinfo {author} {\bibfnamefont
  {V.}~\bibnamefont {Jacobson}}, \bibinfo {author} {\bibfnamefont
  {A.}~\bibnamefont {Mis}}, \bibinfo {author} {\bibfnamefont {G.~L.}\
  \bibnamefont {Brennecka}}, \bibinfo {author} {\bibfnamefont {J.~D.}\
  \bibnamefont {Perkins}},\ and\ \bibinfo {author} {\bibfnamefont
  {A.}~\bibnamefont {Zakutayev}},\ }\bibfield  {title} {\bibinfo {title}
  {{{COMBIgor}}: {{Data-Analysis Package}} for {{Combinatorial Materials
  Science}}},\ }\href {https://doi.org/10.1021/acscombsci.9b00077} {\bibfield
  {journal} {\bibinfo  {journal} {ACS Combinatorial Science}\ }\textbf
  {\bibinfo {volume} {21}},\ \bibinfo {pages} {537} (\bibinfo {year}
  {2019})}\BibitemShut {NoStop}%
\bibitem [{\citenamefont {Talley}\ \emph {et~al.}(2021)\citenamefont {Talley},
  \citenamefont {White}, \citenamefont {Wunder}, \citenamefont {Eash},
  \citenamefont {Schwarting}, \citenamefont {Evenson}, \citenamefont {Perkins},
  \citenamefont {Tumas}, \citenamefont {Munch}, \citenamefont {Phillips},\ and\
  \citenamefont {Zakutayev}}]{Talley2021a}%
  \BibitemOpen
  \bibfield  {author} {\bibinfo {author} {\bibfnamefont {K.~R.}\ \bibnamefont
  {Talley}}, \bibinfo {author} {\bibfnamefont {R.}~\bibnamefont {White}},
  \bibinfo {author} {\bibfnamefont {N.}~\bibnamefont {Wunder}}, \bibinfo
  {author} {\bibfnamefont {M.}~\bibnamefont {Eash}}, \bibinfo {author}
  {\bibfnamefont {M.}~\bibnamefont {Schwarting}}, \bibinfo {author}
  {\bibfnamefont {D.}~\bibnamefont {Evenson}}, \bibinfo {author} {\bibfnamefont
  {J.~D.}\ \bibnamefont {Perkins}}, \bibinfo {author} {\bibfnamefont
  {W.}~\bibnamefont {Tumas}}, \bibinfo {author} {\bibfnamefont
  {K.}~\bibnamefont {Munch}}, \bibinfo {author} {\bibfnamefont
  {C.}~\bibnamefont {Phillips}},\ and\ \bibinfo {author} {\bibfnamefont
  {A.}~\bibnamefont {Zakutayev}},\ }\bibfield  {title} {\bibinfo {title}
  {Research data infrastructure for high-throughput experimental materials
  science},\ }\href {https://doi.org/10.1016/j.patter.2021.100373} {\bibfield
  {journal} {\bibinfo  {journal} {Patterns}\ }\textbf {\bibinfo {volume} {2}},\
  \bibinfo {pages} {100373} (\bibinfo {year} {2021})}\BibitemShut {NoStop}%
\bibitem [{\citenamefont {Zakutayev}\ \emph {et~al.}(2018)\citenamefont
  {Zakutayev}, \citenamefont {Wunder}, \citenamefont {Schwarting},
  \citenamefont {Perkins}, \citenamefont {White}, \citenamefont {Munch},
  \citenamefont {Tumas},\ and\ \citenamefont {Phillips}}]{Zakutayev2018}%
  \BibitemOpen
  \bibfield  {author} {\bibinfo {author} {\bibfnamefont {A.}~\bibnamefont
  {Zakutayev}}, \bibinfo {author} {\bibfnamefont {N.}~\bibnamefont {Wunder}},
  \bibinfo {author} {\bibfnamefont {M.}~\bibnamefont {Schwarting}}, \bibinfo
  {author} {\bibfnamefont {J.~D.}\ \bibnamefont {Perkins}}, \bibinfo {author}
  {\bibfnamefont {R.}~\bibnamefont {White}}, \bibinfo {author} {\bibfnamefont
  {K.}~\bibnamefont {Munch}}, \bibinfo {author} {\bibfnamefont
  {W.}~\bibnamefont {Tumas}},\ and\ \bibinfo {author} {\bibfnamefont
  {C.}~\bibnamefont {Phillips}},\ }\bibfield  {title} {\bibinfo {title} {An
  open experimental database for exploring inorganic materials},\ }\href
  {https://doi.org/10.1038/sdata.2018.53} {\bibfield  {journal} {\bibinfo
  {journal} {Scientific Data}\ }\textbf {\bibinfo {volume} {5}},\ \bibinfo
  {pages} {180053} (\bibinfo {year} {2018})}\BibitemShut {NoStop}%
\bibitem [{\citenamefont {Schnepf}\ \emph {et~al.}(2022)\citenamefont
  {Schnepf}, \citenamefont {Crovetto}, \citenamefont {Gorai}, \citenamefont
  {Park}, \citenamefont {Holtz}, \citenamefont {Heinselman}, \citenamefont
  {Bauers}, \citenamefont {Brooks~Tellekamp}, \citenamefont {Zakutayev},
  \citenamefont {Greenaway}, \citenamefont {Toberer},\ and\ \citenamefont
  {Tamboli}}]{Schnepf2021}%
  \BibitemOpen
  \bibfield  {author} {\bibinfo {author} {\bibfnamefont {R.~R.}\ \bibnamefont
  {Schnepf}}, \bibinfo {author} {\bibfnamefont {A.}~\bibnamefont {Crovetto}},
  \bibinfo {author} {\bibfnamefont {P.}~\bibnamefont {Gorai}}, \bibinfo
  {author} {\bibfnamefont {A.}~\bibnamefont {Park}}, \bibinfo {author}
  {\bibfnamefont {M.}~\bibnamefont {Holtz}}, \bibinfo {author} {\bibfnamefont
  {K.~N.}\ \bibnamefont {Heinselman}}, \bibinfo {author} {\bibfnamefont
  {S.~R.}\ \bibnamefont {Bauers}}, \bibinfo {author} {\bibfnamefont
  {M.}~\bibnamefont {Brooks~Tellekamp}}, \bibinfo {author} {\bibfnamefont
  {A.}~\bibnamefont {Zakutayev}}, \bibinfo {author} {\bibfnamefont {A.~L.}\
  \bibnamefont {Greenaway}}, \bibinfo {author} {\bibfnamefont {E.~S.}\
  \bibnamefont {Toberer}},\ and\ \bibinfo {author} {\bibfnamefont {A.~C.}\
  \bibnamefont {Tamboli}},\ }\bibfield  {title} {\bibinfo {title} {Reactive
  phosphine combinatorial co-sputtering of cation disordered ZnGeP$_2$
  films},\ }\href {https://doi.org/10.1039/D1TC04695K} {\bibfield  {journal}
  {\bibinfo  {journal} {Journal of Materials Chemistry C}\ }\textbf {\bibinfo
  {volume} {10}},\ \bibinfo {pages} {870} (\bibinfo {year} {2022})}\BibitemShut
  {NoStop}%
\bibitem [{\citenamefont {Crovetto}\ \emph
  {et~al.}(2022{\natexlab{b}})\citenamefont {Crovetto}, \citenamefont
  {Adamczyk}, \citenamefont {Schnepf}, \citenamefont {Perkins}, \citenamefont
  {Hempel}, \citenamefont {Bauers}, \citenamefont {Toberer}, \citenamefont
  {Tamboli}, \citenamefont {Unold},\ and\ \citenamefont
  {Zakutayev}}]{Crovetto2022}%
  \BibitemOpen
  \bibfield  {author} {\bibinfo {author} {\bibfnamefont {A.}~\bibnamefont
  {Crovetto}}, \bibinfo {author} {\bibfnamefont {J.~M.}\ \bibnamefont
  {Adamczyk}}, \bibinfo {author} {\bibfnamefont {R.~R.}\ \bibnamefont
  {Schnepf}}, \bibinfo {author} {\bibfnamefont {C.~L.}\ \bibnamefont
  {Perkins}}, \bibinfo {author} {\bibfnamefont {H.}~\bibnamefont {Hempel}},
  \bibinfo {author} {\bibfnamefont {S.~R.}\ \bibnamefont {Bauers}}, \bibinfo
  {author} {\bibfnamefont {E.~S.}\ \bibnamefont {Toberer}}, \bibinfo {author}
  {\bibfnamefont {A.~C.}\ \bibnamefont {Tamboli}}, \bibinfo {author}
  {\bibfnamefont {T.}~\bibnamefont {Unold}},\ and\ \bibinfo {author}
  {\bibfnamefont {A.}~\bibnamefont {Zakutayev}},\ }\bibfield  {title} {\bibinfo
  {title} {Boron {{Phosphide Films}} by {{Reactive Sputtering}}: {{Searching}}
  for a {{P}}-{{Type Transparent Conductor}}},\ }\href
  {https://doi.org/10.1002/admi.202200031} {\bibfield  {journal} {\bibinfo
  {journal} {Advanced Materials Interfaces}\ }\textbf {\bibinfo {volume} {9}},\
  \bibinfo {pages} {2200031} (\bibinfo {year}
  {2022}{\natexlab{b}})}\BibitemShut {NoStop}%
\bibitem [{\citenamefont {Willis}\ \emph {et~al.}(2022)\citenamefont {Willis},
  \citenamefont {Bravi{\'c}}, \citenamefont {Schnepf}, \citenamefont
  {Heinselman}, \citenamefont {Monserrat}, \citenamefont {Unold}, \citenamefont
  {Zakutayev}, \citenamefont {Scanlon},\ and\ \citenamefont
  {Crovetto}}]{Willis2022}%
  \BibitemOpen
  \bibfield  {author} {\bibinfo {author} {\bibfnamefont {J.}~\bibnamefont
  {Willis}}, \bibinfo {author} {\bibfnamefont {I.}~\bibnamefont {Bravi{\'c}}},
  \bibinfo {author} {\bibfnamefont {R.~R.}\ \bibnamefont {Schnepf}}, \bibinfo
  {author} {\bibfnamefont {K.~N.}\ \bibnamefont {Heinselman}}, \bibinfo
  {author} {\bibfnamefont {B.}~\bibnamefont {Monserrat}}, \bibinfo {author}
  {\bibfnamefont {T.}~\bibnamefont {Unold}}, \bibinfo {author} {\bibfnamefont
  {A.}~\bibnamefont {Zakutayev}}, \bibinfo {author} {\bibfnamefont {D.~O.}\
  \bibnamefont {Scanlon}},\ and\ \bibinfo {author} {\bibfnamefont
  {A.}~\bibnamefont {Crovetto}},\ }\bibfield  {title} {\bibinfo {title}
  {Prediction and realisation of high mobility and degenerate p-type
  conductivity in {{CaCuP}} thin films},\ }\href
  {https://doi.org/10.1039/D2SC01538B} {\bibfield  {journal} {\bibinfo
  {journal} {Chemical Science}\ }\textbf {\bibinfo {volume} {13}},\ \bibinfo
  {pages} {5872} (\bibinfo {year} {2022})}\BibitemShut {NoStop}%
\bibitem [{\citenamefont {Bruno}\ \emph {et~al.}(1995)\citenamefont {Bruno},
  \citenamefont {Losurdo},\ and\ \citenamefont {Capezzuto}}]{Bruno1995}%
  \BibitemOpen
  \bibfield  {author} {\bibinfo {author} {\bibfnamefont {G.}~\bibnamefont
  {Bruno}}, \bibinfo {author} {\bibfnamefont {M.}~\bibnamefont {Losurdo}},\
  and\ \bibinfo {author} {\bibfnamefont {P.}~\bibnamefont {Capezzuto}},\
  }\bibfield  {title} {\bibinfo {title} {Study of the phosphine plasma
  decomposition and its formation by ablation of red phosphorus in hydrogen
  plasma},\ }\href {https://doi.org/10.1116/1.579364} {\bibfield  {journal}
  {\bibinfo  {journal} {Journal of Vacuum Science and Technology A: Vacuum,
  Surfaces, and Films}\ }\textbf {\bibinfo {volume} {13}},\ \bibinfo {pages}
  {349} (\bibinfo {year} {1995})}\BibitemShut {NoStop}%
\bibitem [{\citenamefont {Ruck}\ \emph {et~al.}(2005)\citenamefont {Ruck},
  \citenamefont {Hoppe}, \citenamefont {Wahl}, \citenamefont {Simon},
  \citenamefont {Wang},\ and\ \citenamefont {Seifert}}]{Ruck2005}%
  \BibitemOpen
  \bibfield  {author} {\bibinfo {author} {\bibfnamefont {M.}~\bibnamefont
  {Ruck}}, \bibinfo {author} {\bibfnamefont {D.}~\bibnamefont {Hoppe}},
  \bibinfo {author} {\bibfnamefont {B.}~\bibnamefont {Wahl}}, \bibinfo {author}
  {\bibfnamefont {P.}~\bibnamefont {Simon}}, \bibinfo {author} {\bibfnamefont
  {Y.}~\bibnamefont {Wang}},\ and\ \bibinfo {author} {\bibfnamefont
  {G.}~\bibnamefont {Seifert}},\ }\bibfield  {title} {\bibinfo {title} {Fibrous
  {{Red Phosphorus}}},\ }\href {https://doi.org/10.1002/anie.200503017}
  {\bibfield  {journal} {\bibinfo  {journal} {Angewandte Chemie International
  Edition}\ }\textbf {\bibinfo {volume} {44}},\ \bibinfo {pages} {7616}
  (\bibinfo {year} {2005})}\BibitemShut {NoStop}%
\bibitem [{\citenamefont {Schlenger}\ \emph {et~al.}(1971)\citenamefont
  {Schlenger}, \citenamefont {Jacobs},\ and\ \citenamefont
  {Juza}}]{Schlenger1971}%
  \BibitemOpen
  \bibfield  {author} {\bibinfo {author} {\bibfnamefont {H.}~\bibnamefont
  {Schlenger}}, \bibinfo {author} {\bibfnamefont {H.}~\bibnamefont {Jacobs}},\
  and\ \bibinfo {author} {\bibfnamefont {R.}~\bibnamefont {Juza}},\ }\bibfield
  {title} {\bibinfo {title} {Tern\"are {{Phasen}} des {{Lithiums}} mit
  {{Kupfer}} und {{Phosphor}}},\ }\href
  {https://doi.org/10.1002/zaac.19713850302} {\bibfield  {journal} {\bibinfo
  {journal} {Zeitschrift f\"ur anorganische und allgemeine Chemie}\ }\textbf
  {\bibinfo {volume} {385}},\ \bibinfo {pages} {177} (\bibinfo {year}
  {1971})}\BibitemShut {NoStop}%
\bibitem [{\citenamefont {Tsai}\ \emph {et~al.}(2016)\citenamefont {Tsai},
  \citenamefont {Nie}, \citenamefont {Blancon}, \citenamefont {Stoumpos},
  \citenamefont {Asadpour}, \citenamefont {Harutyunyan}, \citenamefont
  {Neukirch}, \citenamefont {Verduzco}, \citenamefont {Crochet}, \citenamefont
  {Tretiak}, \citenamefont {Pedesseau}, \citenamefont {Even}, \citenamefont
  {Alam}, \citenamefont {Gupta}, \citenamefont {Lou}, \citenamefont {Ajayan},
  \citenamefont {Bedzyk}, \citenamefont {Kanatzidis},\ and\ \citenamefont
  {Mohite}}]{Tsai2016}%
  \BibitemOpen
  \bibfield  {author} {\bibinfo {author} {\bibfnamefont {H.}~\bibnamefont
  {Tsai}}, \bibinfo {author} {\bibfnamefont {W.}~\bibnamefont {Nie}}, \bibinfo
  {author} {\bibfnamefont {J.-C.}\ \bibnamefont {Blancon}}, \bibinfo {author}
  {\bibfnamefont {C.~C.}\ \bibnamefont {Stoumpos}}, \bibinfo {author}
  {\bibfnamefont {R.}~\bibnamefont {Asadpour}}, \bibinfo {author}
  {\bibfnamefont {B.}~\bibnamefont {Harutyunyan}}, \bibinfo {author}
  {\bibfnamefont {A.~J.}\ \bibnamefont {Neukirch}}, \bibinfo {author}
  {\bibfnamefont {R.}~\bibnamefont {Verduzco}}, \bibinfo {author}
  {\bibfnamefont {J.~J.}\ \bibnamefont {Crochet}}, \bibinfo {author}
  {\bibfnamefont {S.}~\bibnamefont {Tretiak}}, \bibinfo {author} {\bibfnamefont
  {L.}~\bibnamefont {Pedesseau}}, \bibinfo {author} {\bibfnamefont
  {J.}~\bibnamefont {Even}}, \bibinfo {author} {\bibfnamefont {M.~A.}\
  \bibnamefont {Alam}}, \bibinfo {author} {\bibfnamefont {G.}~\bibnamefont
  {Gupta}}, \bibinfo {author} {\bibfnamefont {J.}~\bibnamefont {Lou}}, \bibinfo
  {author} {\bibfnamefont {P.~M.}\ \bibnamefont {Ajayan}}, \bibinfo {author}
  {\bibfnamefont {M.~J.}\ \bibnamefont {Bedzyk}}, \bibinfo {author}
  {\bibfnamefont {M.~G.}\ \bibnamefont {Kanatzidis}},\ and\ \bibinfo {author}
  {\bibfnamefont {A.~D.}\ \bibnamefont {Mohite}},\ }\bibfield  {title}
  {\bibinfo {title} {High-efficiency two-dimensional
  {{Ruddlesden}}\textendash{{Popper}} perovskite solar cells},\ }\href
  {https://doi.org/10.1038/nature18306} {\bibfield  {journal} {\bibinfo
  {journal} {Nature}\ }\textbf {\bibinfo {volume} {536}},\ \bibinfo {pages}
  {312} (\bibinfo {year} {2016})}\BibitemShut {NoStop}%
\bibitem [{\citenamefont {Crovetto}\ \emph
  {et~al.}(2020{\natexlab{b}})\citenamefont {Crovetto}, \citenamefont
  {Hajijafarassar}, \citenamefont {Hansen}, \citenamefont {Seger},
  \citenamefont {Chorkendorff},\ and\ \citenamefont {Vesborg}}]{Crovetto2020a}%
  \BibitemOpen
  \bibfield  {author} {\bibinfo {author} {\bibfnamefont {A.}~\bibnamefont
  {Crovetto}}, \bibinfo {author} {\bibfnamefont {A.}~\bibnamefont
  {Hajijafarassar}}, \bibinfo {author} {\bibfnamefont {O.}~\bibnamefont
  {Hansen}}, \bibinfo {author} {\bibfnamefont {B.}~\bibnamefont {Seger}},
  \bibinfo {author} {\bibfnamefont {I.}~\bibnamefont {Chorkendorff}},\ and\
  \bibinfo {author} {\bibfnamefont {P.~C.~K.}\ \bibnamefont {Vesborg}},\
  }\bibfield  {title} {\bibinfo {title} {Parallel {{Evaluation}} of the BiI$_3$, {{BiOI}}, and Ag$_3$BiI$_6$ {{Layered Photoabsorbers}}},\ }\href
  {https://doi.org/10.1021/acs.chemmater.9b04925} {\bibfield  {journal}
  {\bibinfo  {journal} {Chemistry of Materials}\ }\textbf {\bibinfo {volume}
  {32}},\ \bibinfo {pages} {3385} (\bibinfo {year}
  {2020}{\natexlab{b}})}\BibitemShut {NoStop}%
\bibitem [{\citenamefont {Crovetto}\ \emph {et~al.}(2016)\citenamefont
  {Crovetto}, \citenamefont {Ottsen}, \citenamefont {Stamate}, \citenamefont
  {Kj{\ae}r}, \citenamefont {Schou},\ and\ \citenamefont
  {Hansen}}]{Crovetto2016a}%
  \BibitemOpen
  \bibfield  {author} {\bibinfo {author} {\bibfnamefont {A.}~\bibnamefont
  {Crovetto}}, \bibinfo {author} {\bibfnamefont {T.~S.}\ \bibnamefont
  {Ottsen}}, \bibinfo {author} {\bibfnamefont {E.}~\bibnamefont {Stamate}},
  \bibinfo {author} {\bibfnamefont {D.}~\bibnamefont {Kj{\ae}r}}, \bibinfo
  {author} {\bibfnamefont {J.}~\bibnamefont {Schou}},\ and\ \bibinfo {author}
  {\bibfnamefont {O.}~\bibnamefont {Hansen}},\ }\bibfield  {title} {\bibinfo
  {title} {On performance limitations and property correlations of {{Al-doped
  ZnO}} deposited by radio-frequency sputtering},\ }\href
  {https://doi.org/10.1088/0022-3727/49/29/295101} {\bibfield  {journal}
  {\bibinfo  {journal} {Journal of Physics D: Applied Physics}\ }\textbf
  {\bibinfo {volume} {49}},\ \bibinfo {pages} {295101} (\bibinfo {year}
  {2016})}\BibitemShut {NoStop}%
\bibitem [{\citenamefont {Dingle}(1955)}]{Dingle1955}%
  \BibitemOpen
  \bibfield  {author} {\bibinfo {author} {\bibfnamefont {R.}~\bibnamefont
  {Dingle}},\ }\bibfield  {title} {\bibinfo {title} {Scattering of electrons
  and holes by charged donors and acceptors in semiconductors},\ }\href
  {https://doi.org/10.1080/14786440808561235} {\bibfield  {journal} {\bibinfo
  {journal} {The London, Edinburgh, and Dublin Philosophical Magazine and
  Journal of Science}\ }\textbf {\bibinfo {volume} {46}},\ \bibinfo {pages}
  {831} (\bibinfo {year} {1955})}\BibitemShut {NoStop}%
\bibitem [{\citenamefont {Wiley}\ and\ \citenamefont
  {DiDomenico}(1970)}]{Wiley1970}%
  \BibitemOpen
  \bibfield  {author} {\bibinfo {author} {\bibfnamefont {J.~D.}\ \bibnamefont
  {Wiley}}\ and\ \bibinfo {author} {\bibfnamefont {M.}~\bibnamefont
  {DiDomenico}},\ }\bibfield  {title} {\bibinfo {title} {Lattice {{Mobility}}
  of {{Holes}} in {{III-V Compounds}}},\ }\href
  {https://doi.org/10.1103/PhysRevB.2.427} {\bibfield  {journal} {\bibinfo
  {journal} {Physical Review B}\ }\textbf {\bibinfo {volume} {2}},\ \bibinfo
  {pages} {427} (\bibinfo {year} {1970})}\BibitemShut {NoStop}%
\bibitem [{\citenamefont {Ellmer}(2011)}]{ellmerTransparentConductiveZinc2011}%
  \BibitemOpen
  \bibfield  {author} {\bibinfo {author} {\bibfnamefont {K.}~\bibnamefont
  {Ellmer}},\ }\bibfield  {title} {\bibinfo {title} {Transparent {{Conductive
  Zinc Oxide}} and {{Its Derivatives}}},\ }in\ \href
  {https://doi.org/10.1007/978-1-4419-1638-9_7} {\emph {\bibinfo {booktitle}
  {Handbook of {{Transparent Conductors}}}}},\ \bibinfo {editor} {edited by\
  \bibinfo {editor} {\bibfnamefont {D.~S.}\ \bibnamefont {Ginley}}}\ (\bibinfo
  {publisher} {{Springer US}},\ \bibinfo {address} {{Boston, MA}},\ \bibinfo
  {year} {2011})\ pp.\ \bibinfo {pages} {193--263}\BibitemShut {NoStop}%
\bibitem [{\citenamefont {Ravich}\ \emph {et~al.}(1971)\citenamefont {Ravich},
  \citenamefont {Efimova},\ and\ \citenamefont
  {Tamarchenko}}]{ravichScatteringCurrentCarriers1971a}%
  \BibitemOpen
  \bibfield  {author} {\bibinfo {author} {\bibfnamefont {Y.~I.}\ \bibnamefont
  {Ravich}}, \bibinfo {author} {\bibfnamefont {B.~A.}\ \bibnamefont
  {Efimova}},\ and\ \bibinfo {author} {\bibfnamefont {V.~I.}\ \bibnamefont
  {Tamarchenko}},\ }\bibfield  {title} {\bibinfo {title} {Scattering of
  {{Current Carriers}} and {{Transport Phenomena}} in {{Lead Chalcogenides
  II}}. {{Experiment}}},\ }\href {https://doi.org/10.1002/pssb.2220430202}
  {\bibfield  {journal} {\bibinfo  {journal} {physica status solidi (b)}\
  }\textbf {\bibinfo {volume} {43}},\ \bibinfo {pages} {453} (\bibinfo {year}
  {1971})}\BibitemShut {NoStop}%
\bibitem [{\citenamefont {Perdew}\ \emph {et~al.}(1996)\citenamefont {Perdew},
  \citenamefont {Burke},\ and\ \citenamefont {Ernzerhof}}]{Perdew1996}%
  \BibitemOpen
  \bibfield  {author} {\bibinfo {author} {\bibfnamefont {J.~P.}\ \bibnamefont
  {Perdew}}, \bibinfo {author} {\bibfnamefont {K.}~\bibnamefont {Burke}},\ and\
  \bibinfo {author} {\bibfnamefont {M.}~\bibnamefont {Ernzerhof}},\ }\bibfield
  {title} {\bibinfo {title} {Generalized {{Gradient Approximation Made
  Simple}}},\ }\href {https://doi.org/10.1103/PhysRevLett.77.3865} {\bibfield
  {journal} {\bibinfo  {journal} {Physical Review Letters}\ }\textbf {\bibinfo
  {volume} {77}},\ \bibinfo {pages} {3865} (\bibinfo {year}
  {1996})}\BibitemShut {NoStop}%
\bibitem [{\citenamefont {Jain}\ \emph {et~al.}(2013)\citenamefont {Jain},
  \citenamefont {Ong}, \citenamefont {Hautier}, \citenamefont {Chen},
  \citenamefont {Richards}, \citenamefont {Dacek}, \citenamefont {Cholia},
  \citenamefont {Gunter}, \citenamefont {Skinner}, \citenamefont {Ceder},\ and\
  \citenamefont {Persson}}]{Jain2013}%
  \BibitemOpen
  \bibfield  {author} {\bibinfo {author} {\bibfnamefont {A.}~\bibnamefont
  {Jain}}, \bibinfo {author} {\bibfnamefont {S.~P.}\ \bibnamefont {Ong}},
  \bibinfo {author} {\bibfnamefont {G.}~\bibnamefont {Hautier}}, \bibinfo
  {author} {\bibfnamefont {W.}~\bibnamefont {Chen}}, \bibinfo {author}
  {\bibfnamefont {W.~D.}\ \bibnamefont {Richards}}, \bibinfo {author}
  {\bibfnamefont {S.}~\bibnamefont {Dacek}}, \bibinfo {author} {\bibfnamefont
  {S.}~\bibnamefont {Cholia}}, \bibinfo {author} {\bibfnamefont
  {D.}~\bibnamefont {Gunter}}, \bibinfo {author} {\bibfnamefont
  {D.}~\bibnamefont {Skinner}}, \bibinfo {author} {\bibfnamefont
  {G.}~\bibnamefont {Ceder}},\ and\ \bibinfo {author} {\bibfnamefont {K.~A.}\
  \bibnamefont {Persson}},\ }\bibfield  {title} {\bibinfo {title} {Commentary:
  {{The Materials Project}}: {{A}} materials genome approach to accelerating
  materials innovation},\ }\href {https://doi.org/10.1063/1.4812323} {\bibfield
   {journal} {\bibinfo  {journal} {APL Materials}\ }\textbf {\bibinfo {volume}
  {1}},\ \bibinfo {pages} {011002} (\bibinfo {year} {2013})}\BibitemShut
  {NoStop}%
\bibitem [{\citenamefont {Sham}\ and\ \citenamefont
  {Schl{\"u}ter}(1985)}]{Sham1985}%
  \BibitemOpen
  \bibfield  {author} {\bibinfo {author} {\bibfnamefont {L.~J.}\ \bibnamefont
  {Sham}}\ and\ \bibinfo {author} {\bibfnamefont {M.}~\bibnamefont
  {Schl{\"u}ter}},\ }\bibfield  {title} {\bibinfo {title} {Density-functional
  theory of the band gap},\ }\href {https://doi.org/10.1103/PhysRevB.32.3883}
  {\bibfield  {journal} {\bibinfo  {journal} {Physical Review B}\ }\textbf
  {\bibinfo {volume} {32}},\ \bibinfo {pages} {3883} (\bibinfo {year}
  {1985})}\BibitemShut {NoStop}%
\bibitem [{\citenamefont {Kim}\ \emph {et~al.}(2020)\citenamefont {Kim},
  \citenamefont {Lee}, \citenamefont {Hong}, \citenamefont {Yoon},
  \citenamefont {An}, \citenamefont {Lee}, \citenamefont {Jeong}, \citenamefont
  {Yoo}, \citenamefont {Kang}, \citenamefont {Youn},\ and\ \citenamefont
  {Han}}]{Kim2020b}%
  \BibitemOpen
  \bibfield  {author} {\bibinfo {author} {\bibfnamefont {S.}~\bibnamefont
  {Kim}}, \bibinfo {author} {\bibfnamefont {M.}~\bibnamefont {Lee}}, \bibinfo
  {author} {\bibfnamefont {C.}~\bibnamefont {Hong}}, \bibinfo {author}
  {\bibfnamefont {Y.}~\bibnamefont {Yoon}}, \bibinfo {author} {\bibfnamefont
  {H.}~\bibnamefont {An}}, \bibinfo {author} {\bibfnamefont {D.}~\bibnamefont
  {Lee}}, \bibinfo {author} {\bibfnamefont {W.}~\bibnamefont {Jeong}}, \bibinfo
  {author} {\bibfnamefont {D.}~\bibnamefont {Yoo}}, \bibinfo {author}
  {\bibfnamefont {Y.}~\bibnamefont {Kang}}, \bibinfo {author} {\bibfnamefont
  {Y.}~\bibnamefont {Youn}},\ and\ \bibinfo {author} {\bibfnamefont
  {S.}~\bibnamefont {Han}},\ }\bibfield  {title} {\bibinfo {title} {A band-gap
  database for semiconducting inorganic materials calculated with hybrid
  functional},\ }\href {https://doi.org/10.1038/s41597-020-00723-8} {\bibfield
  {journal} {\bibinfo  {journal} {Scientific Data}\ }\textbf {\bibinfo {volume}
  {7}},\ \bibinfo {pages} {1} (\bibinfo {year} {2020})}\BibitemShut {NoStop}%
\bibitem [{\citenamefont {Heyd}\ \emph {et~al.}(2005)\citenamefont {Heyd},
  \citenamefont {Peralta}, \citenamefont {Scuseria},\ and\ \citenamefont
  {Martin}}]{Heyd2005}%
  \BibitemOpen
  \bibfield  {author} {\bibinfo {author} {\bibfnamefont {J.}~\bibnamefont
  {Heyd}}, \bibinfo {author} {\bibfnamefont {J.~E.}\ \bibnamefont {Peralta}},
  \bibinfo {author} {\bibfnamefont {G.~E.}\ \bibnamefont {Scuseria}},\ and\
  \bibinfo {author} {\bibfnamefont {R.~L.}\ \bibnamefont {Martin}},\ }\bibfield
   {title} {\bibinfo {title} {Energy band gaps and lattice parameters evaluated
  with the {{Heyd-Scuseria-Ernzerhof}} screened hybrid functional},\ }\href
  {https://doi.org/10.1063/1.2085170} {\bibfield  {journal} {\bibinfo
  {journal} {The Journal of Chemical Physics}\ }\textbf {\bibinfo {volume}
  {123}},\ \bibinfo {pages} {174101} (\bibinfo {year} {2005})}\BibitemShut
  {NoStop}%
\bibitem [{\citenamefont {Ashcroft}\ and\ \citenamefont
  {Mermin}(1976)}]{Ashcroft1976}%
  \BibitemOpen
  \bibfield  {author} {\bibinfo {author} {\bibfnamefont {N.~W.}\ \bibnamefont
  {Ashcroft}}\ and\ \bibinfo {author} {\bibfnamefont {N.~D.}\ \bibnamefont
  {Mermin}},\ }\href@noop {} {\emph {\bibinfo {title} {Solid {{State
  Physics}}}}}\ (\bibinfo  {publisher} {{Holt, Rinehart, and Winston}},\
  \bibinfo {year} {1976})\BibitemShut {NoStop}%
\bibitem [{\citenamefont {Zhang}\ \emph {et~al.}(2005)\citenamefont {Zhang},
  \citenamefont {Tan}, \citenamefont {Stormer},\ and\ \citenamefont
  {Kim}}]{Zhang2005}%
  \BibitemOpen
  \bibfield  {author} {\bibinfo {author} {\bibfnamefont {Y.}~\bibnamefont
  {Zhang}}, \bibinfo {author} {\bibfnamefont {Y.-W.}\ \bibnamefont {Tan}},
  \bibinfo {author} {\bibfnamefont {H.~L.}\ \bibnamefont {Stormer}},\ and\
  \bibinfo {author} {\bibfnamefont {P.}~\bibnamefont {Kim}},\ }\bibfield
  {title} {\bibinfo {title} {Experimental observation of the quantum {{Hall}}
  effect and {{Berry}}'s phase in graphene},\ }\href
  {https://doi.org/10.1038/nature04235} {\bibfield  {journal} {\bibinfo
  {journal} {Nature}\ }\textbf {\bibinfo {volume} {438}},\ \bibinfo {pages}
  {201} (\bibinfo {year} {2005})}\BibitemShut {NoStop}%
\bibitem [{\citenamefont {Madsen}\ \emph {et~al.}(2018)\citenamefont {Madsen},
  \citenamefont {Carrete},\ and\ \citenamefont {Verstraete}}]{Madsen2018}%
  \BibitemOpen
  \bibfield  {author} {\bibinfo {author} {\bibfnamefont {G.~K.}\ \bibnamefont
  {Madsen}}, \bibinfo {author} {\bibfnamefont {J.}~\bibnamefont {Carrete}},\
  and\ \bibinfo {author} {\bibfnamefont {M.~J.}\ \bibnamefont {Verstraete}},\
  }\bibfield  {title} {\bibinfo {title} {{{BoltzTraP2}}, a program for
  interpolating band structures and calculating semi-classical transport
  coefficients},\ }\href {https://doi.org/10.1016/j.cpc.2018.05.010} {\bibfield
   {journal} {\bibinfo  {journal} {Computer Physics Communications}\ }\textbf
  {\bibinfo {volume} {231}},\ \bibinfo {pages} {140} (\bibinfo {year}
  {2018})}\BibitemShut {NoStop}%
\bibitem [{\citenamefont {Ricci}\ \emph {et~al.}(2017)\citenamefont {Ricci},
  \citenamefont {Chen}, \citenamefont {Aydemir}, \citenamefont {Snyder},
  \citenamefont {Rignanese}, \citenamefont {Jain},\ and\ \citenamefont
  {Hautier}}]{Ricci2017}%
  \BibitemOpen
  \bibfield  {author} {\bibinfo {author} {\bibfnamefont {F.}~\bibnamefont
  {Ricci}}, \bibinfo {author} {\bibfnamefont {W.}~\bibnamefont {Chen}},
  \bibinfo {author} {\bibfnamefont {U.}~\bibnamefont {Aydemir}}, \bibinfo
  {author} {\bibfnamefont {G.~J.}\ \bibnamefont {Snyder}}, \bibinfo {author}
  {\bibfnamefont {G.-M.}\ \bibnamefont {Rignanese}}, \bibinfo {author}
  {\bibfnamefont {A.}~\bibnamefont {Jain}},\ and\ \bibinfo {author}
  {\bibfnamefont {G.}~\bibnamefont {Hautier}},\ }\bibfield  {title} {\bibinfo
  {title} {An ab initio electronic transport database for inorganic
  materials},\ }\href {https://doi.org/10.1038/sdata.2017.85} {\bibfield
  {journal} {\bibinfo  {journal} {Scientific Data}\ }\textbf {\bibinfo {volume}
  {4}},\ \bibinfo {pages} {170085} (\bibinfo {year} {2017})}\BibitemShut
  {NoStop}%
\bibitem [{\citenamefont {May}\ and\ \citenamefont {Snyder}(2012)}]{May2012}%
  \BibitemOpen
  \bibfield  {author} {\bibinfo {author} {\bibfnamefont {A.~F.}\ \bibnamefont
  {May}}\ and\ \bibinfo {author} {\bibfnamefont {G.~J.}\ \bibnamefont
  {Snyder}},\ }\bibfield  {title} {\bibinfo {title} {Introduction to modeling
  thermoelectric transport at high temperatures},\ }in\ \href
  {https://doi.org/10.1201/b11891} {\emph {\bibinfo {booktitle}
  {Thermoelectrics and Its {{Energy Harvesting}}}}},\ \bibinfo {editor} {edited
  by\ \bibinfo {editor} {\bibfnamefont {D.~M.}\ \bibnamefont {Rowe}}}\
  (\bibinfo  {publisher} {{CRC Press}},\ \bibinfo {year} {2012})\BibitemShut
  {NoStop}%
\bibitem [{\citenamefont {Crovetto}\ and\ \citenamefont
  {Hansen}(2017)}]{Crovetto2017b}%
  \BibitemOpen
  \bibfield  {author} {\bibinfo {author} {\bibfnamefont {A.}~\bibnamefont
  {Crovetto}}\ and\ \bibinfo {author} {\bibfnamefont {O.}~\bibnamefont
  {Hansen}},\ }\bibfield  {title} {\bibinfo {title} {What is the band alignment
  of {{Cu}}$_2${{ZnSn}}({{S}},{{Se}})$_4$ solar cells?},\ }\href
  {https://doi.org/10.1016/j.solmat.2017.05.008} {\bibfield  {journal}
  {\bibinfo  {journal} {Solar Energy Materials and Solar Cells}\ }\textbf
  {\bibinfo {volume} {169}},\ \bibinfo {pages} {177} (\bibinfo {year}
  {2017})}\BibitemShut {NoStop}%
\bibitem [{\citenamefont {Mortensen}\ \emph {et~al.}(2005)\citenamefont
  {Mortensen}, \citenamefont {Hansen},\ and\ \citenamefont
  {Jacobsen}}]{Mortensen2005}%
  \BibitemOpen
  \bibfield  {author} {\bibinfo {author} {\bibfnamefont {J.}~\bibnamefont
  {Mortensen}}, \bibinfo {author} {\bibfnamefont {L.}~\bibnamefont {Hansen}},\
  and\ \bibinfo {author} {\bibfnamefont {K.}~\bibnamefont {Jacobsen}},\
  }\bibfield  {title} {\bibinfo {title} {Real-space grid implementation of the
  projector augmented wave method},\ }\href
  {https://doi.org/10.1103/PhysRevB.71.035109} {\bibfield  {journal} {\bibinfo
  {journal} {Physical Review B}\ }\textbf {\bibinfo {volume} {71}},\ \bibinfo
  {pages} {035109} (\bibinfo {year} {2005})}\BibitemShut {NoStop}%
\bibitem [{\citenamefont {Bertoni}\ \emph {et~al.}(2019)\citenamefont
  {Bertoni}, \citenamefont {Ramasse}, \citenamefont {Brescia}, \citenamefont
  {De~Trizio}, \citenamefont {De~Donato},\ and\ \citenamefont
  {Manna}}]{Bertoni2019}%
  \BibitemOpen
  \bibfield  {author} {\bibinfo {author} {\bibfnamefont {G.}~\bibnamefont
  {Bertoni}}, \bibinfo {author} {\bibfnamefont {Q.}~\bibnamefont {Ramasse}},
  \bibinfo {author} {\bibfnamefont {R.}~\bibnamefont {Brescia}}, \bibinfo
  {author} {\bibfnamefont {L.}~\bibnamefont {De~Trizio}}, \bibinfo {author}
  {\bibfnamefont {F.}~\bibnamefont {De~Donato}},\ and\ \bibinfo {author}
  {\bibfnamefont {L.}~\bibnamefont {Manna}},\ }\bibfield  {title} {\bibinfo
  {title} {Direct {{Quantification}} of {{Cu Vacancies}} and {{Spatial
  Localization}} of {{Surface Plasmon Resonances}} in {{Copper Phosphide
  Nanocrystals}}},\ }\href {https://doi.org/10.1021/acsmaterialslett.9b00412}
  {\bibfield  {journal} {\bibinfo  {journal} {ACS Materials Letters}\ }\textbf
  {\bibinfo {volume} {1}},\ \bibinfo {pages} {665} (\bibinfo {year}
  {2019})}\BibitemShut {NoStop}%
\bibitem [{\citenamefont {Jasperson}\ and\ \citenamefont
  {Schnatterly}(1969)}]{Jasperson1969}%
  \BibitemOpen
  \bibfield  {author} {\bibinfo {author} {\bibfnamefont {S.~N.}\ \bibnamefont
  {Jasperson}}\ and\ \bibinfo {author} {\bibfnamefont {S.~E.}\ \bibnamefont
  {Schnatterly}},\ }\bibfield  {title} {\bibinfo {title}
  {Photon-{{Surface-Plasmon Coupling}} in {{Thick Ag Foils}}},\ }\href
  {https://doi.org/10.1103/PhysRev.188.759} {\bibfield  {journal} {\bibinfo
  {journal} {Physical Review}\ }\textbf {\bibinfo {volume} {188}},\ \bibinfo
  {pages} {759} (\bibinfo {year} {1969})}\BibitemShut {NoStop}%
\bibitem [{\citenamefont {Kaspar}\ and\ \citenamefont
  {Kreibig}(1977)}]{Kaspar1977}%
  \BibitemOpen
  \bibfield  {author} {\bibinfo {author} {\bibfnamefont {W.}~\bibnamefont
  {Kaspar}}\ and\ \bibinfo {author} {\bibfnamefont {U.}~\bibnamefont
  {Kreibig}},\ }\bibfield  {title} {\bibinfo {title} {Surface structure
  influences on the absorptance of thick silver films},\ }\href
  {https://doi.org/10.1016/0039-6028(77)90137-6} {\bibfield  {journal}
  {\bibinfo  {journal} {Surface Science}\ }\textbf {\bibinfo {volume} {69}},\
  \bibinfo {pages} {619} (\bibinfo {year} {1977})}\BibitemShut {NoStop}%
\bibitem [{\citenamefont {Burstein}(1954)}]{Burstein1954}%
  \BibitemOpen
  \bibfield  {author} {\bibinfo {author} {\bibfnamefont {E.}~\bibnamefont
  {Burstein}},\ }\bibfield  {title} {\bibinfo {title} {Anomalous {{Optical
  Absorption Limit}} in {{InSb}}},\ }\href
  {https://doi.org/10.1103/PhysRev.93.632} {\bibfield  {journal} {\bibinfo
  {journal} {Physical Review}\ }\textbf {\bibinfo {volume} {93}},\ \bibinfo
  {pages} {632} (\bibinfo {year} {1954})}\BibitemShut {NoStop}%
\bibitem [{\citenamefont {Sernelius}\ \emph {et~al.}(1988)\citenamefont
  {Sernelius}, \citenamefont {Berggren}, \citenamefont {Jin}, \citenamefont
  {Hamberg},\ and\ \citenamefont {Granqvist}}]{Sernelius1988}%
  \BibitemOpen
  \bibfield  {author} {\bibinfo {author} {\bibfnamefont {B.~E.}\ \bibnamefont
  {Sernelius}}, \bibinfo {author} {\bibfnamefont {K.-F.}\ \bibnamefont
  {Berggren}}, \bibinfo {author} {\bibfnamefont {Z.-C.}\ \bibnamefont {Jin}},
  \bibinfo {author} {\bibfnamefont {I.}~\bibnamefont {Hamberg}},\ and\ \bibinfo
  {author} {\bibfnamefont {C.~G.}\ \bibnamefont {Granqvist}},\ }\bibfield
  {title} {\bibinfo {title} {Band-gap tailoring of {{ZnO}} by means of heavy
  {{Al}} doping},\ }\href {https://doi.org/10.1103/PhysRevB.37.10244}
  {\bibfield  {journal} {\bibinfo  {journal} {Physical Review B}\ }\textbf
  {\bibinfo {volume} {37}},\ \bibinfo {pages} {10244} (\bibinfo {year}
  {1988})}\BibitemShut {NoStop}%
\bibitem [{\citenamefont {Momma}\ and\ \citenamefont
  {Izumi}(2011)}]{Momma2011}%
  \BibitemOpen
  \bibfield  {author} {\bibinfo {author} {\bibfnamefont {K.}~\bibnamefont
  {Momma}}\ and\ \bibinfo {author} {\bibfnamefont {F.}~\bibnamefont {Izumi}},\
  }\bibfield  {title} {\bibinfo {title} {{{VESTA}} 3 for three-dimensional
  visualization of crystal, volumetric and morphology data},\ }\href
  {https://doi.org/10.1107/S0021889811038970} {\bibfield  {journal} {\bibinfo
  {journal} {Journal of Applied Crystallography}\ }\textbf {\bibinfo {volume}
  {44}},\ \bibinfo {pages} {1272} (\bibinfo {year} {2011})}\BibitemShut
  {NoStop}%
\bibitem [{\citenamefont {Perdew}\ \emph {et~al.}(2008)\citenamefont {Perdew},
  \citenamefont {Ruzsinszky}, \citenamefont {Csonka}, \citenamefont {Vydrov},
  \citenamefont {Scuseria}, \citenamefont {Constantin}, \citenamefont {Zhou},\
  and\ \citenamefont {Burke}}]{Perdew2008}%
  \BibitemOpen
  \bibfield  {author} {\bibinfo {author} {\bibfnamefont {J.~P.}\ \bibnamefont
  {Perdew}}, \bibinfo {author} {\bibfnamefont {A.}~\bibnamefont {Ruzsinszky}},
  \bibinfo {author} {\bibfnamefont {G.~I.}\ \bibnamefont {Csonka}}, \bibinfo
  {author} {\bibfnamefont {O.~A.}\ \bibnamefont {Vydrov}}, \bibinfo {author}
  {\bibfnamefont {G.~E.}\ \bibnamefont {Scuseria}}, \bibinfo {author}
  {\bibfnamefont {L.~A.}\ \bibnamefont {Constantin}}, \bibinfo {author}
  {\bibfnamefont {X.}~\bibnamefont {Zhou}},\ and\ \bibinfo {author}
  {\bibfnamefont {K.}~\bibnamefont {Burke}},\ }\bibfield  {title} {\bibinfo
  {title} {Restoring the {{Density-Gradient Expansion}} for {{Exchange}} in
  {{Solids}} and {{Surfaces}}},\ }\href
  {https://doi.org/10.1103/PhysRevLett.100.136406} {\bibfield  {journal}
  {\bibinfo  {journal} {Physical Review Letters}\ }\textbf {\bibinfo {volume}
  {100}},\ \bibinfo {pages} {136406} (\bibinfo {year} {2008})}\BibitemShut
  {NoStop}%
\bibitem [{\citenamefont {Weber}\ \emph {et~al.}(1994)\citenamefont {Weber},
  \citenamefont {Sutter},\ and\ \citenamefont {{von K{\"a}nel}}}]{Weber1998}%
  \BibitemOpen
  \bibfield  {author} {\bibinfo {author} {\bibfnamefont {A.}~\bibnamefont
  {Weber}}, \bibinfo {author} {\bibfnamefont {P.}~\bibnamefont {Sutter}},\ and\
  \bibinfo {author} {\bibfnamefont {H.}~\bibnamefont {{von K{\"a}nel}}},\
  }\bibfield  {title} {\bibinfo {title} {Optical, electrical, and
  photoelectrical properties of sputtered thin amorphous {{Zn}}$_3${{P}}$_2$
  films},\ }\href {https://doi.org/10.1063/1.356613} {\bibfield  {journal}
  {\bibinfo  {journal} {Journal of Applied Physics}\ }\textbf {\bibinfo
  {volume} {75}},\ \bibinfo {pages} {7448} (\bibinfo {year}
  {1994})}\BibitemShut {NoStop}%
\bibitem [{\citenamefont {Zawadzki}\ \emph {et~al.}(2015)\citenamefont
  {Zawadzki}, \citenamefont {Zakutayev},\ and\ \citenamefont
  {Lany}}]{Zawadzki2015}%
  \BibitemOpen
  \bibfield  {author} {\bibinfo {author} {\bibfnamefont {P.}~\bibnamefont
  {Zawadzki}}, \bibinfo {author} {\bibfnamefont {A.}~\bibnamefont
  {Zakutayev}},\ and\ \bibinfo {author} {\bibfnamefont {S.}~\bibnamefont
  {Lany}},\ }\bibfield  {title} {\bibinfo {title} {Entropy-{{Driven
  Clustering}} in {{Tetrahedrally Bonded Multinary Materials}}},\ }\href
  {https://doi.org/10.1103/PhysRevApplied.3.034007} {\bibfield  {journal}
  {\bibinfo  {journal} {Physical Review Applied}\ }\textbf {\bibinfo {volume}
  {3}},\ \bibinfo {pages} {034007} (\bibinfo {year} {2015})}\BibitemShut
  {NoStop}%
\bibitem [{\citenamefont {Fang}\ \emph {et~al.}(1997)\citenamefont {Fang},
  \citenamefont {{de Groot}},\ and\ \citenamefont {Haas}}]{Fang1997}%
  \BibitemOpen
  \bibfield  {author} {\bibinfo {author} {\bibfnamefont {C.~M.}\ \bibnamefont
  {Fang}}, \bibinfo {author} {\bibfnamefont {R.~A.}\ \bibnamefont {{de
  Groot}}},\ and\ \bibinfo {author} {\bibfnamefont {C.}~\bibnamefont {Haas}},\
  }\bibfield  {title} {\bibinfo {title} {Bulk and surface electronic structure
  of {{1T-TiS$_2$}} and {{1T-TiSe$_2$}}},\ }\href
  {https://doi.org/10.1103/PhysRevB.56.4455} {\bibfield  {journal} {\bibinfo
  {journal} {Physical Review B}\ }\textbf {\bibinfo {volume} {56}},\ \bibinfo
  {pages} {4455} (\bibinfo {year} {1997})}\BibitemShut {NoStop}%
\bibitem [{\citenamefont {Shockley}\ and\ \citenamefont
  {Queisser}(1961)}]{Shockley1961}%
  \BibitemOpen
  \bibfield  {author} {\bibinfo {author} {\bibfnamefont {W.}~\bibnamefont
  {Shockley}}\ and\ \bibinfo {author} {\bibfnamefont {H.~J.}\ \bibnamefont
  {Queisser}},\ }\bibfield  {title} {\bibinfo {title} {Detailed {{Balance
  Limit}} of {{Efficiency}} of p-n {{Junction Solar Cells}}},\ }\href
  {https://doi.org/10.1063/1.1736034} {\bibfield  {journal} {\bibinfo
  {journal} {Journal of Applied Physics}\ }\textbf {\bibinfo {volume} {32}},\
  \bibinfo {pages} {510} (\bibinfo {year} {1961})}\BibitemShut {NoStop}%
\bibitem [{\citenamefont {Crovetto}\ \emph {et~al.}(2018)\citenamefont
  {Crovetto}, \citenamefont {Cazzaniga}, \citenamefont {Ettlinger},
  \citenamefont {Schou},\ and\ \citenamefont {Hansen}}]{Crovetto2018a}%
  \BibitemOpen
  \bibfield  {author} {\bibinfo {author} {\bibfnamefont {A.}~\bibnamefont
  {Crovetto}}, \bibinfo {author} {\bibfnamefont {A.}~\bibnamefont {Cazzaniga}},
  \bibinfo {author} {\bibfnamefont {R.~B.}\ \bibnamefont {Ettlinger}}, \bibinfo
  {author} {\bibfnamefont {J.}~\bibnamefont {Schou}},\ and\ \bibinfo {author}
  {\bibfnamefont {O.}~\bibnamefont {Hansen}},\ }\bibfield  {title} {\bibinfo
  {title} {Large process-dependent variations in band alignment and interface
  band gaps of {{Cu$_2$ZnSnS$_4$}}/{{CdS}} solar cells},\ }\href
  {https://doi.org/10.1016/j.solmat.2018.08.005} {\bibfield  {journal}
  {\bibinfo  {journal} {Solar Energy Materials and Solar Cells}\ }\textbf
  {\bibinfo {volume} {187}},\ \bibinfo {pages} {233} (\bibinfo {year}
  {2018})}\BibitemShut {NoStop}%
\bibitem [{\citenamefont {Gaspari}\ \emph {et~al.}(2016)\citenamefont
  {Gaspari}, \citenamefont {Labat}, \citenamefont {Manna}, \citenamefont
  {Adamo},\ and\ \citenamefont
  {Cavalli}}]{gaspariSemiconductingOpticalProperties2016}%
  \BibitemOpen
  \bibfield  {author} {\bibinfo {author} {\bibfnamefont {R.}~\bibnamefont
  {Gaspari}}, \bibinfo {author} {\bibfnamefont {F.}~\bibnamefont {Labat}},
  \bibinfo {author} {\bibfnamefont {L.}~\bibnamefont {Manna}}, \bibinfo
  {author} {\bibfnamefont {C.}~\bibnamefont {Adamo}},\ and\ \bibinfo {author}
  {\bibfnamefont {A.}~\bibnamefont {Cavalli}},\ }\bibfield  {title} {\bibinfo
  {title} {Semiconducting and optical properties of selected binary compounds
  by linear response {{DFT}}+{{U}} and hybrid functional methods},\ }\href
  {https://doi.org/10.1007/s00214-016-1833-9} {\bibfield  {journal} {\bibinfo
  {journal} {Theoretical Chemistry Accounts}\ }\textbf {\bibinfo {volume}
  {135}},\ \bibinfo {pages} {73} (\bibinfo {year} {2016})}\BibitemShut
  {NoStop}%
\bibitem [{\citenamefont {Anisimov}\ \emph {et~al.}(1991)\citenamefont
  {Anisimov}, \citenamefont {Zaanen},\ and\ \citenamefont
  {Andersen}}]{Anisimov1991}%
  \BibitemOpen
  \bibfield  {author} {\bibinfo {author} {\bibfnamefont {V.~I.}\ \bibnamefont
  {Anisimov}}, \bibinfo {author} {\bibfnamefont {J.}~\bibnamefont {Zaanen}},\
  and\ \bibinfo {author} {\bibfnamefont {O.~K.}\ \bibnamefont {Andersen}},\
  }\bibfield  {title} {\bibinfo {title} {Band theory and {{Mott}} insulators:
  {{Hubbard U}} instead of {{Stoner I}}},\ }\href
  {https://doi.org/10.1103/PhysRevB.44.943} {\bibfield  {journal} {\bibinfo
  {journal} {Physical Review B}\ }\textbf {\bibinfo {volume} {44}},\ \bibinfo
  {pages} {943} (\bibinfo {year} {1991})}\BibitemShut {NoStop}%
\end{thebibliography}%


\begin{thebibliography}{16}%
\makeatletter
\providecommand \@ifxundefined [1]{%
 \@ifx{#1\undefined}
}%
\providecommand \@ifnum [1]{%
 \ifnum #1\expandafter \@firstoftwo
 \else \expandafter \@secondoftwo
 \fi
}%
\providecommand \@ifx [1]{%
 \ifx #1\expandafter \@firstoftwo
 \else \expandafter \@secondoftwo
 \fi
}%
\providecommand \natexlab [1]{#1}%
\providecommand \enquote  [1]{``#1''}%
\providecommand \bibnamefont  [1]{#1}%
\providecommand \bibfnamefont [1]{#1}%
\providecommand \citenamefont [1]{#1}%
\providecommand \href@noop [0]{\@secondoftwo}%
\providecommand \href [0]{\begingroup \@sanitize@url \@href}%
\providecommand \@href[1]{\@@startlink{#1}\@@href}%
\providecommand \@@href[1]{\endgroup#1\@@endlink}%
\providecommand \@sanitize@url [0]{\catcode `\\12\catcode `\$12\catcode
  `\&12\catcode `\#12\catcode `\^12\catcode `\_12\catcode `\%12\relax}%
\providecommand \@@startlink[1]{}%
\providecommand \@@endlink[0]{}%
\providecommand \url  [0]{\begingroup\@sanitize@url \@url }%
\providecommand \@url [1]{\endgroup\@href {#1}{\urlprefix }}%
\providecommand \urlprefix  [0]{URL }%
\providecommand \Eprint [0]{\href }%
\providecommand \doibase [0]{https://doi.org/}%
\providecommand \selectlanguage [0]{\@gobble}%
\providecommand \bibinfo  [0]{\@secondoftwo}%
\providecommand \bibfield  [0]{\@secondoftwo}%
\providecommand \translation [1]{[#1]}%
\providecommand \BibitemOpen [0]{}%
\providecommand \bibitemStop [0]{}%
\providecommand \bibitemNoStop [0]{.\EOS\space}%
\providecommand \EOS [0]{\spacefactor3000\relax}%
\providecommand \BibitemShut  [1]{\csname bibitem#1\endcsname}%
\let\auto@bib@innerbib\@empty
\bibitem [{\citenamefont {Talley}\ \emph {et~al.}(2019)\citenamefont {Talley},
  \citenamefont {Bauers}, \citenamefont {Melamed}, \citenamefont {Papac},
  \citenamefont {Heinselman}, \citenamefont {Khan}, \citenamefont {Roberts},
  \citenamefont {Jacobson}, \citenamefont {Mis}, \citenamefont {Brennecka},
  \citenamefont {Perkins},\ and\ \citenamefont {Zakutayev}}]{SITalley2019}%
  \BibitemOpen
  \bibfield  {author} {\bibinfo {author} {\bibfnamefont {K.~R.}\ \bibnamefont
  {Talley}}, \bibinfo {author} {\bibfnamefont {S.~R.}\ \bibnamefont {Bauers}},
  \bibinfo {author} {\bibfnamefont {C.~L.}\ \bibnamefont {Melamed}}, \bibinfo
  {author} {\bibfnamefont {M.~C.}\ \bibnamefont {Papac}}, \bibinfo {author}
  {\bibfnamefont {K.~N.}\ \bibnamefont {Heinselman}}, \bibinfo {author}
  {\bibfnamefont {I.}~\bibnamefont {Khan}}, \bibinfo {author} {\bibfnamefont
  {D.~M.}\ \bibnamefont {Roberts}}, \bibinfo {author} {\bibfnamefont
  {V.}~\bibnamefont {Jacobson}}, \bibinfo {author} {\bibfnamefont
  {A.}~\bibnamefont {Mis}}, \bibinfo {author} {\bibfnamefont {G.~L.}\
  \bibnamefont {Brennecka}}, \bibinfo {author} {\bibfnamefont {J.~D.}\
  \bibnamefont {Perkins}},\ and\ \bibinfo {author} {\bibfnamefont
  {A.}~\bibnamefont {Zakutayev}},\ }\bibfield  {title} {\bibinfo {title}
  {{COMBIgor: Data-Analysis Package for Combinatorial Materials Science}},\
  }\href {https://doi.org/10.1021/acscombsci.9b00077} {\bibfield  {journal}
  {\bibinfo  {journal} {ACS Combinatorial Science}\ }\textbf {\bibinfo {volume}
  {21}},\ \bibinfo {pages} {537} (\bibinfo {year} {2019})}\BibitemShut
  {NoStop}%
\bibitem [{\citenamefont {Talley}\ \emph {et~al.}(2021)\citenamefont {Talley},
  \citenamefont {White}, \citenamefont {Wunder}, \citenamefont {Eash},
  \citenamefont {Schwarting}, \citenamefont {Evenson}, \citenamefont {Perkins},
  \citenamefont {Tumas}, \citenamefont {Munch}, \citenamefont {Phillips},\ and\
  \citenamefont {Zakutayev}}]{SITalley2021a}%
  \BibitemOpen
  \bibfield  {author} {\bibinfo {author} {\bibfnamefont {K.~R.}\ \bibnamefont
  {Talley}}, \bibinfo {author} {\bibfnamefont {R.}~\bibnamefont {White}},
  \bibinfo {author} {\bibfnamefont {N.}~\bibnamefont {Wunder}}, \bibinfo
  {author} {\bibfnamefont {M.}~\bibnamefont {Eash}}, \bibinfo {author}
  {\bibfnamefont {M.}~\bibnamefont {Schwarting}}, \bibinfo {author}
  {\bibfnamefont {D.}~\bibnamefont {Evenson}}, \bibinfo {author} {\bibfnamefont
  {J.~D.}\ \bibnamefont {Perkins}}, \bibinfo {author} {\bibfnamefont
  {W.}~\bibnamefont {Tumas}}, \bibinfo {author} {\bibfnamefont
  {K.}~\bibnamefont {Munch}}, \bibinfo {author} {\bibfnamefont
  {C.}~\bibnamefont {Phillips}},\ and\ \bibinfo {author} {\bibfnamefont
  {A.}~\bibnamefont {Zakutayev}},\ }\bibfield  {title} {\bibinfo {title}
  {{Research data infrastructure for high-throughput experimental materials
  science}},\ }\href {https://doi.org/10.1016/j.patter.2021.100373} {\bibfield
  {journal} {\bibinfo  {journal} {Patterns}\ }\textbf {\bibinfo {volume} {2}},\
  \bibinfo {pages} {100373} (\bibinfo {year} {2021})} \BibitemShut {NoStop}%
\bibitem [{\citenamefont {Zakutayev}\ \emph {et~al.}(2018)\citenamefont
  {Zakutayev}, \citenamefont {Wunder}, \citenamefont {Schwarting},
  \citenamefont {Perkins}, \citenamefont {White}, \citenamefont {Munch},
  \citenamefont {Tumas},\ and\ \citenamefont {Phillips}}]{SIZakutayev2018}%
  \BibitemOpen
  \bibfield  {author} {\bibinfo {author} {\bibfnamefont {A.}~\bibnamefont
  {Zakutayev}}, \bibinfo {author} {\bibfnamefont {N.}~\bibnamefont {Wunder}},
  \bibinfo {author} {\bibfnamefont {M.}~\bibnamefont {Schwarting}}, \bibinfo
  {author} {\bibfnamefont {J.~D.}\ \bibnamefont {Perkins}}, \bibinfo {author}
  {\bibfnamefont {R.}~\bibnamefont {White}}, \bibinfo {author} {\bibfnamefont
  {K.}~\bibnamefont {Munch}}, \bibinfo {author} {\bibfnamefont
  {W.}~\bibnamefont {Tumas}},\ and\ \bibinfo {author} {\bibfnamefont
  {C.}~\bibnamefont {Phillips}},\ }\bibfield  {title} {\bibinfo {title} {{An
  open experimental database for exploring inorganic materials}},\ }\href
  {https://doi.org/10.1038/sdata.2018.53} {\bibfield  {journal} {\bibinfo
  {journal} {Scientific Data}\ }\textbf {\bibinfo {volume} {5}},\ \bibinfo
  {pages} {180053} (\bibinfo {year} {2018})}\BibitemShut {NoStop}%
\bibitem [{\citenamefont {Crovetto}\ \emph
  {et~al.}(2022{\natexlab{a}})\citenamefont {Crovetto}, \citenamefont {Kojda},
  \citenamefont {Yi}, \citenamefont {Heinselman}, \citenamefont {LaVan},
  \citenamefont {Habicht}, \citenamefont {Unold},\ and\ \citenamefont
  {Zakutayev}}]{SICrovetto2022b}%
  \BibitemOpen
  \bibfield  {author} {\bibinfo {author} {\bibfnamefont {A.}~\bibnamefont
  {Crovetto}}, \bibinfo {author} {\bibfnamefont {D.}~\bibnamefont {Kojda}},
  \bibinfo {author} {\bibfnamefont {F.}~\bibnamefont {Yi}}, \bibinfo {author}
  {\bibfnamefont {K.~N.}\ \bibnamefont {Heinselman}}, \bibinfo {author}
  {\bibfnamefont {D.~A.}\ \bibnamefont {LaVan}}, \bibinfo {author}
  {\bibfnamefont {K.}~\bibnamefont {Habicht}}, \bibinfo {author} {\bibfnamefont
  {T.}~\bibnamefont {Unold}},\ and\ \bibinfo {author} {\bibfnamefont
  {A.}~\bibnamefont {Zakutayev}},\ }\bibfield  {title} {\bibinfo {title}
  {Crystallize {{It}} before {{It Diffuses}}: {{Kinetic Stabilization}} of
  {{Thin-Film Phosphorus-Rich Semiconductor CuP$_2$}}},\ }\href
  {https://doi.org/10.1021/jacs.2c04868} {\bibfield  {journal} {\bibinfo
  {journal} {Journal of the American Chemical Society}\ }\textbf {\bibinfo
  {volume} {144}},\ \bibinfo {pages} {13334} (\bibinfo {year}
  {2022}{\natexlab{a}})}\BibitemShut {NoStop}%
\bibitem [{\citenamefont {Verma}(2007)}]{SIVerma2007}%
  \BibitemOpen
  \bibfield  {author} {\bibinfo {author} {\bibfnamefont {H.~R.}\ \bibnamefont
  {Verma}},\ }\href {https://doi.org/10.1007/978-3-540-30279-7/COVER} {\emph
  {\bibinfo {title} {{Atomic and nuclear analytical methods: XRF,
  m{\"{o}}ssbauer, XPS, NAA and ion-beam spectroscopic techniques}}}}\
  (\bibinfo  {publisher} {Springer},\ \bibinfo {address} {Berlin Heidelberg},\
  \bibinfo {year} {2007})\BibitemShut {NoStop}%
\bibitem [{\citenamefont {Bl{\"{o}}chl}(1994)}]{SIBlochl1994}%
  \BibitemOpen
  \bibfield  {author} {\bibinfo {author} {\bibfnamefont {P.~E.}\ \bibnamefont
  {Bl{\"{o}}chl}},\ }\bibfield  {title} {\bibinfo {title} {{Projector
  augmented-wave method}},\ }\href {https://doi.org/10.1103/PhysRevB.50.17953}
  {\bibfield  {journal} {\bibinfo  {journal} {Physical Review B}\ }\textbf
  {\bibinfo {volume} {50}},\ \bibinfo {pages} {17953} (\bibinfo {year}
  {1994})}\BibitemShut {NoStop}%
\bibitem [{\citenamefont {Mortensen}\ \emph {et~al.}(2005)\citenamefont
  {Mortensen}, \citenamefont {Hansen},\ and\ \citenamefont
  {Jacobsen}}]{SIMortensen2005}%
  \BibitemOpen
  \bibfield  {author} {\bibinfo {author} {\bibfnamefont {J.}~\bibnamefont
  {Mortensen}}, \bibinfo {author} {\bibfnamefont {L.}~\bibnamefont {Hansen}},\
  and\ \bibinfo {author} {\bibfnamefont {K.}~\bibnamefont {Jacobsen}},\
  }\bibfield  {title} {\bibinfo {title} {{Real-space grid implementation of the
  projector augmented wave method}},\ }\href
  {https://doi.org/10.1103/PhysRevB.71.035109} {\bibfield  {journal} {\bibinfo
  {journal} {Physical Review B}\ }\textbf {\bibinfo {volume} {71}},\ \bibinfo
  {pages} {035109} (\bibinfo {year} {2005})}\BibitemShut {NoStop}%
\bibitem [{\citenamefont {Enkovaara}\ \emph {et~al.}(2010)\citenamefont
  {Enkovaara}, \citenamefont {Rostgaard}, \citenamefont {Mortensen},
  \citenamefont {Chen}, \citenamefont {Du{\l}ak}, \citenamefont {Ferrighi},
  \citenamefont {Gavnholt}, \citenamefont {Glinsvad}, \citenamefont {Haikola},
  \citenamefont {Hansen}, \citenamefont {Kristoffersen}, \citenamefont
  {Kuisma}, \citenamefont {Larsen}, \citenamefont {Lehtovaara}, \citenamefont
  {Ljungberg}, \citenamefont {Lopez-Acevedo}, \citenamefont {Moses},
  \citenamefont {Ojanen}, \citenamefont {Olsen}, \citenamefont {Petzold},
  \citenamefont {Romero}, \citenamefont {Stausholm-M{\o}ller}, \citenamefont
  {Strange}, \citenamefont {Tritsaris}, \citenamefont {Vanin}, \citenamefont
  {Walter}, \citenamefont {Hammer}, \citenamefont {H{\"{a}}kkinen},
  \citenamefont {Madsen}, \citenamefont {Nieminen}, \citenamefont {N{\o}rskov},
  \citenamefont {Puska}, \citenamefont {Rantala}, \citenamefont {Schi{\o}tz},
  \citenamefont {Thygesen},\ and\ \citenamefont {Jacobsen}}]{SIEnkovaara2010}%
  \BibitemOpen
  \bibfield  {author} {\bibinfo {author} {\bibfnamefont {J.}~\bibnamefont
  {Enkovaara}}, \bibinfo {author} {\bibfnamefont {C.}~\bibnamefont
  {Rostgaard}}, \bibinfo {author} {\bibfnamefont {J.~J.}\ \bibnamefont
  {Mortensen}}, \bibinfo {author} {\bibfnamefont {J.}~\bibnamefont {Chen}},
  \bibinfo {author} {\bibfnamefont {M.}~\bibnamefont {Du{\l}ak}}, \bibinfo
  {author} {\bibfnamefont {L.}~\bibnamefont {Ferrighi}}, \bibinfo {author}
  {\bibfnamefont {J.}~\bibnamefont {Gavnholt}}, \bibinfo {author}
  {\bibfnamefont {C.}~\bibnamefont {Glinsvad}}, \bibinfo {author}
  {\bibfnamefont {V.}~\bibnamefont {Haikola}}, \bibinfo {author} {\bibfnamefont
  {H.~A.}\ \bibnamefont {Hansen}}, \bibinfo {author} {\bibfnamefont {H.~H.}\
  \bibnamefont {Kristoffersen}}, \bibinfo {author} {\bibfnamefont
  {M.}~\bibnamefont {Kuisma}}, \bibinfo {author} {\bibfnamefont {A.~H.}\
  \bibnamefont {Larsen}}, \bibinfo {author} {\bibfnamefont {L.}~\bibnamefont
  {Lehtovaara}}, \bibinfo {author} {\bibfnamefont {M.}~\bibnamefont
  {Ljungberg}}, \bibinfo {author} {\bibfnamefont {O.}~\bibnamefont
  {Lopez-Acevedo}}, \bibinfo {author} {\bibfnamefont {P.~G.}\ \bibnamefont
  {Moses}}, \bibinfo {author} {\bibfnamefont {J.}~\bibnamefont {Ojanen}},
  \bibinfo {author} {\bibfnamefont {T.}~\bibnamefont {Olsen}}, \bibinfo
  {author} {\bibfnamefont {V.}~\bibnamefont {Petzold}}, \bibinfo {author}
  {\bibfnamefont {N.~A.}\ \bibnamefont {Romero}}, \bibinfo {author}
  {\bibfnamefont {J.}~\bibnamefont {Stausholm-M{\o}ller}}, \bibinfo {author}
  {\bibfnamefont {M.}~\bibnamefont {Strange}}, \bibinfo {author} {\bibfnamefont
  {G.~A.}\ \bibnamefont {Tritsaris}}, \bibinfo {author} {\bibfnamefont
  {M.}~\bibnamefont {Vanin}}, \bibinfo {author} {\bibfnamefont
  {M.}~\bibnamefont {Walter}}, \bibinfo {author} {\bibfnamefont
  {B.}~\bibnamefont {Hammer}}, \bibinfo {author} {\bibfnamefont
  {H.}~\bibnamefont {H{\"{a}}kkinen}}, \bibinfo {author} {\bibfnamefont
  {G.~K.~H.}\ \bibnamefont {Madsen}}, \bibinfo {author} {\bibfnamefont {R.~M.}\
  \bibnamefont {Nieminen}}, \bibinfo {author} {\bibfnamefont {J.~K.}\
  \bibnamefont {N{\o}rskov}}, \bibinfo {author} {\bibfnamefont
  {M.}~\bibnamefont {Puska}}, \bibinfo {author} {\bibfnamefont {T.~T.}\
  \bibnamefont {Rantala}}, \bibinfo {author} {\bibfnamefont {J.}~\bibnamefont
  {Schi{\o}tz}}, \bibinfo {author} {\bibfnamefont {K.~S.}\ \bibnamefont
  {Thygesen}},\ and\ \bibinfo {author} {\bibfnamefont {K.~W.}\ \bibnamefont
  {Jacobsen}},\ }\bibfield  {title} {\bibinfo {title} {{Electronic structure
  calculations with GPAW: a real-space implementation of the projector
  augmented-wave method}},\ }\href
  {https://doi.org/10.1088/0953-8984/22/25/253202} {\bibfield  {journal}
  {\bibinfo  {journal} {Journal of Physics: Condensed Matter}\ }\textbf
  {\bibinfo {volume} {22}},\ \bibinfo {pages} {253202} (\bibinfo {year}
  {2010})}\BibitemShut {NoStop}%
\bibitem [{\citenamefont {Larsen}\ \emph {et~al.}(2017)\citenamefont {Larsen},
  \citenamefont {Mortensen}, \citenamefont {Blomqvist}, \citenamefont
  {Castelli}, \citenamefont {Christensen}, \citenamefont {Du{\l}ak},
  \citenamefont {Friis}, \citenamefont {Groves}, \citenamefont {Hammer},
  \citenamefont {Hargus}, \citenamefont {Hermes}, \citenamefont {Jennings},
  \citenamefont {{Bjerre Jensen}}, \citenamefont {Kermode}, \citenamefont
  {Kitchin}, \citenamefont {{Leonhard Kolsbjerg}}, \citenamefont {Kubal},
  \citenamefont {Kaasbjerg}, \citenamefont {Lysgaard}, \citenamefont {{Bergmann
  Maronsson}}, \citenamefont {Maxson}, \citenamefont {Olsen}, \citenamefont
  {Pastewka}, \citenamefont {Peterson}, \citenamefont {Rostgaard},
  \citenamefont {Schi{\o}tz}, \citenamefont {Sch{\"{u}}tt}, \citenamefont
  {Strange}, \citenamefont {Thygesen}, \citenamefont {Vegge}, \citenamefont
  {Vilhelmsen}, \citenamefont {Walter}, \citenamefont {Zeng},\ and\
  \citenamefont {Jacobsen}}]{SILarsen2017}%
  \BibitemOpen
  \bibfield  {author} {\bibinfo {author} {\bibfnamefont {A.~H.}\ \bibnamefont
  {Larsen}}, \bibinfo {author} {\bibfnamefont {J.~J.}\ \bibnamefont
  {Mortensen}}, \bibinfo {author} {\bibfnamefont {J.}~\bibnamefont
  {Blomqvist}}, \bibinfo {author} {\bibfnamefont {I.~E.}\ \bibnamefont
  {Castelli}}, \bibinfo {author} {\bibfnamefont {R.}~\bibnamefont
  {Christensen}}, \bibinfo {author} {\bibfnamefont {M.}~\bibnamefont
  {Du{\l}ak}}, \bibinfo {author} {\bibfnamefont {J.}~\bibnamefont {Friis}},
  \bibinfo {author} {\bibfnamefont {M.~N.}\ \bibnamefont {Groves}}, \bibinfo
  {author} {\bibfnamefont {B.}~\bibnamefont {Hammer}}, \bibinfo {author}
  {\bibfnamefont {C.}~\bibnamefont {Hargus}}, \bibinfo {author} {\bibfnamefont
  {E.~D.}\ \bibnamefont {Hermes}}, \bibinfo {author} {\bibfnamefont {P.~C.}\
  \bibnamefont {Jennings}}, \bibinfo {author} {\bibfnamefont {P.}~\bibnamefont
  {{Bjerre Jensen}}}, \bibinfo {author} {\bibfnamefont {J.}~\bibnamefont
  {Kermode}}, \bibinfo {author} {\bibfnamefont {J.~R.}\ \bibnamefont
  {Kitchin}}, \bibinfo {author} {\bibfnamefont {E.}~\bibnamefont {{Leonhard
  Kolsbjerg}}}, \bibinfo {author} {\bibfnamefont {J.}~\bibnamefont {Kubal}},
  \bibinfo {author} {\bibfnamefont {K.}~\bibnamefont {Kaasbjerg}}, \bibinfo
  {author} {\bibfnamefont {S.}~\bibnamefont {Lysgaard}}, \bibinfo {author}
  {\bibfnamefont {J.}~\bibnamefont {{Bergmann Maronsson}}}, \bibinfo {author}
  {\bibfnamefont {T.}~\bibnamefont {Maxson}}, \bibinfo {author} {\bibfnamefont
  {T.}~\bibnamefont {Olsen}}, \bibinfo {author} {\bibfnamefont
  {L.}~\bibnamefont {Pastewka}}, \bibinfo {author} {\bibfnamefont
  {A.}~\bibnamefont {Peterson}}, \bibinfo {author} {\bibfnamefont
  {C.}~\bibnamefont {Rostgaard}}, \bibinfo {author} {\bibfnamefont
  {J.}~\bibnamefont {Schi{\o}tz}}, \bibinfo {author} {\bibfnamefont
  {O.}~\bibnamefont {Sch{\"{u}}tt}}, \bibinfo {author} {\bibfnamefont
  {M.}~\bibnamefont {Strange}}, \bibinfo {author} {\bibfnamefont {K.~S.}\
  \bibnamefont {Thygesen}}, \bibinfo {author} {\bibfnamefont {T.}~\bibnamefont
  {Vegge}}, \bibinfo {author} {\bibfnamefont {L.}~\bibnamefont {Vilhelmsen}},
  \bibinfo {author} {\bibfnamefont {M.}~\bibnamefont {Walter}}, \bibinfo
  {author} {\bibfnamefont {Z.}~\bibnamefont {Zeng}},\ and\ \bibinfo {author}
  {\bibfnamefont {K.~W.}\ \bibnamefont {Jacobsen}},\ }\bibfield  {title}
  {\bibinfo {title} {{The atomic simulation environment—a Python library for
  working with atoms}},\ }\href {https://doi.org/10.1088/1361-648X/aa680e}
  {\bibfield  {journal} {\bibinfo  {journal} {Journal of Physics: Condensed
  Matter}\ }\textbf {\bibinfo {volume} {29}},\ \bibinfo {pages} {273002}
  (\bibinfo {year} {2017})}\BibitemShut {NoStop}%
\bibitem [{\citenamefont {Perdew}\ \emph {et~al.}(1996)\citenamefont {Perdew},
  \citenamefont {Burke},\ and\ \citenamefont {Ernzerhof}}]{SIPerdew1996}%
  \BibitemOpen
  \bibfield  {author} {\bibinfo {author} {\bibfnamefont {J.~P.}\ \bibnamefont
  {Perdew}}, \bibinfo {author} {\bibfnamefont {K.}~\bibnamefont {Burke}},\ and\
  \bibinfo {author} {\bibfnamefont {M.}~\bibnamefont {Ernzerhof}},\ }\bibfield
  {title} {\bibinfo {title} {{Generalized Gradient Approximation Made
  Simple}},\ }\href {https://doi.org/10.1103/PhysRevLett.77.3865} {\bibfield
  {journal} {\bibinfo  {journal} {Physical Review Letters}\ }\textbf {\bibinfo
  {volume} {77}},\ \bibinfo {pages} {3865} (\bibinfo {year}
  {1996})}\BibitemShut {NoStop}%
\bibitem [{\citenamefont {Olofsson}(1972)}]{SIOlofsson1972}%
  \BibitemOpen
  \bibfield  {author} {\bibinfo {author} {\bibfnamefont {O.}~\bibnamefont
  {Olofsson}},\ }\bibfield  {title} {\bibinfo {title} {{The crystal structure
  of Cu$_3$P}},\ }\href
  {http://actachemscand.org/pdf/acta_vol_26_p2777-2787.pdf} {\bibfield
  {journal} {\bibinfo  {journal} {Acta Chemica Scandinava}\ }\textbf {\bibinfo
  {volume} {26}},\ \bibinfo {pages} {2777} (\bibinfo {year}
  {1972})}\BibitemShut {NoStop}%
\bibitem [{\citenamefont {Kuisma}\ \emph {et~al.}(2010)\citenamefont {Kuisma},
  \citenamefont {Ojanen}, \citenamefont {Enkovaara},\ and\ \citenamefont
  {Rantala}}]{SIKuisma2010}%
  \BibitemOpen
  \bibfield  {author} {\bibinfo {author} {\bibfnamefont {M.}~\bibnamefont
  {Kuisma}}, \bibinfo {author} {\bibfnamefont {J.}~\bibnamefont {Ojanen}},
  \bibinfo {author} {\bibfnamefont {J.}~\bibnamefont {Enkovaara}},\ and\
  \bibinfo {author} {\bibfnamefont {T.~T.}\ \bibnamefont {Rantala}},\
  }\bibfield  {title} {\bibinfo {title} {{Kohn-Sham potential with
  discontinuity for band gap materials}},\ }\href
  {https://doi.org/10.1103/PhysRevB.82.115106} {\bibfield  {journal} {\bibinfo
  {journal} {Physical Review B}\ }\textbf {\bibinfo {volume} {82}},\ \bibinfo
  {pages} {115106} (\bibinfo {year} {2010})}\BibitemShut {NoStop}%
\bibitem [{\citenamefont {Gritsenko}\ \emph {et~al.}(1995)\citenamefont
  {Gritsenko}, \citenamefont {van Leeuwen}, \citenamefont {van Lenthe},\ and\
  \citenamefont {Baerends}}]{SIGritsenko1995}%
  \BibitemOpen
  \bibfield  {author} {\bibinfo {author} {\bibfnamefont {O.}~\bibnamefont
  {Gritsenko}}, \bibinfo {author} {\bibfnamefont {R.}~\bibnamefont {van
  Leeuwen}}, \bibinfo {author} {\bibfnamefont {E.}~\bibnamefont {van Lenthe}},\
  and\ \bibinfo {author} {\bibfnamefont {E.~J.}\ \bibnamefont {Baerends}},\
  }\bibfield  {title} {\bibinfo {title} {{Self-consistent approximation to the
  Kohn-Sham exchange potential}},\ }\href
  {https://doi.org/10.1103/PhysRevA.51.1944} {\bibfield  {journal} {\bibinfo
  {journal} {Physical Review A}\ }\textbf {\bibinfo {volume} {51}},\ \bibinfo
  {pages} {1944} (\bibinfo {year} {1995})}\BibitemShut {NoStop}%
\bibitem [{\citenamefont {Madsen}\ \emph {et~al.}(2018)\citenamefont {Madsen},
  \citenamefont {Carrete},\ and\ \citenamefont {Verstraete}}]{SIMadsen2018}%
  \BibitemOpen
  \bibfield  {author} {\bibinfo {author} {\bibfnamefont {G.~K.}\ \bibnamefont
  {Madsen}}, \bibinfo {author} {\bibfnamefont {J.}~\bibnamefont {Carrete}},\
  and\ \bibinfo {author} {\bibfnamefont {M.~J.}\ \bibnamefont {Verstraete}},\
  }\bibfield  {title} {\bibinfo {title} {{BoltzTraP2, a program for
  interpolating band structures and calculating semi-classical transport
  coefficients}},\ }\href {https://doi.org/10.1016/j.cpc.2018.05.010}
  {\bibfield  {journal} {\bibinfo  {journal} {Computer Physics Communications}\
  }\textbf {\bibinfo {volume} {231}},\ \bibinfo {pages} {140} (\bibinfo {year}
  {2018})}\BibitemShut {NoStop}%
\bibitem [{\citenamefont {Jain}\ \emph {et~al.}(2013)\citenamefont {Jain},
  \citenamefont {Ong}, \citenamefont {Hautier}, \citenamefont {Chen},
  \citenamefont {Richards}, \citenamefont {Dacek}, \citenamefont {Cholia},
  \citenamefont {Gunter}, \citenamefont {Skinner}, \citenamefont {Ceder},\ and\
  \citenamefont {Persson}}]{SIJain2013}%
  \BibitemOpen
  \bibfield  {author} {\bibinfo {author} {\bibfnamefont {A.}~\bibnamefont
  {Jain}}, \bibinfo {author} {\bibfnamefont {S.~P.}\ \bibnamefont {Ong}},
  \bibinfo {author} {\bibfnamefont {G.}~\bibnamefont {Hautier}}, \bibinfo
  {author} {\bibfnamefont {W.}~\bibnamefont {Chen}}, \bibinfo {author}
  {\bibfnamefont {W.~D.}\ \bibnamefont {Richards}}, \bibinfo {author}
  {\bibfnamefont {S.}~\bibnamefont {Dacek}}, \bibinfo {author} {\bibfnamefont
  {S.}~\bibnamefont {Cholia}}, \bibinfo {author} {\bibfnamefont
  {D.}~\bibnamefont {Gunter}}, \bibinfo {author} {\bibfnamefont
  {D.}~\bibnamefont {Skinner}}, \bibinfo {author} {\bibfnamefont
  {G.}~\bibnamefont {Ceder}},\ and\ \bibinfo {author} {\bibfnamefont {K.~A.}\
  \bibnamefont {Persson}},\ }\bibfield  {title} {\bibinfo {title} {{Commentary:
  The Materials Project: A materials genome approach to accelerating materials
  innovation}},\ }\href {https://doi.org/10.1063/1.4812323} {\bibfield
  {journal} {\bibinfo  {journal} {APL Materials}\ }\textbf {\bibinfo {volume}
  {1}},\ \bibinfo {pages} {011002} (\bibinfo {year} {2013})}\BibitemShut
  {NoStop}%
\bibitem [{\citenamefont {Ong}\ \emph {et~al.}(2015)\citenamefont {Ong},
  \citenamefont {Cholia}, \citenamefont {Jain}, \citenamefont {Brafman},
  \citenamefont {Gunter}, \citenamefont {Ceder},\ and\ \citenamefont
  {Persson}}]{SIOng2015}%
  \BibitemOpen
  \bibfield  {author} {\bibinfo {author} {\bibfnamefont {S.~P.}\ \bibnamefont
  {Ong}}, \bibinfo {author} {\bibfnamefont {S.}~\bibnamefont {Cholia}},
  \bibinfo {author} {\bibfnamefont {A.}~\bibnamefont {Jain}}, \bibinfo {author}
  {\bibfnamefont {M.}~\bibnamefont {Brafman}}, \bibinfo {author} {\bibfnamefont
  {D.}~\bibnamefont {Gunter}}, \bibinfo {author} {\bibfnamefont
  {G.}~\bibnamefont {Ceder}},\ and\ \bibinfo {author} {\bibfnamefont {K.~A.}\
  \bibnamefont {Persson}},\ }\bibfield  {title} {\bibinfo {title} {{The
  Materials Application Programming Interface (API): A simple, flexible and
  efficient API for materials data based on REpresentational State Transfer
  (REST) principles}},\ }\href
  {https://doi.org/10.1016/j.commatsci.2014.10.037} {\bibfield  {journal}
  {\bibinfo  {journal} {Computational Materials Science}\ }\textbf {\bibinfo
  {volume} {97}},\ \bibinfo {pages} {209} (\bibinfo {year} {2015})}\BibitemShut
  {NoStop}%
\bibitem [{\citenamefont {Ong}\ \emph {et~al.}(2013)\citenamefont {Ong},
  \citenamefont {Richards}, \citenamefont {Jain}, \citenamefont {Hautier},
  \citenamefont {Kocher}, \citenamefont {Cholia}, \citenamefont {Gunter},
  \citenamefont {Chevrier}, \citenamefont {Persson},\ and\ \citenamefont
  {Ceder}}]{SIOng2013}%
  \BibitemOpen
  \bibfield  {author} {\bibinfo {author} {\bibfnamefont {S.~P.}\ \bibnamefont
  {Ong}}, \bibinfo {author} {\bibfnamefont {W.~D.}\ \bibnamefont {Richards}},
  \bibinfo {author} {\bibfnamefont {A.}~\bibnamefont {Jain}}, \bibinfo {author}
  {\bibfnamefont {G.}~\bibnamefont {Hautier}}, \bibinfo {author} {\bibfnamefont
  {M.}~\bibnamefont {Kocher}}, \bibinfo {author} {\bibfnamefont
  {S.}~\bibnamefont {Cholia}}, \bibinfo {author} {\bibfnamefont
  {D.}~\bibnamefont {Gunter}}, \bibinfo {author} {\bibfnamefont {V.~L.}\
  \bibnamefont {Chevrier}}, \bibinfo {author} {\bibfnamefont {K.~A.}\
  \bibnamefont {Persson}},\ and\ \bibinfo {author} {\bibfnamefont
  {G.}~\bibnamefont {Ceder}},\ }\bibfield  {title} {\bibinfo {title} {{Python
  Materials Genomics (pymatgen): A robust, open-source python library for
  materials analysis}},\ }\href
  {https://doi.org/10.1016/j.commatsci.2012.10.028} {\bibfield  {journal}
  {\bibinfo  {journal} {Computational Materials Science}\ }\textbf {\bibinfo
  {volume} {68}},\ \bibinfo {pages} {314} (\bibinfo {year} {2013})}\BibitemShut
  {NoStop}%
\bibitem [{\citenamefont {Ricci}\ \emph {et~al.}(2017)\citenamefont {Ricci},
  \citenamefont {Chen}, \citenamefont {Aydemir}, \citenamefont {Snyder},
  \citenamefont {Rignanese}, \citenamefont {Jain},\ and\ \citenamefont
  {Hautier}}]{SIRicci2017}%
  \BibitemOpen
  \bibfield  {author} {\bibinfo {author} {\bibfnamefont {F.}~\bibnamefont
  {Ricci}}, \bibinfo {author} {\bibfnamefont {W.}~\bibnamefont {Chen}},
  \bibinfo {author} {\bibfnamefont {U.}~\bibnamefont {Aydemir}}, \bibinfo
  {author} {\bibfnamefont {G.~J.}\ \bibnamefont {Snyder}}, \bibinfo {author}
  {\bibfnamefont {G.-M.}\ \bibnamefont {Rignanese}}, \bibinfo {author}
  {\bibfnamefont {A.}~\bibnamefont {Jain}},\ and\ \bibinfo {author}
  {\bibfnamefont {G.}~\bibnamefont {Hautier}},\ }\bibfield  {title} {\bibinfo
  {title} {{An ab initio electronic transport database for inorganic
  materials}},\ }\href {https://doi.org/10.1038/sdata.2017.85} {\bibfield
  {journal} {\bibinfo  {journal} {Scientific Data}\ }\textbf {\bibinfo {volume}
  {4}},\ \bibinfo {pages} {170085} (\bibinfo {year} {2017})}\BibitemShut
  {NoStop}%
\bibitem [{\citenamefont {Wolff}\ \emph {et~al.}(2018)\citenamefont {Wolff},
  \citenamefont {Doert}, \citenamefont {Hunger}, \citenamefont {Kaiser},
  \citenamefont {Pallmann}, \citenamefont {Reinhold}, \citenamefont {Yogendra},
  \citenamefont {Giebeler}, \citenamefont {Sichelschmidt}, \citenamefont
  {Schnelle}, \citenamefont {Whiteside}, \citenamefont {Gunaratne},
  \citenamefont {Nockemann}, \citenamefont {Weigand}, \citenamefont {Brunner},\
  and\ \citenamefont {Ruck}}]{SIWolff2018}%
  \BibitemOpen
  \bibfield  {author} {\bibinfo {author} {\bibfnamefont {A.}~\bibnamefont
  {Wolff}}, \bibinfo {author} {\bibfnamefont {T.}~\bibnamefont {Doert}},
  \bibinfo {author} {\bibfnamefont {J.}~\bibnamefont {Hunger}}, \bibinfo
  {author} {\bibfnamefont {M.}~\bibnamefont {Kaiser}}, \bibinfo {author}
  {\bibfnamefont {J.}~\bibnamefont {Pallmann}}, \bibinfo {author}
  {\bibfnamefont {R.}~\bibnamefont {Reinhold}}, \bibinfo {author}
  {\bibfnamefont {S.}~\bibnamefont {Yogendra}}, \bibinfo {author}
  {\bibfnamefont {L.}~\bibnamefont {Giebeler}}, \bibinfo {author}
  {\bibfnamefont {J.}~\bibnamefont {Sichelschmidt}}, \bibinfo {author}
  {\bibfnamefont {W.}~\bibnamefont {Schnelle}}, \bibinfo {author}
  {\bibfnamefont {R.}~\bibnamefont {Whiteside}}, \bibinfo {author}
  {\bibfnamefont {H.~Q.~N.}\ \bibnamefont {Gunaratne}}, \bibinfo {author}
  {\bibfnamefont {P.}~\bibnamefont {Nockemann}}, \bibinfo {author}
  {\bibfnamefont {J.~J.}\ \bibnamefont {Weigand}}, \bibinfo {author}
  {\bibfnamefont {E.}~\bibnamefont {Brunner}},\ and\ \bibinfo {author}
  {\bibfnamefont {M.}~\bibnamefont {Ruck}},\ }\bibfield  {title} {\bibinfo
  {title} {{Low-Temperature Tailoring of Copper-Deficient Cu$_{3–x}$P — Electric Properties, Phase Transitions, and Performance in Lithium-Ion
  Batteries}},\ }\href {https://doi.org/10.1021/acs.chemmater.8b02950}
  {\bibfield  {journal} {\bibinfo  {journal} {Chemistry of Materials}\ }\textbf
  {\bibinfo {volume} {30}},\ \bibinfo {pages} {7111} (\bibinfo {year}
  {2018})}\BibitemShut {NoStop}%
\bibitem [{\citenamefont {{De Trizio}}\ \emph {et~al.}(2015)\citenamefont {{De
  Trizio}}, \citenamefont {Gaspari}, \citenamefont {Bertoni}, \citenamefont
  {Kriegel}, \citenamefont {Moretti}, \citenamefont {Scotognella},
  \citenamefont {Maserati}, \citenamefont {Zhang}, \citenamefont {Messina},
  \citenamefont {Prato}, \citenamefont {Marras}, \citenamefont {Cavalli},\ and\
  \citenamefont {Manna}}]{SIDeTrizio2015}%
  \BibitemOpen
  \bibfield  {author} {\bibinfo {author} {\bibfnamefont {L.}~\bibnamefont {{De
  Trizio}}}, \bibinfo {author} {\bibfnamefont {R.}~\bibnamefont {Gaspari}},
  \bibinfo {author} {\bibfnamefont {G.}~\bibnamefont {Bertoni}}, \bibinfo
  {author} {\bibfnamefont {I.}~\bibnamefont {Kriegel}}, \bibinfo {author}
  {\bibfnamefont {L.}~\bibnamefont {Moretti}}, \bibinfo {author} {\bibfnamefont
  {F.}~\bibnamefont {Scotognella}}, \bibinfo {author} {\bibfnamefont
  {L.}~\bibnamefont {Maserati}}, \bibinfo {author} {\bibfnamefont
  {Y.}~\bibnamefont {Zhang}}, \bibinfo {author} {\bibfnamefont {G.~C.}\
  \bibnamefont {Messina}}, \bibinfo {author} {\bibfnamefont {M.}~\bibnamefont
  {Prato}}, \bibinfo {author} {\bibfnamefont {S.}~\bibnamefont {Marras}},
  \bibinfo {author} {\bibfnamefont {A.}~\bibnamefont {Cavalli}},\ and\ \bibinfo
  {author} {\bibfnamefont {L.}~\bibnamefont {Manna}},\ }\bibfield  {title}
  {\bibinfo {title} {{Cu$_{3–x}$P Nanocrystals as a Material Platform for
  Near-Infrared Plasmonics and Cation Exchange Reactions}},\ }\href
  {https://doi.org/10.1021/cm5044792} {\bibfield  {journal} {\bibinfo
  {journal} {Chemistry of Materials}\ }\textbf {\bibinfo {volume} {27}},\
  \bibinfo {pages} {1120} (\bibinfo {year} {2015})}\BibitemShut {NoStop}%
\bibitem [{\citenamefont {Werner}(2017)}]{SIWerner2017}%
  \BibitemOpen
  \bibfield  {author} {\bibinfo {author} {\bibfnamefont {F.}~\bibnamefont
  {Werner}},\ }\bibfield  {title} {\bibinfo {title} {{Hall measurements on
  low-mobility thin films}},\ }\href {https://doi.org/10.1063/1.4990470}
  {\bibfield  {journal} {\bibinfo  {journal} {Journal of Applied Physics}\
  }\textbf {\bibinfo {volume} {122}},\ \bibinfo {pages} {135306} (\bibinfo
  {year} {2017})}\BibitemShut {NoStop}%
\bibitem [{\citenamefont {Alderson}\ \emph {et~al.}(1968)\citenamefont
  {Alderson}, \citenamefont {Farrell},\ and\ \citenamefont
  {Hurd}}]{SIAlderson1968}%
  \BibitemOpen
  \bibfield  {author} {\bibinfo {author} {\bibfnamefont {J.~E.~A.}\
  \bibnamefont {Alderson}}, \bibinfo {author} {\bibfnamefont {T.}~\bibnamefont
  {Farrell}},\ and\ \bibinfo {author} {\bibfnamefont {C.~M.}\ \bibnamefont
  {Hurd}},\ }\bibfield  {title} {\bibinfo {title} {{Hall Coefficients of Cu,
  Ag, and Au in the Range 4.2-300°K}},\ }\href
  {https://doi.org/10.1103/PhysRev.174.729} {\bibfield  {journal} {\bibinfo
  {journal} {Physical Review}\ }\textbf {\bibinfo {volume} {174}},\ \bibinfo
  {pages} {729} (\bibinfo {year} {1968})}\BibitemShut {NoStop}%
\bibitem [{\citenamefont {Macheda}\ \emph {et~al.}(2020)\citenamefont
  {Macheda}, \citenamefont {Ponc{\'{e}}}, \citenamefont {Giustino},\ and\
  \citenamefont {Bonini}}]{SIMacheda2020}%
  \BibitemOpen
  \bibfield  {author} {\bibinfo {author} {\bibfnamefont {F.}~\bibnamefont
  {Macheda}}, \bibinfo {author} {\bibfnamefont {S.}~\bibnamefont
  {Ponc{\'{e}}}}, \bibinfo {author} {\bibfnamefont {F.}~\bibnamefont
  {Giustino}},\ and\ \bibinfo {author} {\bibfnamefont {N.}~\bibnamefont
  {Bonini}},\ }\bibfield  {title} {\bibinfo {title} {{Theory and Computation of
  Hall Scattering Factor in Graphene}},\ }\href
  {https://doi.org/10.1021/acs.nanolett.0c03874} {\bibfield  {journal}
  {\bibinfo  {journal} {Nano Letters}\ }\textbf {\bibinfo {volume} {20}},\
  \bibinfo {pages} {8861} (\bibinfo {year} {2020})}\BibitemShut {NoStop}%
\bibitem [{\citenamefont {Willis}\ \emph {et~al.}(2022)\citenamefont {Willis},
  \citenamefont {Bravi{\'{c}}}, \citenamefont {Schnepf}, \citenamefont
  {Heinselman}, \citenamefont {Monserrat}, \citenamefont {Unold}, \citenamefont
  {Zakutayev}, \citenamefont {Scanlon},\ and\ \citenamefont
  {Crovetto}}]{SIWillis2022}%
  \BibitemOpen
  \bibfield  {author} {\bibinfo {author} {\bibfnamefont {J.}~\bibnamefont
  {Willis}}, \bibinfo {author} {\bibfnamefont {I.}~\bibnamefont
  {Bravi{\'{c}}}}, \bibinfo {author} {\bibfnamefont {R.~R.}\ \bibnamefont
  {Schnepf}}, \bibinfo {author} {\bibfnamefont {K.~N.}\ \bibnamefont
  {Heinselman}}, \bibinfo {author} {\bibfnamefont {B.}~\bibnamefont
  {Monserrat}}, \bibinfo {author} {\bibfnamefont {T.}~\bibnamefont {Unold}},
  \bibinfo {author} {\bibfnamefont {A.}~\bibnamefont {Zakutayev}}, \bibinfo
  {author} {\bibfnamefont {D.~O.}\ \bibnamefont {Scanlon}},\ and\ \bibinfo
  {author} {\bibfnamefont {A.}~\bibnamefont {Crovetto}},\ }\bibfield  {title}
  {\bibinfo {title} {{Prediction and realisation of high mobility and
  degenerate p-type conductivity in CaCuP thin films}},\ }\href
  {https://doi.org/10.1039/D2SC01538B} {\bibfield  {journal} {\bibinfo
  {journal} {Chemical Science}\ }\textbf {\bibinfo {volume} {13}},\ \bibinfo
  {pages} {5872} (\bibinfo {year} {2022})}\BibitemShut {NoStop}%
\bibitem [{\citenamefont {Ellmer}\ and\ \citenamefont
  {Mientus}(2008)}]{SIEllmer2008a}%
  \BibitemOpen
  \bibfield  {author} {\bibinfo {author} {\bibfnamefont {K.}~\bibnamefont
  {Ellmer}}\ and\ \bibinfo {author} {\bibfnamefont {R.}~\bibnamefont
  {Mientus}},\ }\bibfield  {title} {\bibinfo {title} {{Carrier transport in
  polycrystalline ITO and ZnO:Al II: The influence of grain barriers and
  boundaries}},\ }\href {https://doi.org/10.1016/j.tsf.2007.10.082} {\bibfield
  {journal} {\bibinfo  {journal} {Thin Solid Films}\ }\textbf {\bibinfo
  {volume} {516}},\ \bibinfo {pages} {5829} (\bibinfo {year}
  {2008})}\BibitemShut {NoStop}%
\bibitem [{\citenamefont {Chopra}\ and\ \citenamefont
  {Bahl}(1967)}]{SIChopra1967}%
  \BibitemOpen
  \bibfield  {author} {\bibinfo {author} {\bibfnamefont {K.~L.}\ \bibnamefont
  {Chopra}}\ and\ \bibinfo {author} {\bibfnamefont {S.~K.}\ \bibnamefont
  {Bahl}},\ }\bibfield  {title} {\bibinfo {title} {{Hall Effect in Thin Metal
  Films}},\ }\href {https://doi.org/10.1063/1.1710180} {\bibfield  {journal}
  {\bibinfo  {journal} {Journal of Applied Physics}\ }\textbf {\bibinfo
  {volume} {38}},\ \bibinfo {pages} {3607} (\bibinfo {year}
  {1967})}\BibitemShut {NoStop}%
\bibitem [{\citenamefont {Weber}\ \emph {et~al.}(1994)\citenamefont {Weber},
  \citenamefont {Sutter},\ and\ \citenamefont {{von K{\"a}nel}}}]{SIWeber1998}%
  \BibitemOpen
  \bibfield  {author} {\bibinfo {author} {\bibfnamefont {A.}~\bibnamefont
  {Weber}}, \bibinfo {author} {\bibfnamefont {P.}~\bibnamefont {Sutter}},\ and\
  \bibinfo {author} {\bibfnamefont {H.}~\bibnamefont {{von K{\"a}nel}}},\
  }\bibfield  {title} {\bibinfo {title} {Optical, electrical, and
  photoelectrical properties of sputtered thin amorphous Zn$_3$P$_2$
  films},\ }\href {https://doi.org/10.1063/1.356613} {\bibfield  {journal}
  {\bibinfo  {journal} {Journal of Applied Physics}\ }\textbf {\bibinfo
  {volume} {75}},\ \bibinfo {pages} {7448} (\bibinfo {year}
  {1994})}\BibitemShut {NoStop}%
 \bibitem [{\citenamefont {Pyykk{\"o}}\ and\ \citenamefont
  {Atsumi}(2009)}]{SIPyykko2009}%
  \BibitemOpen
  \bibfield  {author} {\bibinfo {author} {\bibfnamefont {P.}~\bibnamefont
  {Pyykk{\"o}}}\ and\ \bibinfo {author} {\bibfnamefont {M.}~\bibnamefont
  {Atsumi}},\ }\bibfield  {title} {\bibinfo {title} {Molecular {{Single-Bond
  Covalent Radii}} for {{Elements}} 1-118},\ }\href
  {https://doi.org/10.1002/chem.200800987} {\bibfield  {journal} {\bibinfo
  {journal} {Chemistry - A European Journal}\ }\textbf {\bibinfo {volume}
  {15}},\ \bibinfo {pages} {186} (\bibinfo {year} {2009})}\BibitemShut
  {NoStop}%
\bibitem [{\citenamefont {{Van de Walle}}\ and\ \citenamefont
  {Neugebauer}(2006)}]{SIvandewalleHydrogenSemiconductors2006}%
  \BibitemOpen
  \bibfield  {author} {\bibinfo {author} {\bibfnamefont {C.~G.}\ \bibnamefont
  {{Van de Walle}}}\ and\ \bibinfo {author} {\bibfnamefont {J.}~\bibnamefont
  {Neugebauer}},\ }\bibfield  {title} {\bibinfo {title} {Hydrogen in
  {{Semiconductors}}},\ }\href
  {https://doi.org/10.1146/annurev.matsci.36.010705.155428} {\bibfield
  {journal} {\bibinfo  {journal} {Annual Review of Materials Research}\
  }\textbf {\bibinfo {volume} {36}},\ \bibinfo {pages} {179} (\bibinfo {year}
  {2006})}\BibitemShut {NoStop}%
\bibitem [{\citenamefont {Hou}\ \emph {et~al.}(2016)\citenamefont {Hou},
  \citenamefont {Chen}, \citenamefont {Wang}, \citenamefont {Liang},
  \citenamefont {Lin}, \citenamefont {Fu},\ and\ \citenamefont
  {Chen}}]{SIHou2016}%
  \BibitemOpen
  \bibfield  {author} {\bibinfo {author} {\bibfnamefont {C.-C.}\ \bibnamefont
  {Hou}}, \bibinfo {author} {\bibfnamefont {Q.-Q.}\ \bibnamefont {Chen}},
  \bibinfo {author} {\bibfnamefont {C.-J.}\ \bibnamefont {Wang}}, \bibinfo
  {author} {\bibfnamefont {F.}~\bibnamefont {Liang}}, \bibinfo {author}
  {\bibfnamefont {Z.}~\bibnamefont {Lin}}, \bibinfo {author} {\bibfnamefont
  {W.-F.}\ \bibnamefont {Fu}},\ and\ \bibinfo {author} {\bibfnamefont
  {Y.}~\bibnamefont {Chen}},\ }\bibfield  {title} {\bibinfo {title}
  {{Self-Supported Cedarlike Semimetallic Cu$_3$P Nanoarrays as a 3D
  High-Performance Janus Electrode for Both Oxygen and Hydrogen Evolution under
  Basic Conditions}},\ }\href {https://doi.org/10.1021/acsami.6b06251}
  {\bibfield  {journal} {\bibinfo  {journal} {ACS Applied Materials and
  Interfaces}\ }\textbf {\bibinfo {volume} {8}},\ \bibinfo {pages} {23037}
  (\bibinfo {year} {2016})}\BibitemShut {NoStop}%
\bibitem [{\citenamefont {Liu}\ \emph {et~al.}(2016)\citenamefont {Liu},
  \citenamefont {He}, \citenamefont {Zhu}, \citenamefont {Xu},\ and\
  \citenamefont {Tong}}]{SILiu2016d}%
  \BibitemOpen
  \bibfield  {author} {\bibinfo {author} {\bibfnamefont {S.}~\bibnamefont
  {Liu}}, \bibinfo {author} {\bibfnamefont {X.}~\bibnamefont {He}}, \bibinfo
  {author} {\bibfnamefont {J.}~\bibnamefont {Zhu}}, \bibinfo {author}
  {\bibfnamefont {L.}~\bibnamefont {Xu}},\ and\ \bibinfo {author}
  {\bibfnamefont {J.}~\bibnamefont {Tong}},\ }\bibfield  {title} {\bibinfo
  {title} {{Cu$_3$P/RGO nanocomposite as a new anode for lithium-ion batteries}},\
  }\href {https://doi.org/10.1038/srep35189} {\bibfield  {journal} {\bibinfo
  {journal} {Scientific Reports}\ }\textbf {\bibinfo {volume} {6}},\ \bibinfo
  {pages} {1} (\bibinfo {year} {2016})}\BibitemShut {NoStop}%
\bibitem [{\citenamefont {Peng}\ \emph {et~al.}(2021)\citenamefont {Peng},
  \citenamefont {Lv}, \citenamefont {Fu}, \citenamefont {Chen}, \citenamefont
  {Su}, \citenamefont {Li}, \citenamefont {Zhang},\ and\ \citenamefont
  {Zhao}}]{SIPeng2021}%
  \BibitemOpen
  \bibfield  {author} {\bibinfo {author} {\bibfnamefont {X.}~\bibnamefont
  {Peng}}, \bibinfo {author} {\bibfnamefont {Y.}~\bibnamefont {Lv}}, \bibinfo
  {author} {\bibfnamefont {L.}~\bibnamefont {Fu}}, \bibinfo {author}
  {\bibfnamefont {F.}~\bibnamefont {Chen}}, \bibinfo {author} {\bibfnamefont
  {W.}~\bibnamefont {Su}}, \bibinfo {author} {\bibfnamefont {J.}~\bibnamefont
  {Li}}, \bibinfo {author} {\bibfnamefont {Q.}~\bibnamefont {Zhang}},\ and\
  \bibinfo {author} {\bibfnamefont {S.}~\bibnamefont {Zhao}},\ }\bibfield
  {title} {\bibinfo {title} {{Photoluminescence properties of cuprous phosphide
  prepared through phosphating copper with a native oxide layer}},\ }\href
  {https://doi.org/10.1039/D1RA07112B} {\bibfield  {journal} {\bibinfo
  {journal} {RSC Advances}\ }\textbf {\bibinfo {volume} {11}},\ \bibinfo
  {pages} {34095} (\bibinfo {year} {2021})}\BibitemShut {NoStop}%
 \bibitem [{\citenamefont {Peng}\ \emph {et~al.}(2022)\citenamefont {Peng},
  \citenamefont {Lv},\ and\ \citenamefont {Zhao}}]{SIPeng2022}%
  \BibitemOpen
  \bibfield  {author} {\bibinfo {author} {\bibfnamefont {X.}~\bibnamefont
  {Peng}}, \bibinfo {author} {\bibfnamefont {Y.}~\bibnamefont {Lv}},\ and\
  \bibinfo {author} {\bibfnamefont {S.}~\bibnamefont {Zhao}},\ }\bibfield
  {title} {\bibinfo {title} {{Chemical Vapor Deposition and Thermal Oxidation
  of Cuprous Phosphide Nanofilm}},\ }\href
  {https://doi.org/10.3390/coatings12010068} {\bibfield  {journal} {\bibinfo
  {journal} {Coatings}\ }\textbf {\bibinfo {volume} {12}},\ \bibinfo {pages}
  {68} (\bibinfo {year} {2022})}\BibitemShut {NoStop}%
\end{thebibliography}

\newpage
\clearpage


\onecolumngrid

\begin{center}
\begin{Large}
\textbf{SUPPORTING INFORMATION}
\end{Large}
\end{center}


\setcounter{page}{1}
\renewcommand*{\thepage}{S\arabic{page}}
\renewcommand{\thefigure}{S\arabic{figure}}
\renewcommand{\theequation}{S\arabic{equation}}
\setcounter{figure}{0}
\setcounter{equation}{0}


\vspace{1cm}


\section*{Experimental details}

\subsection*{Film growth}
Amorphous and polycrystalline Cu$_{3-x}$P thin films were deposited by radio-frequency (RF) sputtering, either by non-reactive sputtering in pure Ar or by reactive sputtering in PH$_3$/Ar, with PH$_3$ concentration up to 5\%. The sputter system (PVD Products) had a base pressure in the \SI{e-7}{Torr} range. A Cu target (K. J. Lesker Company) and a Cu$_{3-x}$P target (Princeton Scientific) were employed simultaneously. Both targets were \SI{2}{''} in diameter, \SI{0.25}{''} in thickness, and 99.99\% pure. The two targets were co-sputtered at \SI{5}{mTorr} total pressure, with RF powers of \SI{40}{W} (Cu$_{3-x}$P target) and \SI{20}{W} (Cu target). The target-substrate distance was \SI{16}{cm} and the deposition rate was about \SI{0.7}{\angstrom\per\second}. The deposition temperature was measured at the metallic platen onto which the substrates were clamped during deposition.
In each deposition process, a film was simultaneously grown on two Corning Eagle XG borosilicate glass substrates placed next to each other, covering a total area of $10 \times \SI{5}{cm^2}$. The targets were oriented so that one of the short edges of the total substrate area would mainly be coated by the Cu target and the other short edge by the Cu$_{3-x}$P target. In this way, it was possible to obtain combinatorial gradients in film composition (Cu/P ratio) approximately parallel to the long edge. This combinatorial effect is evident in Fig.~\ref{fig:composition}(b) of the main article. Each data point, spectrum, and XRD pattern in the main article corresponds to one specific point in the combinatorial films, which has its unique composition and properties. This combinatorial data was managed with the COMBIgor tool,~\cite{SITalley2019} the Research Data Infrastructure,~\cite{SITalley2021a} and was integrated into the High-Throughput Experimental Materials Database.~\cite{SIZakutayev2018} The thickness of each film varied by about 10--15\% across the long direction (\SI{10}{cm} long). The thickness of all films considered in this study was between \SI{200}{nm} and \SI{300}{nm}.

\subsection*{Film characterization}

Elemental composition and film thickness were determined by x-ray fluorescence (XRF) in a Bruker Tornado M4 instrument at \SI{15}{Torr} pressure using a Rh source. XRF spectra were fitted with the Bruker XMethod analysis program. The XRF data was calibrated by Rutherford backscattering spectroscopy (RBS) measurements of separate Cu-P films of different thicknesses and compositions deposited on Si. These RBS measurements and the calibration were described in a previous publication.~\cite{SICrovetto2022b} The accuracy of RBS in determining composition is around 3\%.~\cite{SIVerma2007} However, there are other possible sources of error. One of them is the extra calibration step required to obtain composition information from XRF spectra. The other is the use of silicon substrates (rather than the glass substrates used for the rest of the characterization) for the calibration. Taking these factors into account, we estimate an overall \textit{systematic} error in the $\pm 5\%$ range for the Cu/P ratios quoted in the main article. We expect the \textit{random} error in the determination of composition by XRF to be much lower than 5\%. The total x-ray counts in XRF measurements were sufficiently high to ensure better than 1\% reproducibility. Variations in film thickness could be a source of "random" error in the extracted composition. However, the films have rather similar thicknesses and there were no noticeable correlations between thickness and the extracted Cu/P ratios.
Finally, a systematic error in the $\pm 5\%$ range can be assumed for the thickness determined by XRF. This error comes from uncertainty in modeling surface roughness when fitting RBS spectra, from film porosity (resulting in lower density than in ideal Cu$_3$P) and from the XRF calibration step. The main consequence of this error is that it propagates to the resistivity. Hence, we also estimate a systematic error in the $\pm 5\%$ range for resistivity, Hall carrier concentration, and Hall mobility.

XRD measurements were conducted with a Bruker D8 diffractometer using Cu K$_\alpha$ radiation and a 2D detector. To cover the desired $2\theta$ range, two frames were collected with the incidence angle $\omega$ fixed at \SI{10}{\degree} and \SI{22.5}{\degree}, and the detector center fixed at $2\theta$ values of \SI{35}{\degree} and \SI{60}{\degree}, respectively. The diffraction intensity at each $2\theta$ angle was integrated over the $\chi$ range measured by the 2D detector. Since reflections from a range of $2\theta$ angles are measured in parallel by the 2D detector, this XRD measurement is not strictly in the Bragg-Brentano configuration and the lattice planes probed by XRD are not strictly parallel to the substrate plane. However, the angles between the probed lattice planes and the substrate plane are rather small (in the \SIrange{0}{19}{\degree} range depending on the value of $2\theta$ with respect to the detector's center). Thus, we make the approximation that the lattice planes corresponding to XRD reflections are parallel to the plane of the substrate. 

The $\textit{c}$-axis texture cofficient ($TC$) was estimated as



\begin{equation}
TC = \frac{\frac{I(113)}{I_0(113)}}{\frac{I(113)}{I_0(113)} + \frac{I(300)}{I_0(300)}}
\end{equation}

where $I(113)$ and $I(300)$ are the integrated intensities of the (113) and (300) peaks in the thin film, and $I_0(113)$ and $I_0(300)$ are the integrated intensities of the (113) and (300) peaks in a randomly oriented Cu$_{3-x}$P powder, taken from Olofsson's work.~\onlinecite{SIOlofsson1972} With this definition, a film with the (113) planes perfectly parallel to the substrate has a texture coefficient of 1, and a film with (300) planes perfectly parallel to the substrate has a texture coefficient of 0. Note that the [001] direction, rather than the [113] direction, corresponds to the $\textit{c}$-axis of the lattice. The reason for using the (113) peak rather than the (002) peak at $\sim 24.9^\circ~2\theta$ for texture analysis is that the (002) reflection is very weak and not detectable in many samples (see Fig.~\ref{fig:xrd} in the main article). Conversely, the (113) reflections are detected in all samples and form a relatively small angle (\SI{18}{\degree}) with the ideal (001) planes.
Considering the two approximations described above, our derived texture coefficient should only be taken as a semi-quantitative estimation, so it is labeled "estimated texture coefficient" in the figures of the main article. Nevertheless, the conclusions drawn in the article on the basis of the texture coefficient are also semi-quantitative. Hence, our approximations do not affect the high-level conclusions on preferential orientation (Fig.~\ref{fig:xrd_extra}(b) of the main article) and direction-dependent conductivity (Fig.~\ref{fig:vacancies}(c) of the main article).

Raman spectra were measured with a Renishaw inVia Raman microscope under laser light excitation at \SI{532}{nm} wavelength with a $100 \times$ objective. Scanning electron microscopy (SEM) images were taken with a Hitachi S-3400N instrument with a field emission gun and \SI{5}{kV} beam voltage.

Sheet resistance was measured in the substrate plane with a collinear four-point probe directly contacting the film. The sheet resistance was derived by multiplying the raw ohmic resistance by $\pi / \ln(2)$, as appropriate for the collinear geometry. The electrical resistivity was derived by multiplying the sheet resistance by the XRF-determined thickness. Temperature-dependent Hall carrier concentration and mobility were measured in the substrate plane with a Lake Shore 8425 DC Hall System using the van der Pauw configuration. The Cu$_{3-x}$P film employed for this measurement was deposited through a shadow mask to obtain a Hall cross shape, and Ti/Pt contacts were evaporated at the edges of the cross. The DC driving current and magnetic fields were \SI{1}{mA} and \SI{2}{T} respectively. The Hall voltage at each temperature was determined as the average of 8 measurements, by reversing sign of the current and of the magnetic field, and by considering two non-equivalent contact geometries. The Seebeck coefficient of an unpatterned sample with about the same electrical resistivity was measured in a custom-built setup using In contacts and four temperature differences in the vicinity of room temperature.

The complex dielectric function was extracted by spectroscopic ellipsometry using a J.A. Woollam M-2000 ellipsometer and three incidence angles. We modeled the system as a glass substrate of known optical functions, a Cu$_{3-x}$P layer of unknown optical functions, and a roughness layer treated with Bruggeman effective medium theory. The optical functions of Cu$_{3-x}$P were represented by a Kramers-Kronig-consistent b-spline function with 0.1 nodes/eV. The samples analyzed in this study were too absorbing and too thick for ellipsometry to yield thickness information (no light reflected from the film/glass interface). Hence, we fixed the Cu$_{3-x}$P thickness to the XRF-measured value in the ellipsometry model. Ellipsometry spectra were fitted with the CompleteEase software (J.A. Woollam). The absorption coefficient was derived from the complex dielectric function using standard optical relations. The absorbance shown in Fig.~\ref{fig:optical}(c) of the main article was derived by measuring transmission $T$ at normal incidence and reflection $R$ at near-normal incidence with a Cary 7000 spectrophotometer. The measurement was performed with an integrating sphere to include the diffuse component of both transmission and reflection. The absorbance $A$ was extracted as

\begin{equation}
A = -\log_{10}\left(\frac{T}{(1-R)}\right)
\end{equation}

\section*{Computational details}

\subsection*{Density functional theory calculations}
First-principles calculations were performed using Density Functional Theory (DFT) within the Projector-Augmented Wave (PAW) formalism~\cite{SIBlochl1994} and a plane-wave basis set as implemented in the GPAW code~\cite{SIMortensen2005,SIEnkovaara2010}, in combination with the Atomic Simulation Environment (ASE)~\cite{SILarsen2017}. The Perdew-Burke-Ernzerhof (PBE) exchange correlation functional~\cite{SIPerdew1996} was employed for structural relaxation. The plane-wave cutoff and \textbf{k}-mesh density were \SI{450}{eV} and 8$\times$8$\times$4, respectively. The structures were relaxed until the forces were less than \SI{0.05}{eV/\angstrom}.
For the calculations on Cu$_{3-x}$P with one Cu vacancy/unit cell, the vacancy was introduced at one of the symmetry-equivalent Cu(1) sites (one of the two inequivalent sites at 6c Wyckoff positions), according to the notation of Olofsson.~\cite{SIOlofsson1972} For the calculations on Cu$_{3-x}$P with half a Cu vacancy/unit cell, a 2$\times$1$\times$1 (48-atom) supercell was constructed with a Cu vacancy at one of the Cu(1) sites.

Ground-state electronic structure calculations were performed with the GLLB-SC exchange correlation functional~\cite{SIKuisma2010,SIGritsenko1995} with a plane-wave cutoff of \SI{450}{eV} and a \textbf{k}-mesh density of 16$\times$16$\times$8.
Kramers-Kronig-consistent dielectric function spectra were calculated by linear response theory within the Random Phase Approximation (RPA) including local field effects, as implemented in GPAW. The absorption coefficient was derived from the dielectric function using standard relations.


\subsection*{Semiclassical transport calculations}
Temperature- and doping density-dependent transport properties of Cu$_{3-x}$P were estimated using Boltzmann transport theory as implemented in BoltzTraP2.~\cite{SIMadsen2018} The input of the calculations was an interpolation of the previously calculated PBE band structure of Cu$_3$P available on the Materials Project database (mp-7463).~\cite{SIJain2013,SIOng2015} The BoltzTraP2 code was interfaced with Materials Project through the pymatgen package.~\cite{SIOng2013,SIRicci2017} A constant carrier scattering time of \SI{10}{fs} was assumed for all bands at all energies.
The effective masses quoted in the main article are the eigenvalues of the conductivity effective mass tensor, calculated at a temperature of \SI{300}{\kelvin} under a net p-type doping density of \SI{3e21}{cm^{-3}}. Assuming a hole mobility $\mu$ of \SI{29}{cm^2/Vs} as measured by Hall effect (Fig.~\ref{fig:electrical}(b), main article), a \SI{10}{fs} scattering time $\tau$ corresponds to a hole effective mass $m^*$ of 0.61~$m_\mathrm{e}$ using the relation $\mu = e \tau / m^*$. This value is close to the calculated effective mass of 0.54~$m_\mathrm{e}$ in the $ab$-plane of doped Cu$_3$P at room temperature. These considerations justify the choice of the carrier scattering time.

The effective mass as a function of texture coefficient $TC$ was estimated by assuming that the $\textit{a}$-axis is always lying on the substrate plane, that the component of the $\textit{b}$-axis vector parallel to the substrate plane is $|b| (TC)$, and that the component of the $\textit{c}$-axis vector parallel to the substrate plane is $|c|(1 - TC)$.
With this description, there are two relevant hole effective masses in the substrate plane (which is the transport plane probed by the resistivity measurement). One is $m^*_{a} = 0.54~m_\mathrm{e}$ and the other is $m^*_{b/c} = (TC) m^*_{b} + (1-TC) m^*_{c}$, where $m^*_{a} = m^*_{b}$ due to symmetry, and $m^*_{c} = 1.28~m_\mathrm{e}$ is the $\textit{c}$-axis effective mass. The overall effective mass in the substrate plane was simply taken as the arithmetic average between $m^*_{a}$ and $m^*_{b/c}$.


\section*{Supplementary discussion}
\subsection*{Cu secondary phases in Cu$_{3-x}$P films}
From the experimental data in Fig.~\ref{fig:composition}(b) of the main article, we note that the range of Cu-rich (Cu/P $>$ 3) compositions free of metallic Cu peaks becomes narrower with increasing PH$_3$ partial pressure. This observation can be explained as follows. A substrate reaction (phosphorization of metallic Cu) is dominant at substrate locations closer to the Cu target. Direct transport of Cu$_{3-x}$P vapor from the target is dominant at substrate locations closer to the Cu$_{3-x}$P target.
As the PH$_3$ partial pressure increases, a given Cu/P ratio in the films is obtained at locations closer and closer to the Cu target. Thus, the fraction of the incoming vapor consisting of metallic Cu rather than already-formed Cu$_{3-x}$P increases. Since an additional reaction at the substrate is necessary to form Cu$_{3-x}$P from metallic Cu, the segregation of Cu as a polycrystalline impurity is more likely.

\subsection*{Interpretation of Hall carrier concentrations in Cu$_{3-x}$P}
Semiclassical transport simulations have revealed that there can be a major discrepancy between the net carrier concentration $(p-n)$ in Cu$_{3-x}$P and the carrier concentration $n_H$ derived from a Hall measurement assuming a single carrier type. The reason is the comparable concentrations of electrons and holes in Cu$_{3-x}$P near its intrinsic Fermi level. The offset between the two quantities is illustrated in Fig.~\ref{fig:boltztrap}(b) in the main article. Because of this discrepancy, the values of $n_H$ as a function of Fermi level predicted by Boltzmann transport theory can be counterintuitive. Here we provide a list of potentially counterintuitive features.

\begin{enumerate}
\item The calculated $|n_H|$ is always above \SI{1.5e21}{cm^{-3}} regardless of the Fermi level position (Fig.~\ref{fig:boltztrap}(b) in the main article). Hence, we expect that a Hall effect measurement on a Cu$_{3-x}$P film will always indicate very high carrier concentrations $n_H$, even in a hypothetical film with a very low net carrier concentration.
\item $n_H$ diverges at the Fermi level when the numerator of $R_H$ approaches zero and changes sign (about \SI{10}{meV} above the intrinsic Fermi level).
\item The Fermi level where $n_H$ changes sign is $\sim$\SI{70}{meV} higher than the Fermi level of zero net carrier concentration. This implies that there is a $\sim$\SI{70}{meV} Fermi level range where Cu$_3$P would appear as p-type from a Hall measurement (positive $n_H$), even though electrons are more abundant than holes. Intrinsic, defect-free Cu$_3$P is one of such cases ($p-n <0$ but $n_H > 0$).
\item If $S_h/S_e \simeq \mu_h/\mu_e$, the numerators of Eq.~\ref{eq:hall} and Eq.~\ref{eq:seebeck} in the main article are zero at about the same net carrier concentration. Thus, intrinsic, defect-free Cu$_3$P would also appear as p-type in a thermovoltage measurement (positive Seebeck coefficient).
\end{enumerate}

\subsection*{Temperature dependence of carrier concentration in Cu$_{3-x}$P films}
The Hall hole concentration of a Cu$_{3-x}$P film decreases by roughly a factor two with decreasing temperature from \SI{300}{K} to \SI{10}{K} (Fig.~\ref{fig:electrical}(a) in the main article). A small decrease in Hall hole concentration with temperature is also predicted by Boltzmann transport theory on Cu$_3$P with a Fermi level at \SI{0.3}{eV} below the intrinsic value (Fig.~\ref{fig:electrical}(a), main article). However, the theoretically predicted decrease is much smaller than the experimental one. Possible reasons for this discrepancy could be: (i) entropy-driven formation of additional Cu vacancies with increasing temperature, as quantified before~\onlinecite{SIDeTrizio2015}; (ii) band structure modifications occurring at such a high defect concentrations; (iii) temperature-driven structural changes accompanied by a rearrangement of Cu vacancies, as previously reported;~\cite{SIWolff2018}
and (iv) a decrease in the Hall scattering factor $r$ with decreasing temperature. When a single type of carrier is dominant, the Hall carrier concentration is defined as $n_H = r/ e R_H$, where $R_H$ is the Hall coefficient, $e$ is the elementary charge, and the Hall scattering factor $r$ is assumed to be equal to 1.~\cite{SIWerner2017} The Hall scattering factor depends on the energy dependence of the carrier scattering time (taken as a constant in our Boltzmann transport calculations). Thus, a decrease in Hall hole concentration by a factor 2 from \SI{300}{K} to \SI{10}{K} could be explained by a decrease in Hall scattering factor in the same temperature range. A decreasing Hall scattering factor with decreasing temperature occurs, for example, in various metals~\cite{SIAlderson1968} and in graphene.~\cite{SIMacheda2020}

\subsection*{Mobility of Cu$_{3-x}$P films in comparison with other thin-film materials with similar doping levels}
The room-temperature mobility of a Cu$_{3-x}$P film (\SI{28.8}{cm^{2}\per Vs}) is not unusual for non-epitaxial polycrystalline thin-film materials with carrier concentrations above \SI{e21}{\cm^{-3}}, such as transparent conductive oxides and elemental metals.
However, the mobility of \SI{276}{cm^{2}\per Vs} measured at \SI{10}{K} is particularly high. Due to the high hole concentration in Cu$_{3-x}$P, we expect ionized impurity scattering to dominate over grain boundary scattering. We estimated the ionized impurity scattering-limited mobility as $\mu_\mathrm{i} = \SI{279}{cm^{2}\per Vs}$ by fitting temperature-dependent mobility data (Fig.~\ref{fig:electrical}(b) of the main article). Here, we compile values of the Hall mobility measured in non-epitaxial polycrystalline thin-film materials with carrier concentrations comparable to Cu$_{3-x}$P.

In the chemically-related compound CaCuP -- a p-type semiconductor degenerately doped by Cu vacancies and also deposited by RF reactive sputtering -- $\mu_\mathrm{i}$ was only \SI{45.2}{cm^{2}\per Vs}.~\cite{SIWillis2022} This is despite the fact that the hole concentration of CaCuP was over an order of magnitude lower than in Cu$_{3-x}$P~ and that its room-temperature mobility was slightly higher (\SI{36.4}{cm^{2}\per Vs}).

In the heavily-doped (n-type) transparent conductive oxides In$_2$O$_3$:Sn (ITO) and ZnO:Al (AZO), $\mu_\mathrm{i}$ was in the \SIrange{10}{50}{cm^{2}\per Vs} range, even though their carrier concentration was lower, in the \SI{e20}{\cm^{-3}} range.~\cite{SIEllmer2008a}

Non-epitaxial, n-type films of the doped narrow-gap semiconductors PbTe  and Bi$_2$Te$_3$ deposited by RF sputtering had mobilities of \SIrange{15}{40}{cm^{2}\per Vs} at room temperature, and these mobilities decreased with decreasing temperature.

The low-temperature mobility of Cu$_{3-x}$P films is even higher than the corresponding mobility of non-epitaxial polycrystalline films of elemental metals. For example, mobilities around \SI{100}{cm^{2}\per Vs} were reported for evaporated Cu and Au films on glass at \SI{77}{K} measurement temperature.~\cite{SIChopra1967} The mobility of Cu$_{3-x}$P films at the same temperature is about \SI{180}{cm^{2}\per Vs} (Fig.~\ref{fig:electrical}(b) of the main article). The mobility measurements on metals were conducted on sufficiently thick films ($>$\SI{100}{nm}), in which the mobility was not negatively affected by the limited thickness. Note that Cu and Au have more than an order of magnitude higher carrier concentrations than Cu$_{3-x}$P, but their carriers are intrinsic instead of being provided by defects.

We emphasize, however, that both elemental metals and doped narrow-gap semiconductors can have much higher low-temperature mobilities (well above \SI{e3}{cm^{2}\per Vs}) if grown epitaxially or as single crystals.

\subsection*{Possible causes of composition-independent resistivity of Cu$_{3-x}$P films near Cu/P = 3}
The electrical resistivity and the overall composition of Cu$_{3-x}$P films in the $2.95 <$ Cu/P $ < 3.05$ range are completely uncorrelated (Fig.~\ref{fig:resistivity}(c), main article). The conceptually simplest explanation for a composition-independent resistivity is the existence of metallic Cu secondary phases for Cu/P $> 2.75$, so even in highly P-rich films. This hypothesis can be summarized as follows: (i) the films consist of a Cu$_{3-x}$P phase and a Cu phase regardless of the Cu/P ratio; (ii) the point defects present in the Cu$_{3-x}$P phase are Cu vacancies; (iii) the composition of the Cu$_{3-x}$P phase is dictated by the V$_\mathrm{Cu}$ concentration but \textit{not} by the overall Cu/P ratio, so for example when 1 V$_\mathrm{Cu}$/unit cell, the composition of the Cu$_{3-x}$P phase is Cu$_{2.83}$P; (iv) the observed variations in the \textit{overall} composition are caused by varying concentrations of metallic Cu phases depending on process conditions.

\begin{figure}[h!]
\centering%
\includegraphics[width=0.8\textwidth]{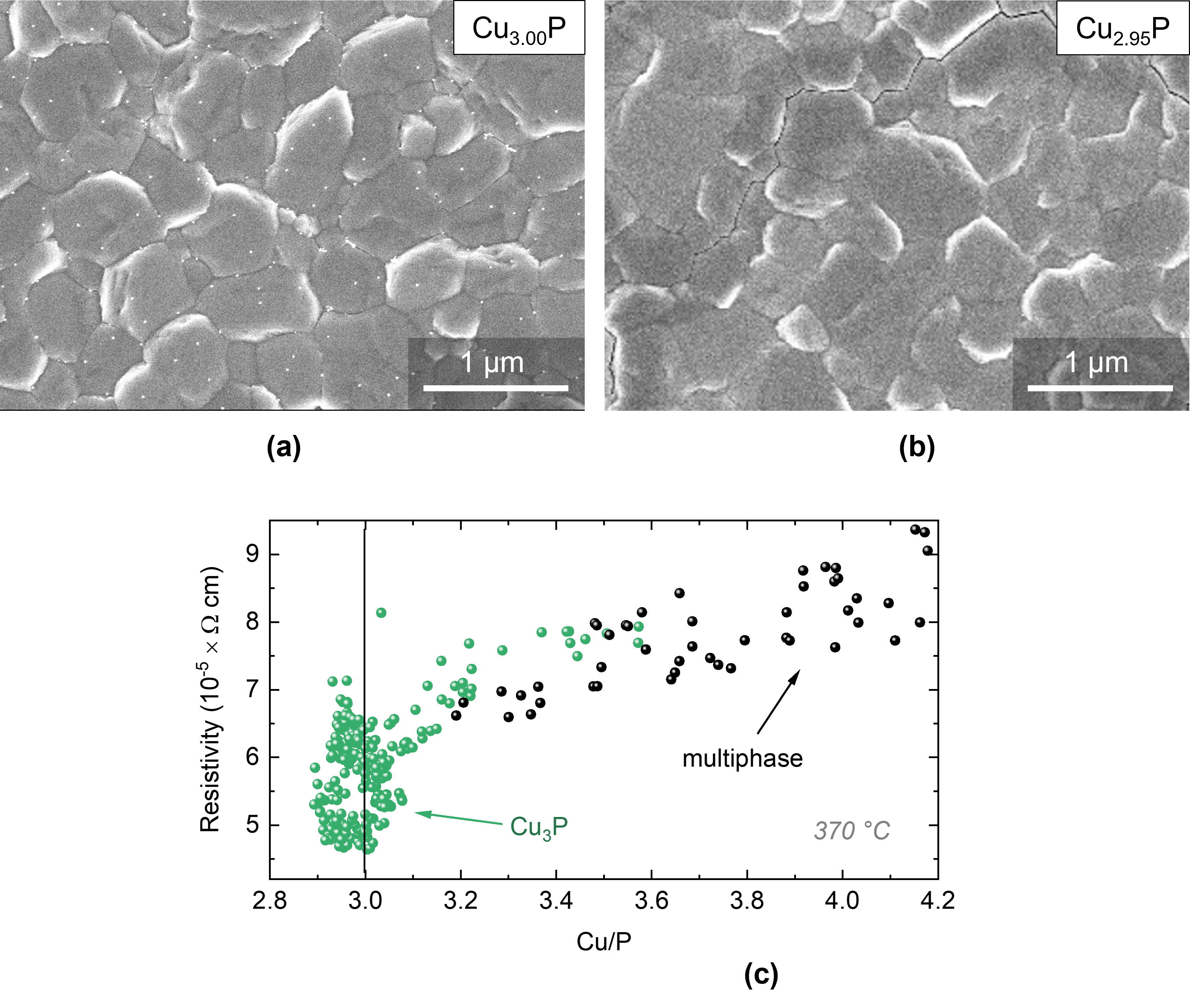}
\caption{\textbf{(a):} SEM image of a film with overall Cu$_{3.00}$P composition, where bright secondary phases are observed throughout the film surface. The diameter of the secondary phases is roughly \SI{10}{nm}. The area coverage of the secondary phases is roughly 0.6\%. Since metallic Cu is detected by XRD at slightly Cu-richer compositions, the phases observed in the SEM are also likely to consist of metallic Cu. This is compatible with their brighter appearance in the SEM, usually a sign of higher average atomic number. \textbf{(b):} SEM image of a film with overall Cu$_{2.95}$P composition, where bright secondary phases are \text{not} observed. The films in (a) and (b) were deposited at \SI{530}{\celsius}. \textbf{(c):} Composition-dependent resistivity of Cu$_{3-x}$P films deposited at \SI{370}{\celsius}. Samples in which no polycrystalline secondary phases were detected by XRD are shown in green. Samples which exhibited both Cu$_{3-x}$P peaks and metallic Cu peaks in their XRD patterns are shown in black. Note that the presence or absence of polycrystalline Cu does not significantly affect the resistivity of films in the $3.2 < \mathrm{Cu/P} < 3.6$ range. This finding confirms that amorphous Cu is very likely to exist in the $3.2 < \mathrm{Cu/P} < 3.6$ range when polycrystalline Cu is not detected.}
\label{fig:sem_phases}
\end{figure}

A problem with this hypothesis is that we have indeed observed metallic Cu by SEM, but only in films with Cu/P $ > 3.0$ (Fig.~\ref{fig:sem_phases}).
Another problem is that a film with overall Cu$_{3.00}$P composition doped with 1.5 V$_\mathrm{Cu}$/unit cell would need to contain 6.5\% metallic Cu by volume, if Cu secondary phases were fully responsible for deviations in the expected stoichiometry. However, the fraction of the top surface of a Cu$_{3.00}$P film that is covered by secondary phases visible in the SEM can be estimated as being only 0.6\% (Fig.~\ref{fig:sem_phases}(b)). Unless metallic Cu preferentially segregates at the bottom of the film, it seems unlikely that all the Cu in excess of the expected Cu$_{3-x}$P composition exists in the form of a secondary phase.

%

A second possible reason for a composition-independent resistivity near Cu/P = 3 is the existence of additional point defects beyond Cu vacancies. Different types of defects could be plausible candidates for explaining the composition-independent resistivity -- either extrinsic impurities or native defects, and either compensating donors or charge-neutral defects.

\subsubsection*{Native defects}
The Cu$_\mathrm{P}$ antisite could be a possible candidate. In the related material CaCuP, defect calculations have shown that Cu$_\mathrm{P}$ can be in the singly or doubly ionized state (donor), or in the neutral state.~\cite{Willis2022} If Cu$_\mathrm{P}$ exists in the singly ionized state in Cu$_{3-x}$P, it may compensate the formation of additional V$_\mathrm{Cu}$ acceptors beyond a certain V$_\mathrm{Cu}$ concentration threshold. Formation of one Cu$_\mathrm{P}$ donor for each V$_\mathrm{Cu}$ acceptor would then keep the resistivity constant, and it would increase the overall Cu/P ratio with increasing defect compensation (one less P atom per pair of compensating defects). Interestingly, in this scenario the Cu/P ratio would \textit{increase} with increasing V$_\mathrm{Cu}$ concentration at high defect concentrations.

Even if Cu$_\mathrm{P}$ is in the neutral state rather than in the ionized state, it could still be a possible candidate for a composition-independent resistivity. With Cu$_\mathrm{P}$ in the neutral state, an increasing concentration of Cu$_\mathrm{P}$ defects would increase the overall Cu/P ratio (one extra Cu atom and one less P atom per defect) at constant V$_\mathrm{Cu}$ concentration (constant resistivity).

V$_\mathrm{P}$ donors could, in principle, be another possible compensating defect that would drive the overall composition towards higher Cu/P ratios. However, compensating V$_\mathrm{P}$ donors would likely result in decreasing unit cell volume with increasing compensation. This would lead to more scattering in the conductivity versus lattice constant data (Fig.~\ref{fig:vacancies}(a)) at high conductivities, which is not observed. Cu interstitials (Cu$_\mathrm{i}$) in the neutral state could also modulate the Cu/P ratio at constant V$_\mathrm{Cu}$ concentration. However, increasing Cu$_\mathrm{i}$ concentrations are likely to lead to increasing unit cell volumes. Thus, neither V$_\mathrm{P}$ donors nor Cu$_\mathrm{i}$ neutrals are likely to cause the composition-independent resistivity of Cu$_{3-x}$P films near the stoichiometric point.

\subsubsection*{Extrinsic defects}
No impurities heavier than Na could be detected in XRF spectra of Cu$_{3-x}$P films. Among lighter elements, the only two impurities that may be present in Cu$_{3-x}$P films at atomic concentrations above 1\% are oxygen (from imperfect vacuum in the deposition chamber) and hydrogen (from PH$_3$ decomposition). A small oxygen peak was detected in energy-dispersive x-ray (EDX) spectra of most of our Cu$_{3-x}$P samples, with overall O concentration estimated at below 2 at.\%. Hydrogen content in Cu$_{3-x}$P films could not be measured with the techniques available to us. A previous study on Zn$_3$P$_2$ films deposited by reactive sputtering at \SI{150}{\celsius} in a PH$_3$-containing atmosphere reported a H content of 4~at.\% using secondary ion mass spectrometry (SIMS).~\cite{SIWeber1998} We expect a lower -- but not insignificant -- hydrogen content in our Cu$_{3-x}$P films due to the higher deposition temperature (\SI{370}{\celsius}), which usually facilitates hydrogen desorption.

Hence, both H and O are likely to be present at low but not negligible concentrations in most our Cu$_{3-x}$P films.
In the case that H and O are incorporated in Cu$_{3-x}$P as point defects, we hypothesize that they are most likely to form interstitials (H$_\mathrm{i}$ and O$_\mathrm{i}$). The reason is their small size compared to both Cu and P, as estimated by their covalent radii.~\cite{SIPyykko2009} As an exception, the H$_\mathrm{Cu}$ antisite may also be energetically favorable, due to the high availability of vacant Cu sites in Cu$_{3-x}$P.

H$_\mathrm{i}$ is typically found to be a compensating donor in p-type materials.~\cite{SIvandewalleHydrogenSemiconductors2006} Thus, it is not implausible that H$_\mathrm{i}$ donors may compensate V$_\mathrm{Cu}$ acceptors above a certain V$_\mathrm{Cu}$ concentration threshold. In this scenario, the decrease in Cu/P ratio from 3.05 to 2.90 in Fig.~\ref{fig:resistivity}(c) of the main article may be due to an increase in V$_\mathrm{Cu}$ concentration, which is however compensated by H$_\mathrm{i}$ formation leading to a constant resistivity. In this case, the decrease in unit cell size by formation of additional V$_\mathrm{Cu}$ defects may be counteracted by an increase in unit cell size due to an equal concentration of interstitials. A change in Cu/P ratio from 3.05 to 2.90 due to increasing V$_\mathrm{Cu}$ concentration would require H concentrations in the film around 4~at.\%, assuming that all H impurities act as H$_\mathrm{i}$ defects. As mentioned above, it cannot be excluded that our films have a H impurity concentration in this range. O$_\mathrm{i}$ is rarely found to be a donor in other materials, so it is less likely to explain our experimental results.

If H$_\mathrm{Cu}$ is a charge-neutral defect (as may be expected from oxidation state arguments), changes in H$_\mathrm{Cu}$ concentration at constant V$_\mathrm{Cu}$ concentration could also explain the composition-independent resistivity close to Cu/P = 3. In this scenario, the V$_\mathrm{Cu}$ concentration would be constant as a function of overall film composition (keeping the resistivity constant), but the Cu/P would decrease as the concentration of H$_\mathrm{Cu}$ defects increases.

\clearpage

\section*{Supplementary figures}

\vspace{1cm}

\begin{figure}[h!]
\centering%
\includegraphics[width=0.7\textwidth]{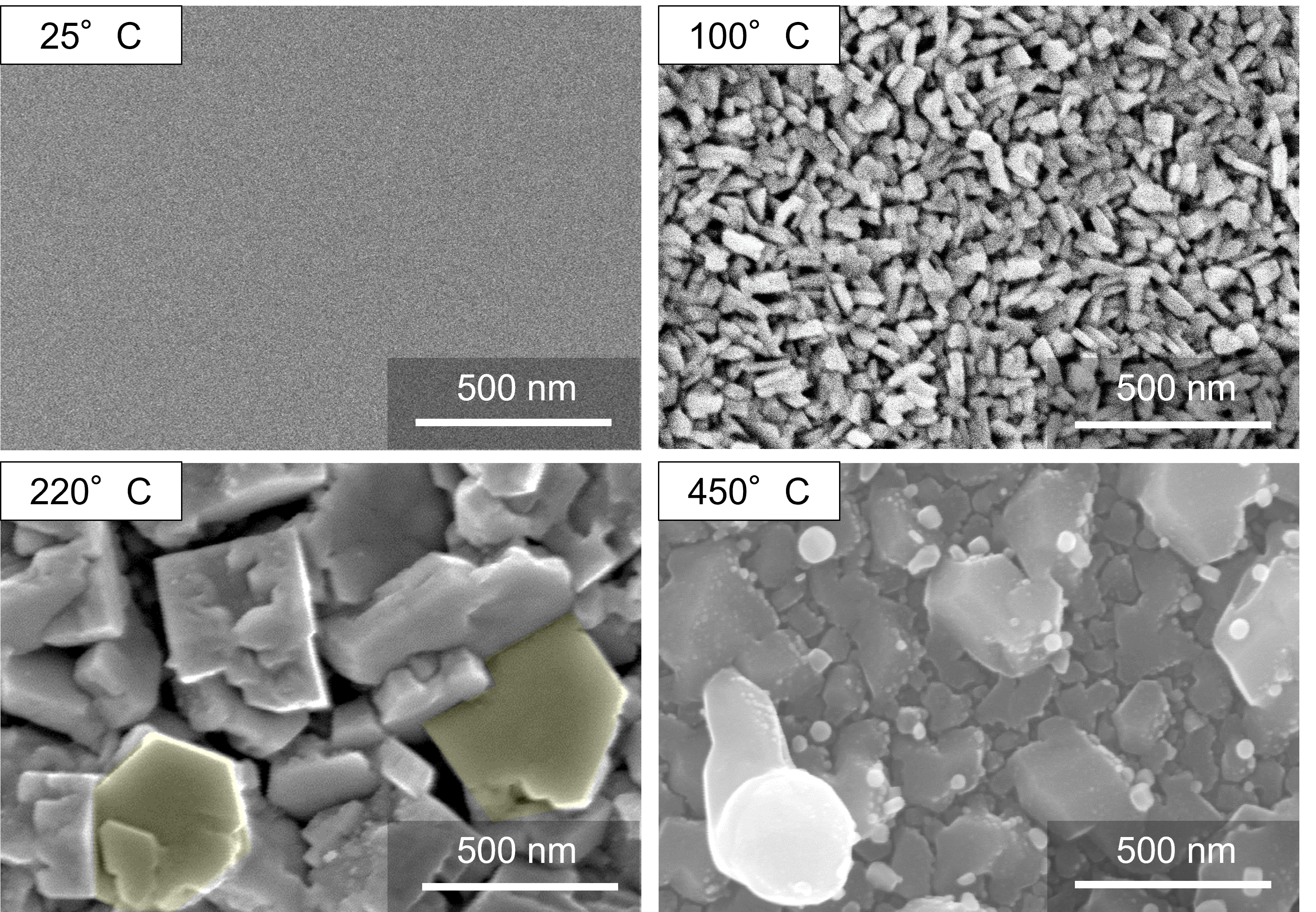}
\caption{SEM images of Cu$_{3-x}$P films sputter-deposited in pure Ar at different substrate temperatures. The films deposited at \SI{25}{\celsius}, \SI{100}{\celsius}, and \SI{220}{\celsius} have composition close to Cu$_3$P. The film deposited at \SI{450}{\celsius} has composition Cu$_{6.45}$P due to phosphorus evaporation from the growing film. Most crystal grains in the film deposited at \SI{220}{\celsius} have hexagonal shapes (highlighted in yellow), in line with the hexagonal symmetry of the P6$_3$cm crystal structure of Cu$_{3-x}$P.}
\label{fig:sem_ar}
\end{figure}

\begin{figure}[h!]
\centering%
\includegraphics[width=0.6\textwidth]{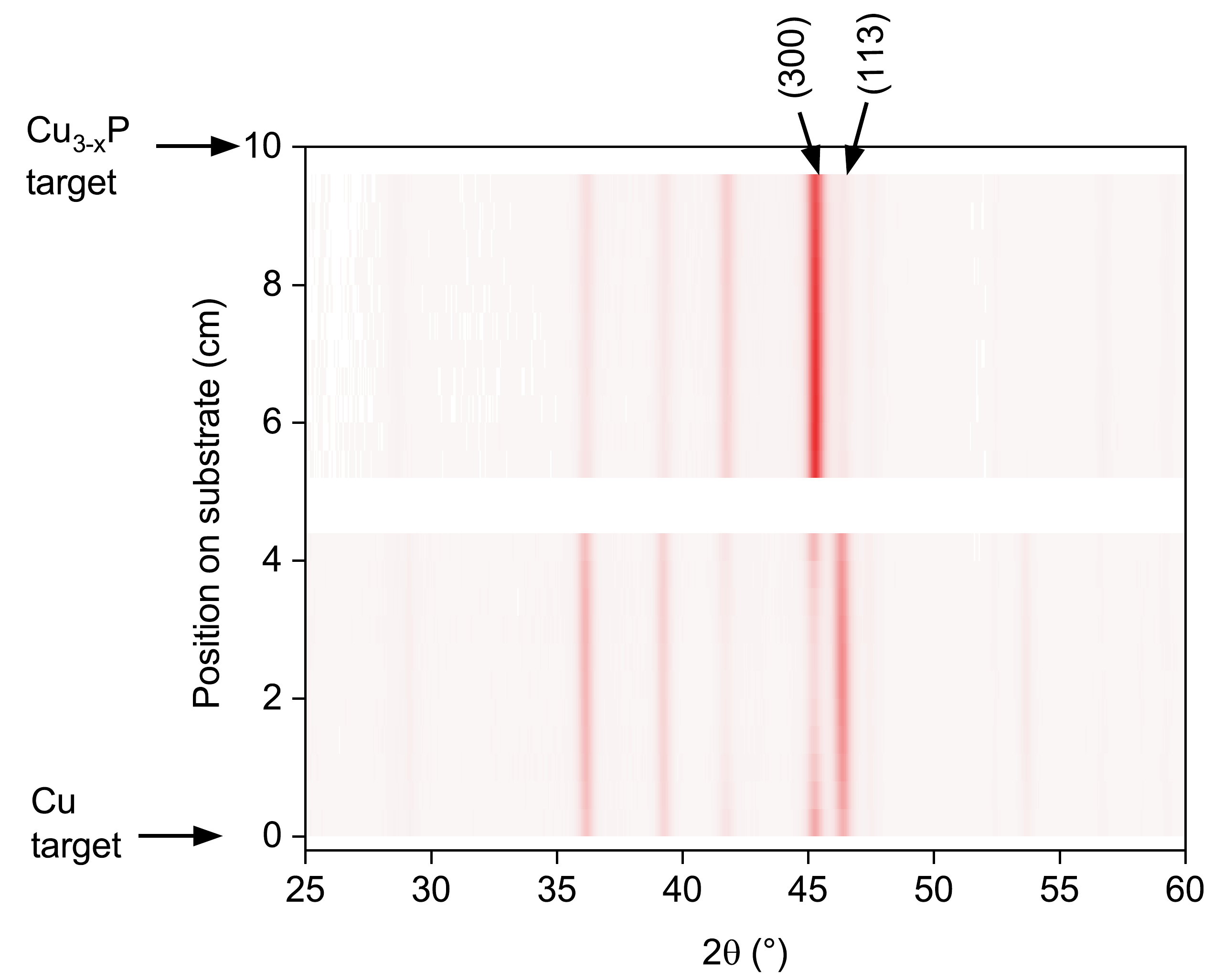}
\caption{Color map representing XRD intensity from a combinatorial Cu$_{3-x}$P film deposited at \SI{220}{\celsius} as a function of position on the substrate with respect to the two sputter targets. At \SI{0}{cm} position, the substrate was mainly coated by the metallic Cu target. At \SI{10}{cm} position, the substrate was mainly coated by the Cu$_{3-x}$P target. This dataset was used for orientation-dependent resistivity measurements because of its wide range of texture coefficients, arising from the different intensity ratios between the (300) peak and the (113) peak at different positions.}
\label{fig:xrd_220c}
\end{figure}

\begin{figure}[h!]
\centering%
\includegraphics[width=0.7\textwidth]{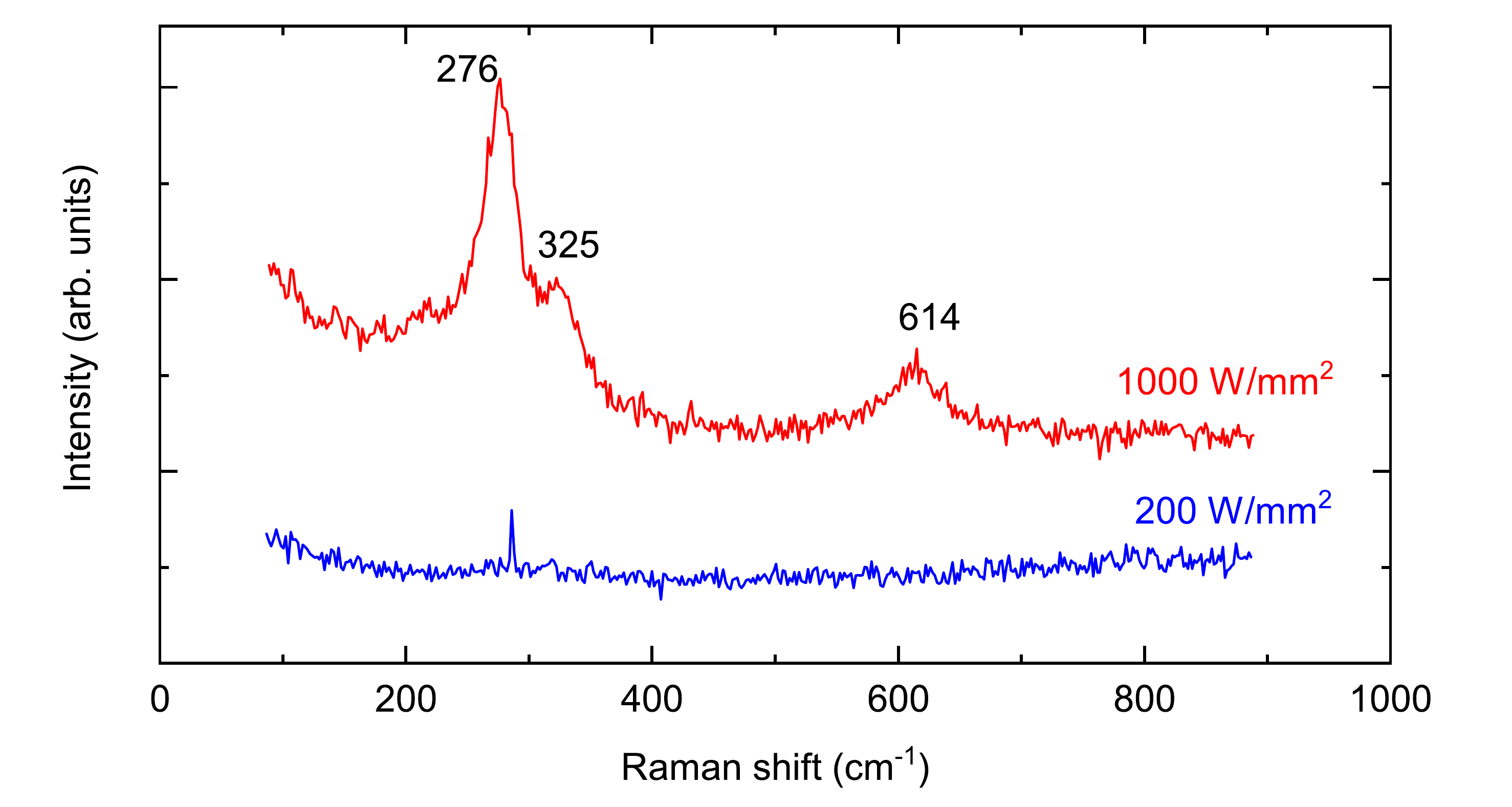}
\caption{Raman spectra of a Cu$_{2.95}$P film deposited at \SI{370}{\celsius}. No peaks are observed at an excitation intensity of \SI{200}{W/mm^{2}}, which is very high for Raman spectroscopy, but still slightly below the ablation threshold of the film. The absence of Raman peaks in Cu$_{3-x}$P is compatible with some of the existing literature.~\cite{SIDeTrizio2015,SIHou2016}
Three peaks compatible with other previous studies~\cite{SILiu2016d,SIPeng2021,SIPeng2022} appear at an excitation intensity of \SI{1000}{W/mm^{2}}. However, this extremely high excitation intensity leads to film ablation, as clearly seen by optical microscopy after the Raman measurement.}
\label{fig:raman}
\end{figure}

\begin{figure}[h!]
\centering%
\includegraphics[width=0.6\textwidth]{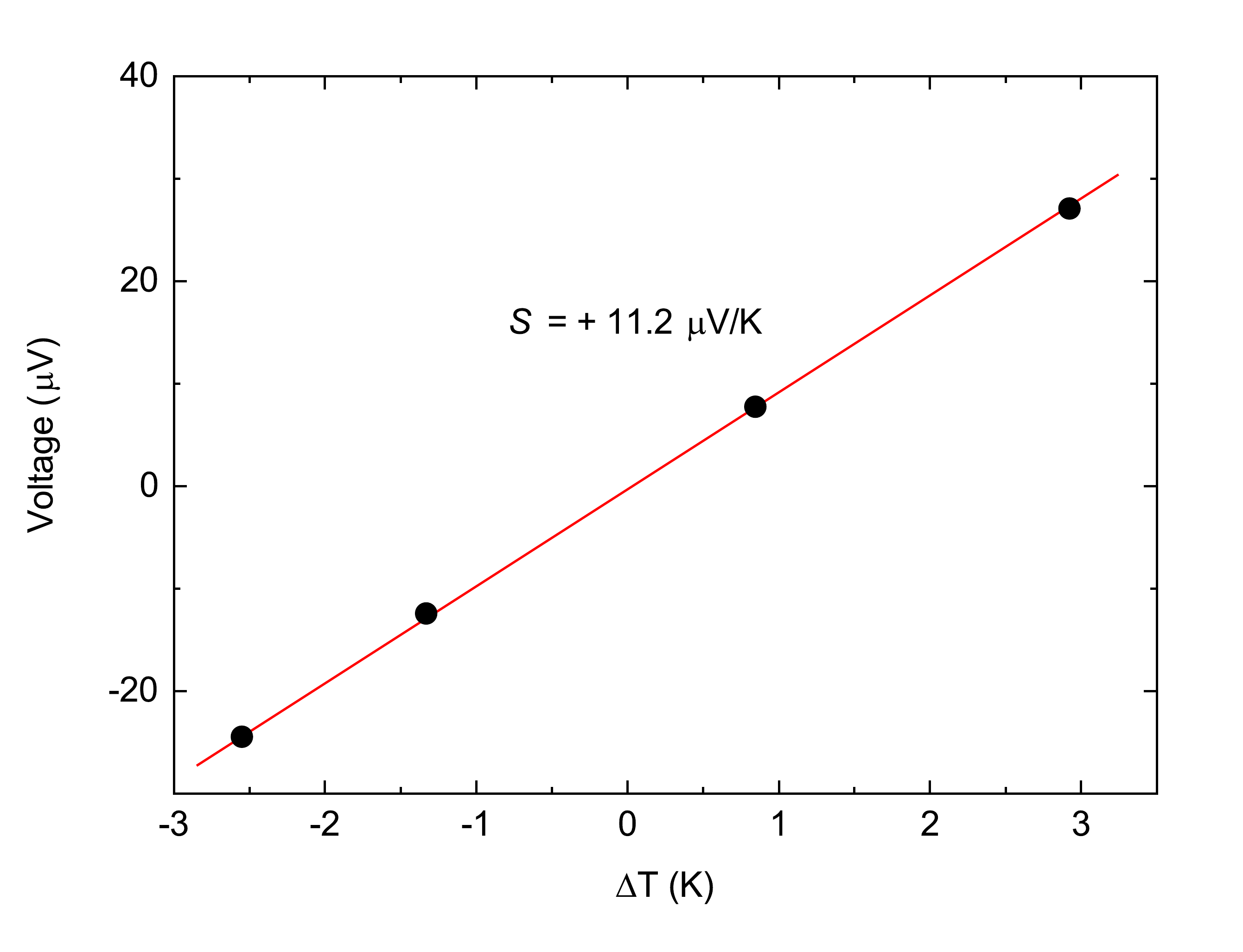}
\caption{Thermovoltage measurements on a Cu$_{2.95}$P film grown in the same way as the sample used for Hall effect measurements (Fig.~\ref{fig:electrical} of the main article) and with similar electrical resistivity. Two corners of the film were contacted with In wire and one contact was heated to create a temperature gradient. The temperature difference and the voltage between the two contacts were measured after temperature stabilization. The Seebeck coefficient $S$ of Cu$_{2.95}$P is extracted as the slope of the thermovoltage versus temperature difference curve, plus the known Seebeck coefficient of In. The positive sign of S confirms that holes are majority charge carriers.}
\label{fig:seebeck}
\end{figure}


\begin{figure}[h!]
\centering%
\includegraphics[width=0.6\textwidth]{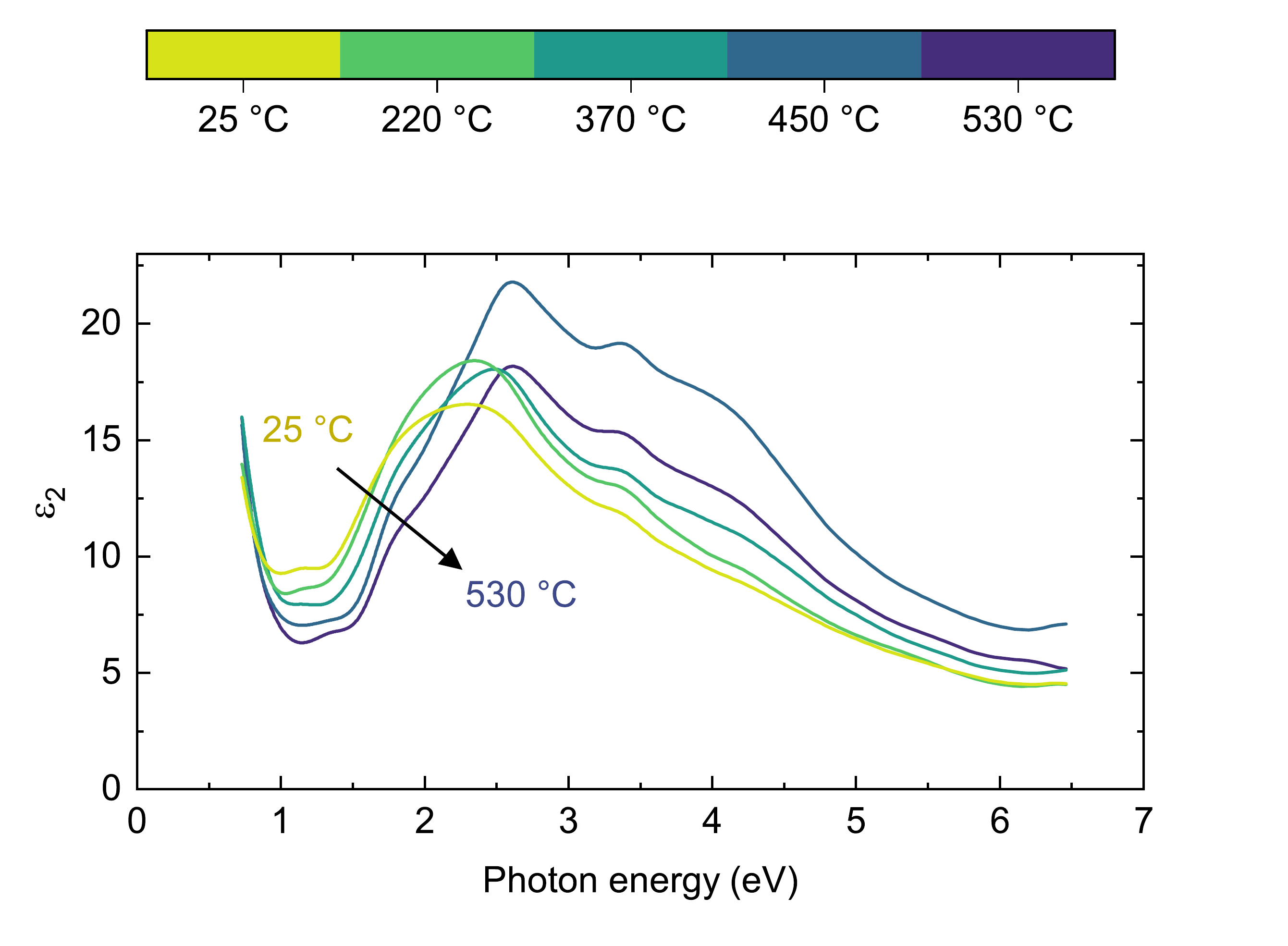}
\caption{Ellipsometry-measured imaginary part of the dielectric function ($\epsilon_2$) for the lowest-resistivity sample at each deposition temperature. The position of the maximum occurring between \SI{2}{eV} and \SI{3}{eV} photon energy is used to plot Fig.~\ref{fig:optical}(d) in the main article. The blue-shift of this maximum, and the blue-shift of the onset of the increase in $\epsilon_2$ around \SI{1.5}{eV}, are indications of a Burstein-Moss effect caused by the increasing hole concentration with increasing deposition temperature.}
\label{fig:epsilon_2}
\end{figure}

%


\clearpage
\section*{Supplementary references}

\end{document}